\documentclass[traditabstract,longauth]{aa}
\usepackage[varg]{txfonts}
\makeatletter
\def\input@path{{./}{SSO4ken/}}
\makeatother 

\usepackage{natbib}
\bibpunct{(}{)}{;}{a}{}{,} 

\usepackage{graphicx}
\usepackage{txfonts}

\usepackage[colorlinks,citecolor=blue,urlcolor=blue,linkcolor=blue]{hyperref} 
\usepackage{multirow}           
\usepackage{units}
\usepackage{fixltx2e}
    
\graphicspath{{./SSO4ken/}{./}} 

\def\setsymbol#1#2{\expandafter\def\csname #1\endcsname{#2}}
\def\getsymbol#1{\csname #1\endcsname}

\def\Planck{\textit{Planck}}

\def\all2013resultspapers{\nocite{planck2013-p01, planck2013-p02, planck2013-p02a, planck2013-p02d, planck2013-p02b, planck2013-p03, planck2013-p03c, planck2013-p03f, planck2013-p03d, planck2013-p03e, planck2013-p01a, planck2013-p06, planck2013-p03a, planck2013-pip88, planck2013-p08, planck2013-p11, planck2013-p12, planck2013-p13, planck2013-p14, planck2013-p15, planck2013-p05b, planck2013-p17, planck2013-p09, planck2013-p09a, planck2013-p20, planck2013-p19, planck2013-pipaberration, planck2013-p05, planck2013-p05a, planck2013-pip56, planck2013-p06b}}

\newbox\tablebox    \newdimen\tablewidth
\def\leaderfil{\leaders\hbox to 5pt{\hss.\hss}\hfil}
\def\endPlancktablewide{\tablewidth=\textwidth 
    $$\hss\copy\tablebox\hss$$
    \vskip-\lastskip\vskip -2pt}
\def\tablenote#1 #2\par{\begingroup \parindent=0.8em
    \abovedisplayshortskip=0pt\belowdisplayshortskip=0pt
    \noindent
    $$\hss\vbox{\hsize\tablewidth \hangindent=\parindent \hangafter=1 \noindent
    \hbox to \parindent{$^#1$\hss}\strut#2\strut\par}\hss$$
    \endgroup}
\def\doubleline{\vskip 3pt\hrule \vskip 1.5pt \hrule \vskip 5pt}

\def\L2{\ifmmode L_2\else $L_2$\fi}

\def\DeltaT{\ifmmode \Delta T\else $\Delta T$\fi}
\def\deltat{\ifmmode \Delta t\else $\Delta t$\fi}
\def\fknee{\ifmmode f_{\rm knee}\else $f_{\rm knee}$\fi}
\def\Fmax{\ifmmode F_{\rm max}\else $F_{\rm max}$\fi}
\def\solar{\ifmmode{\rm M}_{\mathord\odot}\else${\rm M}_{\mathord\odot}$\fi}
\def\Msolar{\ifmmode{\rm M}_{\mathord\odot}\else${\rm M}_{\mathord\odot}$\fi}
\def\Lsolar{\ifmmode{\rm L}_{\mathord\odot}\else${\rm L}_{\mathord\odot}$\fi}

\def\inv{\ifmmode^{-1}\else$^{-1}$\fi}
\def\mo{\ifmmode^{-1}\else$^{-1}$\fi}
\def\sup#1{\ifmmode ^{\rm #1}\else $^{\rm #1}$\fi}
\def\expo#1{\ifmmode \times 10^{#1}\else $\times 10^{#1}$\fi}
\def\,{\thinspace}
\def\lsim{\mathrel{\raise .4ex\hbox{\rlap{$<$}\lower 1.2ex\hbox{$\sim$}}}}
\def\gsim{\mathrel{\raise .4ex\hbox{\rlap{$>$}\lower 1.2ex\hbox{$\sim$}}}}

\def\simprop{\mathrel{\raise .4ex\hbox{\rlap{$\propto$}\lower 1.2ex\hbox{$\sim$}}}}
\def\deg{\ifmmode^\circ\else$^\circ$\fi}
\def\pdeg{\ifmmode $\setbox0=\hbox{$^{\circ}$}\rlap{\hskip.11\wd0 .}$^{\circ}
          \else \setbox0=\hbox{$^{\circ}$}\rlap{\hskip.11\wd0 .}$^{\circ}$\fi}
\def\arcs{\ifmmode {^{\scriptstyle\prime\prime}}
          \else $^{\scriptstyle\prime\prime}$\fi}
\def\arcm{\ifmmode {^{\scriptstyle\prime}}
          \else $^{\scriptstyle\prime}$\fi}
\newdimen\sa  \newdimen\sb
\def\parcs{\sa=.07em \sb=.03em
     \ifmmode \hbox{\rlap{.}}^{\scriptstyle\prime\kern -\sb\prime}\hbox{\kern -\sa}
     \else \rlap{.}$^{\scriptstyle\prime\kern -\sb\prime}$\kern -\sa\fi}
\def\parcm{\sa=.08em \sb=.03em
     \ifmmode \hbox{\rlap{.}\kern\sa}^{\scriptstyle\prime}\hbox{\kern-\sb}
     \else \rlap{.}\kern\sa$^{\scriptstyle\prime}$\kern-\sb\fi}
\def\ra[#1 #2 #3.#4]{#1\sup{h}#2\sup{m}#3\sup{s}\llap.#4}
\def\dec[#1 #2 #3.#4]{#1\deg#2\arcm#3\arcs\llap.#4}
\def\deco[#1 #2 #3]{#1\deg#2\arcm#3\arcs}
\def\rra[#1 #2]{#1\sup{h}#2\sup{m}}

\def\dots{\relax\ifmmode \ldots\else $\ldots$\fi}
\def\WHzsr{\ifmmode $W\,Hz\mo\,sr\mo$\else W\,Hz\mo\,sr\mo\fi}
\def\mHz{\ifmmode $\,mHz$\else \,mHz\fi}
\def\GHz{\ifmmode $\,GHz$\else \,GHz\fi}
\def\mKs{\ifmmode $\,mK\,s$^{1/2}\else \,mK\,s$^{1/2}$\fi}
\def\muKs{\ifmmode \,\mu$K\,s$^{1/2}\else \,$\mu$K\,s$^{1/2}$\fi}
\def\muKRJs{\ifmmode \,\mu$K$_{\rm RJ}$\,s$^{1/2}\else \,$\mu$K$_{\rm RJ}$\,s$^{1/2}$\fi}
\def\muKHz{\ifmmode \,\mu$K\,Hz$^{-1/2}\else \,$\mu$K\,Hz$^{-1/2}$\fi}
\def\MJysr{\ifmmode \,$MJy\,sr\mo$\else \,MJy\,sr\mo\fi}
\def\MJysrmK{\ifmmode \,$MJy\,sr\mo$\,mK$_{\rm CMB}\mo\else \,MJy\,sr\mo\,mK$_{\rm CMB}\mo$\fi}
\def\microns{\ifmmode \,\mu$m$\else \,$\mu$m\fi}
\def\micron{\microns}
\def\muK{\ifmmode \,\mu$K$\else \,$\mu$\hbox{K}\fi}
\def\microK{\ifmmode \,\mu$K$\else \,$\mu$\hbox{K}\fi}
\def\muW{\ifmmode \,\mu$W$\else \,$\mu$\hbox{W}\fi}
\def\kms{\ifmmode $\,km\,s$^{-1}\else \,km\,s$^{-1}$\fi}
\def\kmsMpc{\ifmmode $\,\kms\,Mpc\mo$\else \,\kms\,Mpc\mo\fi}
\setsymbol{LFI:center:frequency:70GHz:units}{70.3\,GHz}
\setsymbol{LFI:center:frequency:44GHz:units}{44.1\,GHz}
\setsymbol{LFI:center:frequency:30GHz:units}{28.5\,GHz}

\setsymbol{LFI:center:frequency:70GHz}{70.3}
\setsymbol{LFI:center:frequency:44GHz}{44.1}
\setsymbol{LFI:center:frequency:30GHz}{28.5}

\setsymbol{LFI:center:frequency:LFI18:Rad:M:units}{71.7\GHz}
\setsymbol{LFI:center:frequency:LFI19:Rad:M:units}{67.5\GHz}
\setsymbol{LFI:center:frequency:LFI20:Rad:M:units}{69.2\GHz}
\setsymbol{LFI:center:frequency:LFI21:Rad:M:units}{70.4\GHz}
\setsymbol{LFI:center:frequency:LFI22:Rad:M:units}{71.5\GHz}
\setsymbol{LFI:center:frequency:LFI23:Rad:M:units}{70.8\GHz}
\setsymbol{LFI:center:frequency:LFI24:Rad:M:units}{44.4\GHz}
\setsymbol{LFI:center:frequency:LFI25:Rad:M:units}{44.0\GHz}
\setsymbol{LFI:center:frequency:LFI26:Rad:M:units}{43.9\GHz}
\setsymbol{LFI:center:frequency:LFI27:Rad:M:units}{28.3\GHz}
\setsymbol{LFI:center:frequency:LFI28:Rad:M:units}{28.8\GHz}
\setsymbol{LFI:center:frequency:LFI18:Rad:S:units}{70.1\GHz}
\setsymbol{LFI:center:frequency:LFI19:Rad:S:units}{69.6\GHz}
\setsymbol{LFI:center:frequency:LFI20:Rad:S:units}{69.5\GHz}
\setsymbol{LFI:center:frequency:LFI21:Rad:S:units}{69.5\GHz}
\setsymbol{LFI:center:frequency:LFI22:Rad:S:units}{72.8\GHz}
\setsymbol{LFI:center:frequency:LFI23:Rad:S:units}{71.3\GHz}
\setsymbol{LFI:center:frequency:LFI24:Rad:S:units}{44.1\GHz}
\setsymbol{LFI:center:frequency:LFI25:Rad:S:units}{44.1\GHz}
\setsymbol{LFI:center:frequency:LFI26:Rad:S:units}{44.1\GHz}
\setsymbol{LFI:center:frequency:LFI27:Rad:S:units}{28.5\GHz}
\setsymbol{LFI:center:frequency:LFI28:Rad:S:units}{28.2\GHz}

\setsymbol{LFI:center:frequency:LFI18:Rad:M}{71.7}
\setsymbol{LFI:center:frequency:LFI19:Rad:M}{67.5}
\setsymbol{LFI:center:frequency:LFI20:Rad:M}{69.2}
\setsymbol{LFI:center:frequency:LFI21:Rad:M}{70.4}
\setsymbol{LFI:center:frequency:LFI22:Rad:M}{71.5}
\setsymbol{LFI:center:frequency:LFI23:Rad:M}{70.8}
\setsymbol{LFI:center:frequency:LFI24:Rad:M}{44.4}
\setsymbol{LFI:center:frequency:LFI25:Rad:M}{44.0}
\setsymbol{LFI:center:frequency:LFI26:Rad:M}{43.9}
\setsymbol{LFI:center:frequency:LFI27:Rad:M}{28.3}
\setsymbol{LFI:center:frequency:LFI28:Rad:M}{28.8}
\setsymbol{LFI:center:frequency:LFI18:Rad:S}{70.1}
\setsymbol{LFI:center:frequency:LFI19:Rad:S}{69.6}
\setsymbol{LFI:center:frequency:LFI20:Rad:S}{69.5}
\setsymbol{LFI:center:frequency:LFI21:Rad:S}{69.5}
\setsymbol{LFI:center:frequency:LFI22:Rad:S}{72.8}
\setsymbol{LFI:center:frequency:LFI23:Rad:S}{71.3}
\setsymbol{LFI:center:frequency:LFI24:Rad:S}{44.1}
\setsymbol{LFI:center:frequency:LFI25:Rad:S}{44.1}
\setsymbol{LFI:center:frequency:LFI26:Rad:S}{44.1}
\setsymbol{LFI:center:frequency:LFI27:Rad:S}{28.5}
\setsymbol{LFI:center:frequency:LFI28:Rad:S}{28.2}

\setsymbol{LFI:white:noise:sensitivity:70GHz:units}{134.7\muKs}
\setsymbol{LFI:white:noise:sensitivity:44GHz:units}{164.7\muKs}
\setsymbol{LFI:white:noise:sensitivity:30GHz:units}{143.4\muKs}

\setsymbol{LFI:white:noise:sensitivity:70GHz}{134.7}
\setsymbol{LFI:white:noise:sensitivity:44GHz}{164.7}
\setsymbol{LFI:white:noise:sensitivity:30GHz}{143.4}

\setsymbol{LFI:white:noise:sensitivity:LFI18:Rad:M:units}{512.0\muKs}
\setsymbol{LFI:white:noise:sensitivity:LFI19:Rad:M:units}{581.4\muKs}
\setsymbol{LFI:white:noise:sensitivity:LFI20:Rad:M:units}{590.8\muKs}
\setsymbol{LFI:white:noise:sensitivity:LFI21:Rad:M:units}{455.2\muKs}
\setsymbol{LFI:white:noise:sensitivity:LFI22:Rad:M:units}{492.0\muKs}
\setsymbol{LFI:white:noise:sensitivity:LFI23:Rad:M:units}{507.7\muKs}
\setsymbol{LFI:white:noise:sensitivity:LFI24:Rad:M:units}{462.2\muKs}
\setsymbol{LFI:white:noise:sensitivity:LFI25:Rad:M:units}{413.6\muKs}
\setsymbol{LFI:white:noise:sensitivity:LFI26:Rad:M:units}{478.6\muKs}
\setsymbol{LFI:white:noise:sensitivity:LFI27:Rad:M:units}{277.7\muKs}
\setsymbol{LFI:white:noise:sensitivity:LFI28:Rad:M:units}{312.3\muKs}
\setsymbol{LFI:white:noise:sensitivity:LFI18:Rad:S:units}{465.7\muKs}
\setsymbol{LFI:white:noise:sensitivity:LFI19:Rad:S:units}{555.6\muKs}
\setsymbol{LFI:white:noise:sensitivity:LFI20:Rad:S:units}{623.2\muKs}
\setsymbol{LFI:white:noise:sensitivity:LFI21:Rad:S:units}{564.1\muKs}
\setsymbol{LFI:white:noise:sensitivity:LFI22:Rad:S:units}{534.4\muKs}
\setsymbol{LFI:white:noise:sensitivity:LFI23:Rad:S:units}{542.4\muKs}
\setsymbol{LFI:white:noise:sensitivity:LFI24:Rad:S:units}{399.2\muKs}
\setsymbol{LFI:white:noise:sensitivity:LFI25:Rad:S:units}{392.6\muKs}
\setsymbol{LFI:white:noise:sensitivity:LFI26:Rad:S:units}{418.6\muKs}
\setsymbol{LFI:white:noise:sensitivity:LFI27:Rad:S:units}{302.9\muKs}
\setsymbol{LFI:white:noise:sensitivity:LFI28:Rad:S:units}{285.3\muKs}

\setsymbol{LFI:white:noise:sensitivity:LFI18:Rad:M}{512.0}
\setsymbol{LFI:white:noise:sensitivity:LFI19:Rad:M}{581.4}
\setsymbol{LFI:white:noise:sensitivity:LFI20:Rad:M}{590.8}
\setsymbol{LFI:white:noise:sensitivity:LFI21:Rad:M}{455.2}
\setsymbol{LFI:white:noise:sensitivity:LFI22:Rad:M}{492.0}
\setsymbol{LFI:white:noise:sensitivity:LFI23:Rad:M}{507.7}
\setsymbol{LFI:white:noise:sensitivity:LFI24:Rad:M}{462.2}
\setsymbol{LFI:white:noise:sensitivity:LFI25:Rad:M}{413.6}
\setsymbol{LFI:white:noise:sensitivity:LFI26:Rad:M}{478.6}
\setsymbol{LFI:white:noise:sensitivity:LFI27:Rad:M}{277.7}
\setsymbol{LFI:white:noise:sensitivity:LFI28:Rad:M}{312.3}
\setsymbol{LFI:white:noise:sensitivity:LFI18:Rad:S}{465.7}
\setsymbol{LFI:white:noise:sensitivity:LFI19:Rad:S}{555.6}
\setsymbol{LFI:white:noise:sensitivity:LFI20:Rad:S}{623.2}
\setsymbol{LFI:white:noise:sensitivity:LFI21:Rad:S}{564.1}
\setsymbol{LFI:white:noise:sensitivity:LFI22:Rad:S}{534.4}
\setsymbol{LFI:white:noise:sensitivity:LFI23:Rad:S}{542.4}
\setsymbol{LFI:white:noise:sensitivity:LFI24:Rad:S}{399.2}
\setsymbol{LFI:white:noise:sensitivity:LFI25:Rad:S}{392.6}
\setsymbol{LFI:white:noise:sensitivity:LFI26:Rad:S}{418.6}
\setsymbol{LFI:white:noise:sensitivity:LFI27:Rad:S}{302.9}
\setsymbol{LFI:white:noise:sensitivity:LFI28:Rad:S}{285.3}

\setsymbol{LFI:knee:frequency:70GHz:units}{29.5\mHz}
\setsymbol{LFI:knee:frequency:44GHz:units}{56.2\mHz}
\setsymbol{LFI:knee:frequency:30GHz:units}{113.7\mHz}

\setsymbol{LFI:knee:frequency:70GHz}{29.5}
\setsymbol{LFI:knee:frequency:44GHz}{56.2}
\setsymbol{LFI:knee:frequency:30GHz}{113.7}

\setsymbol{LFI:knee:frequency:LFI18:Rad:M:units}{16.3\mHz}
\setsymbol{LFI:knee:frequency:LFI19:Rad:M:units}{15.1\mHz}
\setsymbol{LFI:knee:frequency:LFI20:Rad:M:units}{18.7\mHz}
\setsymbol{LFI:knee:frequency:LFI21:Rad:M:units}{37.2\mHz}
\setsymbol{LFI:knee:frequency:LFI22:Rad:M:units}{12.7\mHz}
\setsymbol{LFI:knee:frequency:LFI23:Rad:M:units}{34.6\mHz}
\setsymbol{LFI:knee:frequency:LFI24:Rad:M:units}{46.2\mHz}
\setsymbol{LFI:knee:frequency:LFI25:Rad:M:units}{24.9\mHz}
\setsymbol{LFI:knee:frequency:LFI26:Rad:M:units}{67.6\mHz}
\setsymbol{LFI:knee:frequency:LFI27:Rad:M:units}{187.4\mHz}
\setsymbol{LFI:knee:frequency:LFI28:Rad:M:units}{122.2\mHz}
\setsymbol{LFI:knee:frequency:LFI18:Rad:S:units}{17.7\mHz}
\setsymbol{LFI:knee:frequency:LFI19:Rad:S:units}{22.0\mHz}
\setsymbol{LFI:knee:frequency:LFI20:Rad:S:units}{8.7\mHz}
\setsymbol{LFI:knee:frequency:LFI21:Rad:S:units}{25.9\mHz}
\setsymbol{LFI:knee:frequency:LFI22:Rad:S:units}{15.8\mHz}
\setsymbol{LFI:knee:frequency:LFI23:Rad:S:units}{129.8\mHz}
\setsymbol{LFI:knee:frequency:LFI24:Rad:S:units}{100.9\mHz}
\setsymbol{LFI:knee:frequency:LFI25:Rad:S:units}{38.9\mHz}
\setsymbol{LFI:knee:frequency:LFI26:Rad:S:units}{58.9\mHz}
\setsymbol{LFI:knee:frequency:LFI27:Rad:S:units}{104.4\mHz}
\setsymbol{LFI:knee:frequency:LFI28:Rad:S:units}{40.7\mHz}

\setsymbol{LFI:knee:frequency:LFI18:Rad:M}{16.3}
\setsymbol{LFI:knee:frequency:LFI19:Rad:M}{15.1}
\setsymbol{LFI:knee:frequency:LFI20:Rad:M}{18.7}
\setsymbol{LFI:knee:frequency:LFI21:Rad:M}{37.2}
\setsymbol{LFI:knee:frequency:LFI22:Rad:M}{12.7}
\setsymbol{LFI:knee:frequency:LFI23:Rad:M}{34.6}
\setsymbol{LFI:knee:frequency:LFI24:Rad:M}{46.2}
\setsymbol{LFI:knee:frequency:LFI25:Rad:M}{24.9}
\setsymbol{LFI:knee:frequency:LFI26:Rad:M}{67.6}
\setsymbol{LFI:knee:frequency:LFI27:Rad:M}{187.4}
\setsymbol{LFI:knee:frequency:LFI28:Rad:M}{122.2}
\setsymbol{LFI:knee:frequency:LFI18:Rad:S}{17.7}
\setsymbol{LFI:knee:frequency:LFI19:Rad:S}{22.0}
\setsymbol{LFI:knee:frequency:LFI20:Rad:S}{8.7}
\setsymbol{LFI:knee:frequency:LFI21:Rad:S}{25.9}
\setsymbol{LFI:knee:frequency:LFI22:Rad:S}{15.8}
\setsymbol{LFI:knee:frequency:LFI23:Rad:S}{129.8}
\setsymbol{LFI:knee:frequency:LFI24:Rad:S}{100.9}
\setsymbol{LFI:knee:frequency:LFI25:Rad:S}{38.9}
\setsymbol{LFI:knee:frequency:LFI26:Rad:S}{58.9}
\setsymbol{LFI:knee:frequency:LFI27:Rad:S}{104.4}
\setsymbol{LFI:knee:frequency:LFI28:Rad:S}{40.7}

\setsymbol{LFI:slope:70GHz:units}{$-1.03$\mHz}
\setsymbol{LFI:slope:44GHz:units}{$-0.89$\mHz}
\setsymbol{LFI:slope:30GHz:units}{$-0.87$\mHz}

\setsymbol{LFI:slope:70GHz}{$-1.03$}
\setsymbol{LFI:slope:44GHz}{$-0.89$}
\setsymbol{LFI:slope:30GHz}{$-0.87$}

\setsymbol{LFI:slope:LFI18:Rad:M:units}{$-1.04$\mHz}
\setsymbol{LFI:slope:LFI19:Rad:M:units}{$-1.09$\mHz}
\setsymbol{LFI:slope:LFI20:Rad:M:units}{$-0.69$\mHz}
\setsymbol{LFI:slope:LFI21:Rad:M:units}{$-1.56$\mHz}
\setsymbol{LFI:slope:LFI22:Rad:M:units}{$-1.01$\mHz}
\setsymbol{LFI:slope:LFI23:Rad:M:units}{$-0.96$\mHz}
\setsymbol{LFI:slope:LFI24:Rad:M:units}{$-0.83$\mHz}
\setsymbol{LFI:slope:LFI25:Rad:M:units}{$-0.91$\mHz}
\setsymbol{LFI:slope:LFI26:Rad:M:units}{$-0.95$\mHz}
\setsymbol{LFI:slope:LFI27:Rad:M:units}{$-0.87$\mHz}
\setsymbol{LFI:slope:LFI28:Rad:M:units}{$-0.88$\mHz}
\setsymbol{LFI:slope:LFI18:Rad:S:units}{$-1.15$\mHz}
\setsymbol{LFI:slope:LFI19:Rad:S:units}{$-1.00$\mHz}
\setsymbol{LFI:slope:LFI20:Rad:S:units}{$-0.95$\mHz}
\setsymbol{LFI:slope:LFI21:Rad:S:units}{$-0.92$\mHz}
\setsymbol{LFI:slope:LFI22:Rad:S:units}{$-1.01$\mHz}
\setsymbol{LFI:slope:LFI23:Rad:S:units}{$-0.95$\mHz}
\setsymbol{LFI:slope:LFI24:Rad:S:units}{$-0.73$\mHz}
\setsymbol{LFI:slope:LFI25:Rad:S:units}{$-1.16$\mHz}
\setsymbol{LFI:slope:LFI26:Rad:S:units}{$-0.79$\mHz}
\setsymbol{LFI:slope:LFI27:Rad:S:units}{$-0.82$\mHz}
\setsymbol{LFI:slope:LFI28:Rad:S:units}{$-0.91$\mHz}

\setsymbol{LFI:slope:LFI18:Rad:M}{$-1.04$}
\setsymbol{LFI:slope:LFI19:Rad:M}{$-1.09$}
\setsymbol{LFI:slope:LFI20:Rad:M}{$-0.69$}
\setsymbol{LFI:slope:LFI21:Rad:M}{$-1.56$}
\setsymbol{LFI:slope:LFI22:Rad:M}{$-1.01$}
\setsymbol{LFI:slope:LFI23:Rad:M}{$-0.96$}
\setsymbol{LFI:slope:LFI24:Rad:M}{$-0.83$}
\setsymbol{LFI:slope:LFI25:Rad:M}{$-0.91$}
\setsymbol{LFI:slope:LFI26:Rad:M}{$-0.95$}
\setsymbol{LFI:slope:LFI27:Rad:M}{$-0.87$}
\setsymbol{LFI:slope:LFI28:Rad:M}{$-0.88$}
\setsymbol{LFI:slope:LFI18:Rad:S}{$-1.15$}
\setsymbol{LFI:slope:LFI19:Rad:S}{$-1.00$}
\setsymbol{LFI:slope:LFI20:Rad:S}{$-0.95$}
\setsymbol{LFI:slope:LFI21:Rad:S}{$-0.92$}
\setsymbol{LFI:slope:LFI22:Rad:S}{$-1.01$}
\setsymbol{LFI:slope:LFI23:Rad:S}{$-0.95$}
\setsymbol{LFI:slope:LFI24:Rad:S}{$-0.73$}
\setsymbol{LFI:slope:LFI25:Rad:S}{$-1.16$}
\setsymbol{LFI:slope:LFI26:Rad:S}{$-0.79$}
\setsymbol{LFI:slope:LFI27:Rad:S}{$-0.82$}
\setsymbol{LFI:slope:LFI28:Rad:S}{$-0.91$}

\setsymbol{LFI:FWHM:70GHz:units}{13\parcm01}
\setsymbol{LFI:FWHM:44GHz:units}{27\parcm92}
\setsymbol{LFI:FWHM:30GHz:units}{32\parcm65}

\setsymbol{LFI:FWHM:70GHz}{13.01}
\setsymbol{LFI:FWHM:44GHz}{27.92}
\setsymbol{LFI:FWHM:30GHz}{32.65}

\setsymbol{LFI:FWHM:LFI18:units}{13\parcm39}
\setsymbol{LFI:FWHM:LFI19:units}{13\parcm01}
\setsymbol{LFI:FWHM:LFI20:units}{12\parcm75}
\setsymbol{LFI:FWHM:LFI21:units}{12\parcm74}
\setsymbol{LFI:FWHM:LFI22:units}{12\parcm87}
\setsymbol{LFI:FWHM:LFI23:units}{13\parcm27}
\setsymbol{LFI:FWHM:LFI24:units}{22\parcm98}
\setsymbol{LFI:FWHM:LFI25:units}{30\parcm46}
\setsymbol{LFI:FWHM:LFI26:units}{30\parcm31}
\setsymbol{LFI:FWHM:LFI27:units}{32\parcm65}
\setsymbol{LFI:FWHM:LFI28:units}{32\parcm66}

\setsymbol{LFI:FWHM:LFI18}{13.39}
\setsymbol{LFI:FWHM:LFI19}{13.01}
\setsymbol{LFI:FWHM:LFI20}{12.75}
\setsymbol{LFI:FWHM:LFI21}{12.74}
\setsymbol{LFI:FWHM:LFI22}{12.87}
\setsymbol{LFI:FWHM:LFI23}{13.27}
\setsymbol{LFI:FWHM:LFI24}{22.98}
\setsymbol{LFI:FWHM:LFI25}{30.46}
\setsymbol{LFI:FWHM:LFI26}{30.31}
\setsymbol{LFI:FWHM:LFI27}{32.65}
\setsymbol{LFI:FWHM:LFI28}{32.66}

\setsymbol{LFI:FWHM:uncertainty:LFI18:units}{0.170\arcm}
\setsymbol{LFI:FWHM:uncertainty:LFI19:units}{0.174\arcm}
\setsymbol{LFI:FWHM:uncertainty:LFI20:units}{0.170\arcm}
\setsymbol{LFI:FWHM:uncertainty:LFI21:units}{0.156\arcm}
\setsymbol{LFI:FWHM:uncertainty:LFI22:units}{0.164\arcm}
\setsymbol{LFI:FWHM:uncertainty:LFI23:units}{0.171\arcm}
\setsymbol{LFI:FWHM:uncertainty:LFI24:units}{0.652\arcm}
\setsymbol{LFI:FWHM:uncertainty:LFI25:units}{1.075\arcm}
\setsymbol{LFI:FWHM:uncertainty:LFI26:units}{1.131\arcm}
\setsymbol{LFI:FWHM:uncertainty:LFI27:units}{1.266\arcm}
\setsymbol{LFI:FWHM:uncertainty:LFI28:units}{1.287\arcm}

\setsymbol{LFI:FWHM:uncertainty:LFI18}{0.170}
\setsymbol{LFI:FWHM:uncertainty:LFI19}{0.174}
\setsymbol{LFI:FWHM:uncertainty:LFI20}{0.170}
\setsymbol{LFI:FWHM:uncertainty:LFI21}{0.156}
\setsymbol{LFI:FWHM:uncertainty:LFI22}{0.164}
\setsymbol{LFI:FWHM:uncertainty:LFI23}{0.171}
\setsymbol{LFI:FWHM:uncertainty:LFI24}{0.652}
\setsymbol{LFI:FWHM:uncertainty:LFI25}{1.075}
\setsymbol{LFI:FWHM:uncertainty:LFI26}{1.131}
\setsymbol{LFI:FWHM:uncertainty:LFI27}{1.266}
\setsymbol{LFI:FWHM:uncertainty:LFI28}{1.287}

\setsymbol{HFI:center:frequency:100GHz:units}{100\,GHz}
\setsymbol{HFI:center:frequency:143GHz:units}{143\,GHz}
\setsymbol{HFI:center:frequency:217GHz:units}{217\,GHz}
\setsymbol{HFI:center:frequency:353GHz:units}{353\,GHz}
\setsymbol{HFI:center:frequency:545GHz:units}{545\,GHz}
\setsymbol{HFI:center:frequency:857GHz:units}{857\,GHz}

\setsymbol{HFI:center:frequency:100GHz}{100}
\setsymbol{HFI:center:frequency:143GHz}{143}
\setsymbol{HFI:center:frequency:217GHz}{217}
\setsymbol{HFI:center:frequency:353GHz}{353}
\setsymbol{HFI:center:frequency:545GHz}{545}
\setsymbol{HFI:center:frequency:857GHz}{857}

\setsymbol{HFI:Ndetectors:100GHz}{8}
\setsymbol{HFI:Ndetectors:143GHz}{11}
\setsymbol{HFI:Ndetectors:217GHz}{12}
\setsymbol{HFI:Ndetectors:353GHz}{12}
\setsymbol{HFI:Ndetectors:545GHz}{3}
\setsymbol{HFI:Ndetectors:857GHz}{4}

\setsymbol{HFI:FWHM:Maps:100GHz:units}{9\parcm88}
\setsymbol{HFI:FWHM:Maps:143GHz:units}{7\parcm18}
\setsymbol{HFI:FWHM:Maps:217GHz:units}{4\parcm87}
\setsymbol{HFI:FWHM:Maps:353GHz:units}{4\parcm65}
\setsymbol{HFI:FWHM:Maps:545GHz:units}{4\parcm72}
\setsymbol{HFI:FWHM:Maps:857GHz:units}{4\parcm39}
\setsymbol{HFI:FWHM:Maps:100GHz}{9.88}
\setsymbol{HFI:FWHM:Maps:143GHz}{7.18}
\setsymbol{HFI:FWHM:Maps:217GHz}{4.87}
\setsymbol{HFI:FWHM:Maps:353GHz}{4.65}
\setsymbol{HFI:FWHM:Maps:545GHz}{4.72}
\setsymbol{HFI:FWHM:Maps:857GHz}{4.39}

\setsymbol{HFI:beam:ellipticity:Maps:100GHz}{1.15}
\setsymbol{HFI:beam:ellipticity:Maps:143GHz}{1.01}
\setsymbol{HFI:beam:ellipticity:Maps:217GHz}{1.06}
\setsymbol{HFI:beam:ellipticity:Maps:353GHz}{1.05}
\setsymbol{HFI:beam:ellipticity:Maps:545GHz}{1.14}
\setsymbol{HFI:beam:ellipticity:Maps:857GHz}{1.19}

\setsymbol{HFI:FWHM:Mars:100GHz:units}{9\parcm37}
\setsymbol{HFI:FWHM:Mars:143GHz:units}{7\parcm04}
\setsymbol{HFI:FWHM:Mars:217GHz:units}{4\parcm68}
\setsymbol{HFI:FWHM:Mars:353GHz:units}{4\parcm43}
\setsymbol{HFI:FWHM:Mars:545GHz:units}{3\parcm80}
\setsymbol{HFI:FWHM:Mars:857GHz:units}{3\parcm67}

\setsymbol{HFI:FWHM:Mars:100GHz}{9.37}
\setsymbol{HFI:FWHM:Mars:143GHz}{7.04}
\setsymbol{HFI:FWHM:Mars:217GHz}{4.68}
\setsymbol{HFI:FWHM:Mars:353GHz}{4.43}
\setsymbol{HFI:FWHM:Mars:545GHz}{3.80}
\setsymbol{HFI:FWHM:Mars:857GHz}{3.67}

\setsymbol{HFI:beam:ellipticity:Mars:100GHz}{1.18}
\setsymbol{HFI:beam:ellipticity:Mars:143GHz}{1.03}
\setsymbol{HFI:beam:ellipticity:Mars:217GHz}{1.14}
\setsymbol{HFI:beam:ellipticity:Mars:353GHz}{1.09}
\setsymbol{HFI:beam:ellipticity:Mars:545GHz}{1.25}
\setsymbol{HFI:beam:ellipticity:Mars:857GHz}{1.03}

\setsymbol{HFI:CMB:relative:calibration:100GHz}{$\lsim 1\%$}
\setsymbol{HFI:CMB:relative:calibration:143GHz}{$\lsim 1\%$}
\setsymbol{HFI:CMB:relative:calibration:217GHz}{$\lsim 1\%$}
\setsymbol{HFI:CMB:relative:calibration:353GHz}{$\lsim 1\%$}
\setsymbol{HFI:CMB:relative:calibration:545GHz}{}
\setsymbol{HFI:CMB:relative:calibration:857GHz}{}

\setsymbol{HFI:CMB:absolute:calibration:100GHz}{$\lsim 2\%$}
\setsymbol{HFI:CMB:absolute:calibration:143GHz}{$\lsim 2\%$}
\setsymbol{HFI:CMB:absolute:calibration:217GHz}{$\lsim 2\%$}
\setsymbol{HFI:CMB:absolute:calibration:353GHz}{$\lsim 2\%$}
\setsymbol{HFI:CMB:absolute:calibration:545GHz}{}
\setsymbol{HFI:CMB:absolute:calibration:857GHz}{}

\setsymbol{HFI:FIRAS:gain:calibration:accuracy:statistical:100GHz}{}
\setsymbol{HFI:FIRAS:gain:calibration:accuracy:statistical:143GHz}{}
\setsymbol{HFI:FIRAS:gain:calibration:accuracy:statistical:217GHz}{}
\setsymbol{HFI:FIRAS:gain:calibration:accuracy:statistical:353GHz}{2.5\%}
\setsymbol{HFI:FIRAS:gain:calibration:accuracy:statistical:545GHz}{1\%}
\setsymbol{HFI:FIRAS:gain:calibration:accuracy:statistical:857GHz}{0.5\%}

\setsymbol{HFI:FIRAS:gain:calibration:accuracy:systematic:100GHz}{}
\setsymbol{HFI:FIRAS:gain:calibration:accuracy:systematic:143GHz}{}
\setsymbol{HFI:FIRAS:gain:calibration:accuracy:systematic:217GHz}{}
\setsymbol{HFI:FIRAS:gain:calibration:accuracy:systematic:353GHz}{}
\setsymbol{HFI:FIRAS:gain:calibration:accuracy:systematic:545GHz}{7\%}
\setsymbol{HFI:FIRAS:gain:calibration:accuracy:systematic:857GHz}{7\%}

\setsymbol{HFI:FIRAS:zero:point:accuracy:100GHz:units}{0.8\MJysr}
\setsymbol{HFI:FIRAS:zero:point:accuracy:143GHz:units}{}
\setsymbol{HFI:FIRAS:zero:point:accuracy:217GHz:units}{}
\setsymbol{HFI:FIRAS:zero:point:accuracy:353GHz:units}{1.4\MJysr}
\setsymbol{HFI:FIRAS:zero:point:accuracy:545GHz:units}{2.2\MJysr}
\setsymbol{HFI:FIRAS:zero:point:accuracy:857GHz:units}{1.7\MJysr}

\setsymbol{HFI:FIRAS:zero:point:accuracy:100GHz}{0.8}
\setsymbol{HFI:FIRAS:zero:point:accuracy:143GHz}{}
\setsymbol{HFI:FIRAS:zero:point:accuracy:217GHz}{}
\setsymbol{HFI:FIRAS:zero:point:accuracy:353GHz}{1.4}
\setsymbol{HFI:FIRAS:zero:point:accuracy:545GHz}{2.2}
\setsymbol{HFI:FIRAS:zero:point:accuracy:857GHz}{1.7}

\setsymbol{HFI:unit:conversion:100GHz:units}{0.2415\MJysrmK}
\setsymbol{HFI:unit:conversion:143GHz:units}{0.3694\MJysrmK}
\setsymbol{HFI:unit:conversion:217GHz:units}{0.4811\MJysrmK}
\setsymbol{HFI:unit:conversion:353GHz:units}{0.2883\MJysrmK}
\setsymbol{HFI:unit:conversion:545GHz:units}{0.05826\MJysrmK}
\setsymbol{HFI:unit:conversion:857GHz:units}{0.002238\MJysrmK}

\setsymbol{HFI:unit:conversion:100GHz}{0.2415}
\setsymbol{HFI:unit:conversion:143GHz}{0.3694}
\setsymbol{HFI:unit:conversion:217GHz}{0.4811}
\setsymbol{HFI:unit:conversion:353GHz}{0.2883}
\setsymbol{HFI:unit:conversion:545GHz}{0.05826}
\setsymbol{HFI:unit:conversion:857GHz}{0.002238}

\setsymbol{HFI:colour:correction:alpha=-2:V1.01:100GHz}{0.9893}
\setsymbol{HFI:colour:correction:alpha=-2:V1.01:143GHz}{0.9759}
\setsymbol{HFI:colour:correction:alpha=-2:V1.01:217GHz}{1.0007}
\setsymbol{HFI:colour:correction:alpha=-2:V1.01:353GHz}{1.0028}
\setsymbol{HFI:colour:correction:alpha=-2:V1.01:545GHz}{1.0019}
\setsymbol{HFI:colour:correction:alpha=-2:V1.01:857GHz}{0.9889}

\setsymbol{HFI:colour:correction:alpha=0:V1.01:100GHz}{1.0008}
\setsymbol{HFI:colour:correction:alpha=0:V1.01:143GHz}{1.0148}
\setsymbol{HFI:colour:correction:alpha=0:V1.01:217GHz}{0.9909}
\setsymbol{HFI:colour:correction:alpha=0:V1.01:353GHz}{0.9888}
\setsymbol{HFI:colour:correction:alpha=0:V1.01:545GHz}{0.9878}
\setsymbol{HFI:colour:correction:alpha=0:V1.01:857GHz}{1.0014}

\providecommand{\sorthelp}[1]{}

\def\all2103resultspapers{\nocite{planck2013-p01, planck2013-p02, planck2013-p02a, planck2013-p02d, planck2013-p02b, planck2013-p03, planck2013-p03c, planck2013-p03f, planck2013-p03d, planck2013-p03e, planck2013-p01a, planck2013-p06, planck2013-p03a, planck2013-pip88, planck2013-p08, planck2013-p11, planck2013-p12, planck2013-p13, planck2013-p14, planck2013-p15, planck2013-p05b, planck2013-p17, planck2013-p09, planck2013-p09a, planck2013-p20, planck2013-p19, planck2013-pipaberration, planck2013-p05, planck2013-p05a, planck2013-pip56, planck2013-p06b}}

\begin{document}

 \title{\Planck\ 2013 results. XIV. Zodiacal emission}
 
\author{\small
Planck Collaboration:
P.~A.~R.~Ade\inst{88}
\and
N.~Aghanim\inst{61}
\and
C.~Armitage-Caplan\inst{92}
\and
M.~Arnaud\inst{74}
\and
M.~Ashdown\inst{71, 6}
\and
F.~Atrio-Barandela\inst{19}
\and
J.~Aumont\inst{61}
\and
C.~Baccigalupi\inst{87}
\and
A.~J.~Banday\inst{95, 10}
\and
R.~B.~Barreiro\inst{68}
% ...
}
\institute{\small
APC, AstroParticule et Cosmologie, Universit\'{e} Paris Diderot, CNRS/IN2P3, CEA/lrfu, Observatoire de Paris, Sorbonne Paris Cit\'{e}, 10, rue Alice Domon et L\'{e}onie Duquet, 75205 Paris Cedex 13, France\\
\and
Aalto University Mets\"{a}hovi Radio Observatory, Mets\"{a}hovintie 114, FIN-02540 Kylm\"{a}l\"{a}, Finland\\
\and
African Institute for Mathematical Sciences, 6-8 Melrose Road, Muizenberg, Cape Town, South Africa\\
\and
Agenzia Spaziale Italiana Science Data Center, Via del Politecnico snc, 00133, Roma, Italy\\
\and
Agenzia Spaziale Italiana, Viale Liegi 26, Roma, Italy\\
\and
Astrophysics Group, Cavendish Laboratory, University of Cambridge, J J Thomson Avenue, Cambridge CB3 0HE, U.K.\\
\and
Astrophysics \& Cosmology Research Unit, School of Mathematics, Statistics \& Computer Science, University of KwaZulu-Natal, Westville Campus, Private Bag X54001, Durban 4000, South Africa\\
\and
Atacama Large Millimeter/submillimeter Array, ALMA Santiago Central Offices, Alonso de Cordova 3107, Vitacura, Casilla 763 0355, Santiago, Chile\\
\and
CITA, University of Toronto, 60 St. George St., Toronto, ON M5S 3H8, Canada\\
\and
CNRS, IRAP, 9 Av. colonel Roche, BP 44346, F-31028 Toulouse cedex 4, France\\
\and
California Institute of Technology, Pasadena, California, U.S.A.\\
\and
Centre for Theoretical Cosmology, DAMTP, University of Cambridge, Wilberforce Road, Cambridge CB3 0WA, U.K.\\
\and
Centro de Estudios de F\'{i}sica del Cosmos de Arag\'{o}n (CEFCA), Plaza San Juan, 1, planta 2, E-44001, Teruel, Spain\\
\and
Computational Cosmology Center, Lawrence Berkeley National Laboratory, Berkeley, California, U.S.A.\\
\and
Consejo Superior de Investigaciones Cient\'{\i}ficas (CSIC), Madrid, Spain\\
\and
DSM/Irfu/SPP, CEA-Saclay, F-91191 Gif-sur-Yvette Cedex, France\\
\and
DTU Space, National Space Institute, Technical University of Denmark, Elektrovej 327, DK-2800 Kgs. Lyngby, Denmark\\
\and
D\'{e}partement de Physique Th\'{e}orique, Universit\'{e} de Gen\`{e}ve, 24, Quai E. Ansermet,1211 Gen\`{e}ve 4, Switzerland\\
\and
Departamento de F\'{\i}sica Fundamental, Facultad de Ciencias, Universidad de Salamanca, 37008 Salamanca, Spain\\
\and
Departamento de F\'{\i}sica, Universidad de Oviedo, Avda. Calvo Sotelo s/n, Oviedo, Spain\\
\and
Department of Astronomy and Astrophysics, University of Toronto, 50 Saint George Street, Toronto, Ontario, Canada\\
\and
Department of Astrophysics/IMAPP, Radboud University Nijmegen, P.O. Box 9010, 6500 GL Nijmegen, The Netherlands\\
\and
Department of Electrical Engineering and Computer Sciences, University of California, Berkeley, California, U.S.A.\\
\and
Department of Physics \& Astronomy, University of British Columbia, 6224 Agricultural Road, Vancouver, British Columbia, Canada\\
\and
Department of Physics and Astronomy, Dana and David Dornsife College of Letter, Arts and Sciences, University of Southern California, Los Angeles, CA 90089, U.S.A.\\
\and
Department of Physics and Astronomy, University College London, London WC1E 6BT, U.K.\\
\and
Department of Physics, Florida State University, Keen Physics Building, 77 Chieftan Way, Tallahassee, Florida, U.S.A.\\
\and
Department of Physics, Gustaf H\"{a}llstr\"{o}min katu 2a, University of Helsinki, Helsinki, Finland\\
\and
Department of Physics, Princeton University, Princeton, New Jersey, U.S.A.\\
\and
Department of Physics, University of California, Berkeley, California, U.S.A.\\
\and
Department of Physics, University of California, One Shields Avenue, Davis, California, U.S.A.\\
\and
Department of Physics, University of California, Santa Barbara, California, U.S.A.\\
\and
Department of Physics, University of Illinois at Urbana-Champaign, 1110 West Green Street, Urbana, Illinois, U.S.A.\\
\and
Dipartimento di Fisica e Astronomia G. Galilei, Universit\`{a} degli Studi di Padova, via Marzolo 8, 35131 Padova, Italy\\
\and
Dipartimento di Fisica e Scienze della Terra, Universit\`{a} di Ferrara, Via Saragat 1, 44122 Ferrara, Italy\\
\and
Dipartimento di Fisica, Universit\`{a} La Sapienza, P. le A. Moro 2, Roma, Italy\\
\and
Dipartimento di Fisica, Universit\`{a} degli Studi di Milano, Via Celoria, 16, Milano, Italy\\
\and
Dipartimento di Fisica, Universit\`{a} degli Studi di Trieste, via A. Valerio 2, Trieste, Italy\\
\and
Dipartimento di Fisica, Universit\`{a} di Roma Tor Vergata, Via della Ricerca Scientifica, 1, Roma, Italy\\
\and
Discovery Center, Niels Bohr Institute, Blegdamsvej 17, Copenhagen, Denmark\\
\and
Dpto. Astrof\'{i}sica, Universidad de La Laguna (ULL), E-38206 La Laguna, Tenerife, Spain\\
\and
European Southern Observatory, ESO Vitacura, Alonso de Cordova 3107, Vitacura, Casilla 19001, Santiago, Chile\\
\and
European Space Agency, ESAC, Planck Science Office, Camino bajo del Castillo, s/n, Urbanizaci\'{o}n Villafranca del Castillo, Villanueva de la Ca\~{n}ada, Madrid, Spain\\
\and
European Space Agency, ESTEC, Keplerlaan 1, 2201 AZ Noordwijk, The Netherlands\\
\and
Helsinki Institute of Physics, Gustaf H\"{a}llstr\"{o}min katu 2, University of Helsinki, Helsinki, Finland\\
\and
INAF - Osservatorio Astrofisico di Catania, Via S. Sofia 78, Catania, Italy\\
\and
INAF - Osservatorio Astronomico di Padova, Vicolo dell'Osservatorio 5, Padova, Italy\\
\and
INAF - Osservatorio Astronomico di Roma, via di Frascati 33, Monte Porzio Catone, Italy\\
\and
INAF - Osservatorio Astronomico di Trieste, Via G.B. Tiepolo 11, Trieste, Italy\\
\and
INAF Istituto di Radioastronomia, Via P. Gobetti 101, 40129 Bologna, Italy\\
\and
INAF/IASF Bologna, Via Gobetti 101, Bologna, Italy\\
\and
INAF/IASF Milano, Via E. Bassini 15, Milano, Italy\\
\and
INFN, Sezione di Bologna, Via Irnerio 46, I-40126, Bologna, Italy\\
\and
INFN, Sezione di Roma 1, Universit\`{a} di Roma Sapienza, Piazzale Aldo Moro 2, 00185, Roma, Italy\\
\and
IPAG: Institut de Plan\'{e}tologie et d'Astrophysique de Grenoble, Universit\'{e} Joseph Fourier, Grenoble 1 / CNRS-INSU, UMR 5274, Grenoble, F-38041, France\\
\and
IUCAA, Post Bag 4, Ganeshkhind, Pune University Campus, Pune 411 007, India\\
\and
Imperial College London, Astrophysics group, Blackett Laboratory, Prince Consort Road, London, SW7 2AZ, U.K.\\
\and
Infrared Processing and Analysis Center, California Institute of Technology, Pasadena, CA 91125, U.S.A.\\
\and
Institut N\'{e}el, CNRS, Universit\'{e} Joseph Fourier Grenoble I, 25 rue des Martyrs, Grenoble, France\\
\and
Institut Universitaire de France, 103, bd Saint-Michel, 75005, Paris, France\\
\and
Institut d'Astrophysique Spatiale, CNRS (UMR8617) Universit\'{e} Paris-Sud 11, B\^{a}timent 121, Orsay, France\\
\and
Institut d'Astrophysique de Paris, CNRS (UMR7095), 98 bis Boulevard Arago, F-75014, Paris, France\\
\and
Institute for Space Sciences, Bucharest-Magurale, Romania\\
\and
Institute of Astronomy and Astrophysics, Academia Sinica, Taipei, Taiwan\\
\and
Institute of Astronomy, University of Cambridge, Madingley Road, Cambridge CB3 0HA, U.K.\\
\and
Institute of Theoretical Astrophysics, University of Oslo, Blindern, Oslo, Norway\\
\and
Instituto de Astrof\'{\i}sica de Canarias, C/V\'{\i}a L\'{a}ctea s/n, La Laguna, Tenerife, Spain\\
\and
Instituto de F\'{\i}sica de Cantabria (CSIC-Universidad de Cantabria), Avda. de los Castros s/n, Santander, Spain\\
\and
Jet Propulsion Laboratory, California Institute of Technology, 4800 Oak Grove Drive, Pasadena, California, U.S.A.\\
\and
Jodrell Bank Centre for Astrophysics, Alan Turing Building, School of Physics and Astronomy, The University of Manchester, Oxford Road, Manchester, M13 9PL, U.K.\\
\and
Kavli Institute for Cosmology Cambridge, Madingley Road, Cambridge, CB3 0HA, U.K.\\
\and
LAL, Universit\'{e} Paris-Sud, CNRS/IN2P3, Orsay, France\\
\and
LERMA, CNRS, Observatoire de Paris, 61 Avenue de l'Observatoire, Paris, France\\
\and
Laboratoire AIM, IRFU/Service d'Astrophysique - CEA/DSM - CNRS - Universit\'{e} Paris Diderot, B\^{a}t. 709, CEA-Saclay, F-91191 Gif-sur-Yvette Cedex, France\\
\and
Laboratoire Traitement et Communication de l'Information, CNRS (UMR 5141) and T\'{e}l\'{e}com ParisTech, 46 rue Barrault F-75634 Paris Cedex 13, France\\
\and
Laboratoire de Physique Subatomique et de Cosmologie, Universit\'{e} Joseph Fourier Grenoble I, CNRS/IN2P3, Institut National Polytechnique de Grenoble, 53 rue des Martyrs, 38026 Grenoble cedex, France\\
\and
Laboratoire de Physique Th\'{e}orique, Universit\'{e} Paris-Sud 11 \& CNRS, B\^{a}timent 210, 91405 Orsay, France\\
\and
Lawrence Berkeley National Laboratory, Berkeley, California, U.S.A.\\
\and
Max-Planck-Institut f\"{u}r Astrophysik, Karl-Schwarzschild-Str. 1, 85741 Garching, Germany\\
\and
McGill Physics, Ernest Rutherford Physics Building, McGill University, 3600 rue University, Montr\'{e}al, QC, H3A 2T8, Canada\\
\and
MilliLab, VTT Technical Research Centre of Finland, Tietotie 3, Espoo, Finland\\
\and
National University of Ireland, Department of Experimental Physics, Maynooth, Co. Kildare, Ireland\\
\and
Niels Bohr Institute, Blegdamsvej 17, Copenhagen, Denmark\\
\and
Observational Cosmology, Mail Stop 367-17, California Institute of Technology, Pasadena, CA, 91125, U.S.A.\\
\and
Optical Science Laboratory, University College London, Gower Street, London, U.K.\\
\and
SB-ITP-LPPC, EPFL, CH-1015, Lausanne, Switzerland\\
\and
SISSA, Astrophysics Sector, via Bonomea 265, 34136, Trieste, Italy\\
\and
School of Physics and Astronomy, Cardiff University, Queens Buildings, The Parade, Cardiff, CF24 3AA, U.K.\\
\and
Space Sciences Laboratory, University of California, Berkeley, California, U.S.A.\\
\and
Special Astrophysical Observatory, Russian Academy of Sciences, Nizhnij Arkhyz, Zelenchukskiy region, Karachai-Cherkessian Republic, 369167, Russia\\
\and
Stanford University, Dept of Physics, Varian Physics Bldg, 382 Via Pueblo Mall, Stanford, California, U.S.A.\\
\and
Sub-Department of Astrophysics, University of Oxford, Keble Road, Oxford OX1 3RH, U.K.\\
\and
Theory Division, PH-TH, CERN, CH-1211, Geneva 23, Switzerland\\
\and
UPMC Univ Paris 06, UMR7095, 98 bis Boulevard Arago, F-75014, Paris, France\\
\and
Universit\'{e} de Toulouse, UPS-OMP, IRAP, F-31028 Toulouse cedex 4, France\\
% ...
}

 \abstract 
 {The \Planck\ satellite provides a set of all-sky maps
  at nine frequencies from 30\,GHz to
  857\,GHz.  Planets, minor bodies, and diffuse interplanetary dust emission (IPD) are all observed. 
  The IPD can be separated from Galactic and other emissions
  because \Planck\ views a given point on the celestial sphere
  multiple times, through different columns of IPD. 
  We use the \Planck\ data to investigate the behaviour of zodiacal 
  emission over the whole sky at sub-millimetre and millimetre wavelengths.
  We fit the \Planck\ data to find the
  emissivities of the various components of the \textit{COBE}\ zodiacal model -- 
  a diffuse cloud, three asteroidal dust bands, a circumsolar ring, and an
  Earth-trailing feature.
  The emissivity of the diffuse cloud decreases with increasing wavelength, 
  as expected from earlier analyses.
  The emissivities of the dust bands, however, decrease less rapidly,
  indicating that the properties of the grains in the bands are 
  different from those in the diffuse cloud. 
  We fit the small amount of Galactic emission
  seen through the telescope's far sidelobes, and place limits on possible contamination 
  of the CMB results from both zodiacal and far-sidelobe emission. 
  When necessary, the results are used in the \Planck\ pipeline
  to make maps with zodiacal emission and far sidelobes removed.
  We show that the zodiacal correction to the CMB maps
  is small compared to the \Planck\ CMB temperature power spectrum and 
  give a list of flux densities for small Solar System bodies. 
 }

 \def\sep{--}
 \keywords{
  Zodiacal Dust \sep\
  Interplanetary Medium \sep\
  (Cosmology): Cosmic Microwave Background 
 }
 
 \authorrunning{\Planck\ Collaboration}
 \titlerunning{Zodiacal Emission}

 \maketitle

 \section{Introduction}\label{sec:introduction}
  
  This paper, one of a set associated with the 2013 release
  of data from the \Planck\footnote{\Planck\ (http://www.esa.int/Planck) is a project
    of the European Space Agency (ESA) with instruments provided by
    two scientific consortia funded by ESA member states (in
    particular the lead countries France and Italy), with
    contributions from NASA (USA) and telescope reflectors provided by
    a collaboration between ESA and a scientific consortium led and
    funded by Denmark.} mission \citep{planck2011-1.1}\all2103resultspapers, describes
  the measurement of zodiacal emission with
  \Planck.
  
  Zodiacal light, the reflection of sunlight by small dust
  particles in our Solar System, can be seen by eye at dawn or dusk in
  dark locations, and contributes significantly to the diffuse sky
  brightness at optical and near-infrared wavelengths. In recent decades the study of
  zodiacal emission, or the thermal re-emission of absorbed energy from these
  interplanetary dust (IPD) particles, has been enabled by the advent
  of infrared astronomical techniques, and it is now known to dominate
  the diffuse brightness of most of the sky between 10 and
  50\micron\ \citep[see, for example,][]{Leinert1997}.
  
  Full-sky, infrared satellite surveys in particular have allowed us to begin
  to determine the structure of the density of the IPD~\citep{Hauser84,kelsall1998,fixsendwek2002,Pyo2010}.
  The full-sky model of zodiacal emission from the Cosmic Background Explorer Diffuse Infrared
  Brightness Experiment (\textit{COBE}/DIRBE) team~\cite[hereafter
    K98]{kelsall1998} is commonly used at longer wavelengths, and is easily
  adapted for \Planck.  Other models are presented in~\cite{Good1986},
  \cite{Rowan-Robinson1990,Rowan-Robinson1991}, \cite{Jones1993},
  \cite{Vrtilek1995}, \cite{Wright1998}, and \cite{Rowan-Robinson2012}.
  The K98 model comprises the well-known Diffuse Cloud, three sets of
  Dust Bands, first discovered by \textit{IRAS}~\citep{Low1984}, and a
  Circumsolar Ring and Earth-Trailing Feature, hinted at in
  \textit{IRAS}\ and confirmed by DIRBE~\citep[called a ``blob'' in
  K98. See][and references therein]{Reach1995}.
  
  \citet{fixsendwek2002} used data from \textit{COBE}'s Far Infrared Absolute
  Spectrophotometer (FIRAS) to extend measurements of the diffuse
  cloud to longer wavelengths, but given FIRAS' modest angular resolution
  and large uncertainties in the submillimetre region, they could not say
  more about zodiacal features on small angular scales. 
  \Planck's sensitivity allows it to detect and measure
  the emissivity of the diffuse zodiacal cloud at long wavelengths, 
  and its angular resolution allows it to 
  characterize the smaller-scale components of the zodiacal emission.
  
  This paper is structured as follows:
  Sect.~\ref{sec:planck:mission} describes relevant aspects of the \Planck\ mission, including its
  observing strategy and data processing; 
  Sect.~\ref{sec:detection} describes how \Planck\ detects zodiacal emission; Sect.~\ref{sec:model} 
  describes the \textit{COBE}\ zodiacal emission model; Sect.~\ref{sec:fit} describes how we fit that model to the \Planck/HFI   
  data; and Sect.~\ref{sec:discussion} gives the results of the fit.  
  Sect.~\ref{sec:conclusion} gives our conclusions.

 \section{The \Planck\ Mission}\label{sec:planck:mission}
 
  \Planck, launched in May 2009, comprised two instruments: the Low Frequency Instrument (LFI), which observed at 30, 44 and 70\,GHz, and the High Frequency Instrument (HFI), which observed at 100, 143, 217, 353, 545, and 857\,GHz.  The mission is described in~\cite{planck2013-p01}.  This work uses data only at frequencies of 100\,GHz and higher. At these frequencies, \Planck's angular resolution ranges from 9\parcm7 to 4\parcm6~\citep{planck2013-p03c}. 
  
 \subsection{Orbit, Scanning Strategy, and Dates of Observation}
  
\begin{figure}[htbp]
  \centering
  \includegraphics[width=44mm]{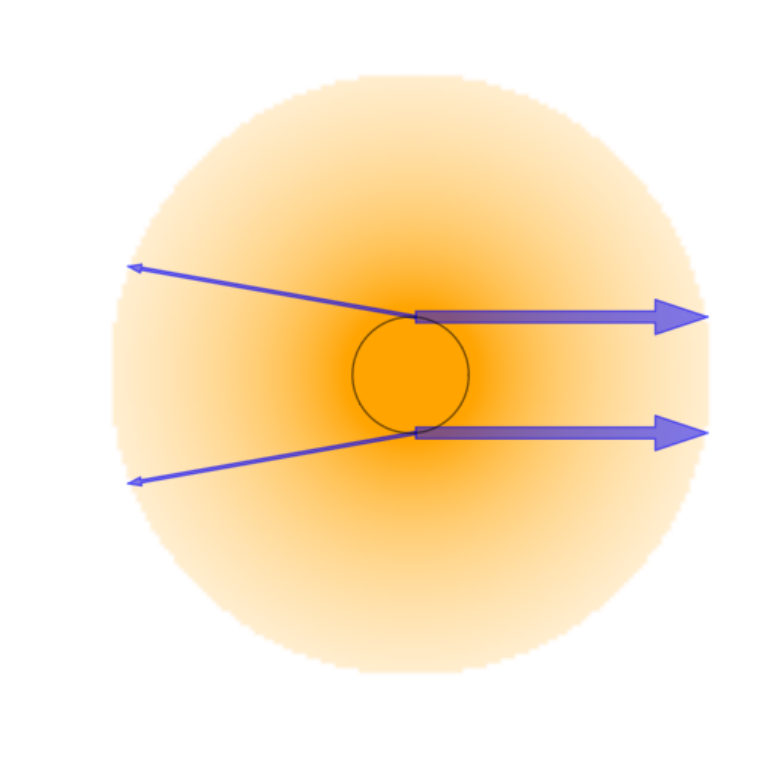}
  \includegraphics[width=44mm]{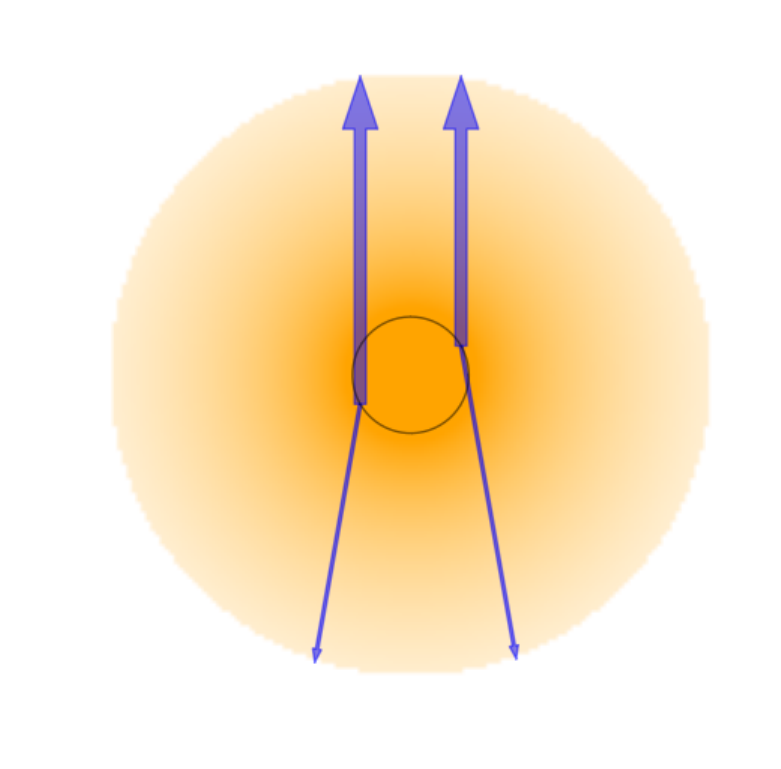}
  \caption{
   Schematic representation of the geometry of \Planck's
   measurements, which shows that it can view different amounts of zodiacal emission
   while looking at the same point on the distant sky. 
   Observations of two points on the sky are shown, each with two measurements
   of the given point. 
   The plane of the Ecliptic is in the plane of the
   diagram. The Sun is in the centre of each panel. The solid black
   ring represents the orbit of the Earth and \Planck. 
   The orange circle represents the IPD cloud, cut off at
   the orbit of Jupiter, beyond which we assume
   there is no contribution to the zodiacal emission. 
   {\it Left\/}: Case where the phase of the scan cycloid and the location
   of the observed point on the sky (towards the right of the page in this case) yield two measurements for
   which the lines of sight through the IPD are roughly equal,
   and the same zodiacal signal is seen. {\it Right\/}: Case where the
   phase of the scan cycloid and the location of the observed point
   on the sky (now towards the top of the page) yield different total columns of IPD along the lines of sight, 
   and thus a different zodiacal signal. 
   \textit{Note that this figure is highly stylized and not to scale.} 
  }
  \label{fig:Scan}
 \end{figure}

  The \Planck\ orbit, scanning strategy, and dates of observation for various subsets of the data are described in detail in Sect.~4.1 of \citet{planck2013-p01}.  The spin axis lies 7\pdeg5 from the Sun-Earth vector, moving around the anti-Sun point in a cycloid of period six months.  This cycloid component results in differing total amounts of IPD in \Planck's line of sight for different observations of the same point on the distant celestial sphere, as shown schematically in Fig.~\ref{fig:Scan}.
 
\begin{figure}[htbp]
  \centering
  \includegraphics[scale=0.85]{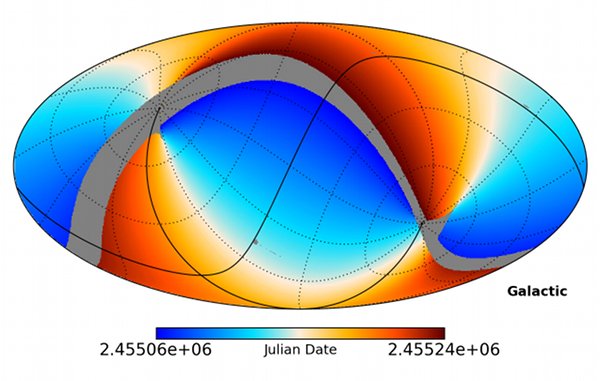}
  \caption{Julian date of observation of pixels on the sky during
   Survey 1, for a single detector, in Galactic coordinates.
    There are only very small differences between maps for different detectors.
    The grid lines show ecliptic coordinates, with the darker lines
    representing the ecliptic plane and the line of zero ecliptic
    longitude. Undefined pixels,
    which were either not observed at all, or which were observed
    multiple times over a period that spanned more than one week and
    are thus not used in this analysis, are shown as the uniform
    grey band.}
  \label{fig:jd}
 \end{figure}

 As detailed in \citet{planck2013-p01}, the \Planck\ beams scan the entire sky exactly twice in one year, but scan only 93\,\% of the sky in six months.  For convenience, we call an approximately six month period one Survey (Table~1 in \citealt{planck2013-p01}), and use that term as shorthand for one coverage of the sky. The ``nominal'' mission comprises the first two Surveys and part of the third.
 During any single one of these Surveys, some pixels near the ecliptic
 poles are observed multiple times, as are the pixels near the
 ecliptic plane that are seen both at the beginning and at the end of the
 Survey. The bulk of the sky, however, is observed only
 during well-defined periods of approximately one 
 week. Figure~\ref{fig:jd} shows the Julian dates of observations
 of those pixels on the sky for which the observation times during Survey~1 spanned one week or less. The
 equivalent figure for Survey~2 is similar.  The scanning strategies for Surveys~3 and 4 were almost
 identical to those of Surveys 1 and 2, respectively.
 
  \subsection{Data Processing}

   HFI data processing is described in~\cite{planck2011-1.7} and~\cite{planck2013-p03}. 
   Given the time-dependent nature of the
   zodiacal signal seen with the \Planck\ scanning strategy, the analysis in this paper
   is based on individual Survey 1--4 maps. This allows us to
   exclude from the analysis regions of the sky and periods of time 
   where the column of IPD viewed by \Planck\ is not constant.  
   
   The HFI instrument has multiple horns at each frequency \citep[Fig. 9]{planck2013-p03c}. 
   Working with individual horn maps, rather than co-added frequency maps,
   allows us to adjust the response of each detector for uniform response to the zodiacal spectrum, rather
   than to the CMB spectrum, as is done in the standard 
   processing~\citep{planck2013-p03d}.  At 100, 143, 217, and 353\,GHz, some
   horns couple to two polarization sensitive bolometers~\citep[PSBs;][]{Jones2003}. 
   As we are not addressing polarization here, we simply average the maps from a PSB pair.
    
   As the evaluation of the model to be presented in Sect.~\ref{sec:model}
   involves calculating emission from a number of points along each
   line of sight and summing them, the computations are time
   consuming. To mitigate this to some extent, we use $13\parcm7\times13\parcm7$
   pixels ({\tt HEALPix} $N_{\rm side} = 256$; \citealt{gorski2005}), rather than the original $1\parcm7\times1\parcm7$ HFI pixels ($N_{\rm side} = 2048$), thereby reducing the number of map pixels from 50~million to less than 800~thousand. 
   Although this reduces our sensitivity to finer-scale structures, it does not hinder comparisons with DIRBE, which had a still larger beam.  Smaller pixels will be used in future work, as
   more detail is teased out of the data.

  Pre-launch estimates of \Planck's ability to detect zodiacal emission and an
  estimate of the possible level of contamination at the highest
  \Planck\ frequencies were presented in~\cite{maris2006}. More
  recent predictions have addressed the possibility of
  zodiacal contamination at lower frequencies~\citep{diego2010}, 
  and speculated that emission from dust in the outer Solar System might
  contribute to the large-scale anomalies reported in data from the Wilkinson Microwave Anisotropy Probe (\textit{WMAP}) at large angular scales~\citep{maris2011,Hansen2012}.

 \section{Detection}\label{sec:detection}

\begin{figure}[htbp]
   \centering
   \includegraphics[width=80mm]{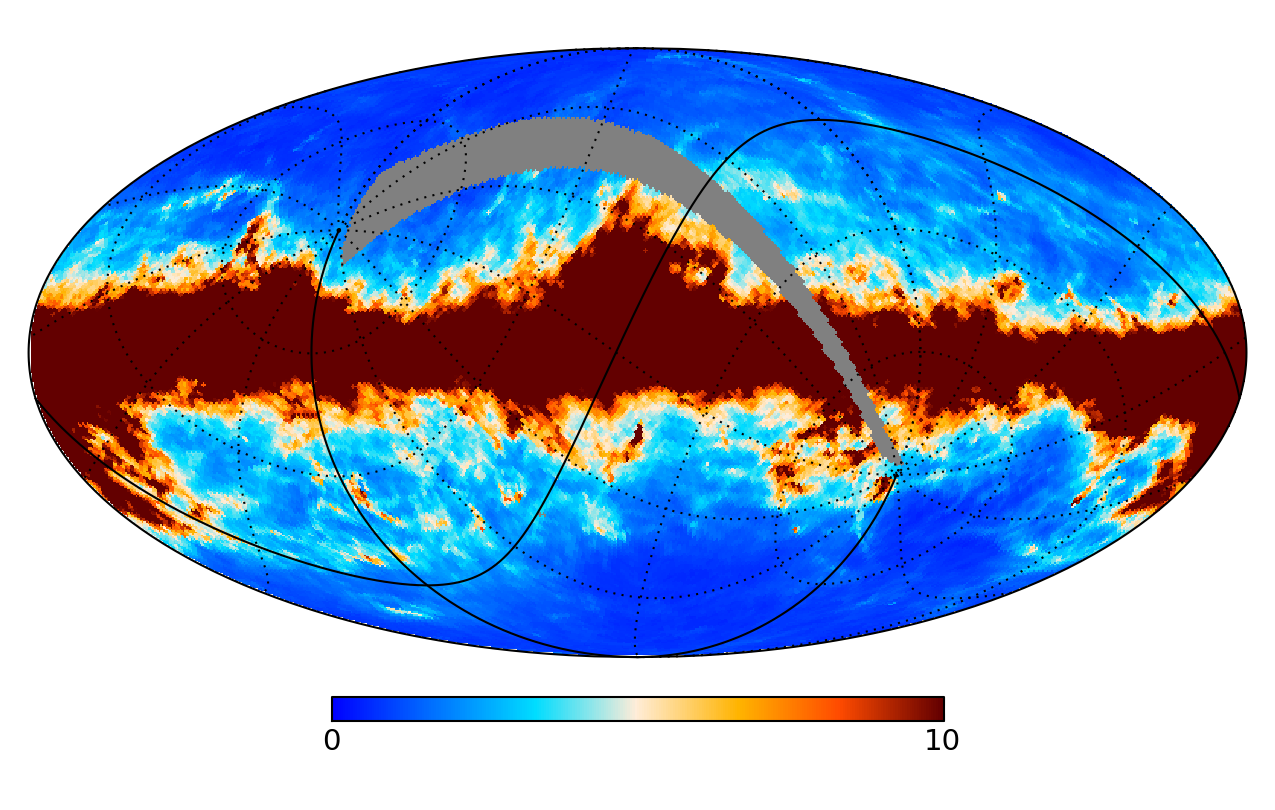} \\
   \includegraphics[width=80mm]{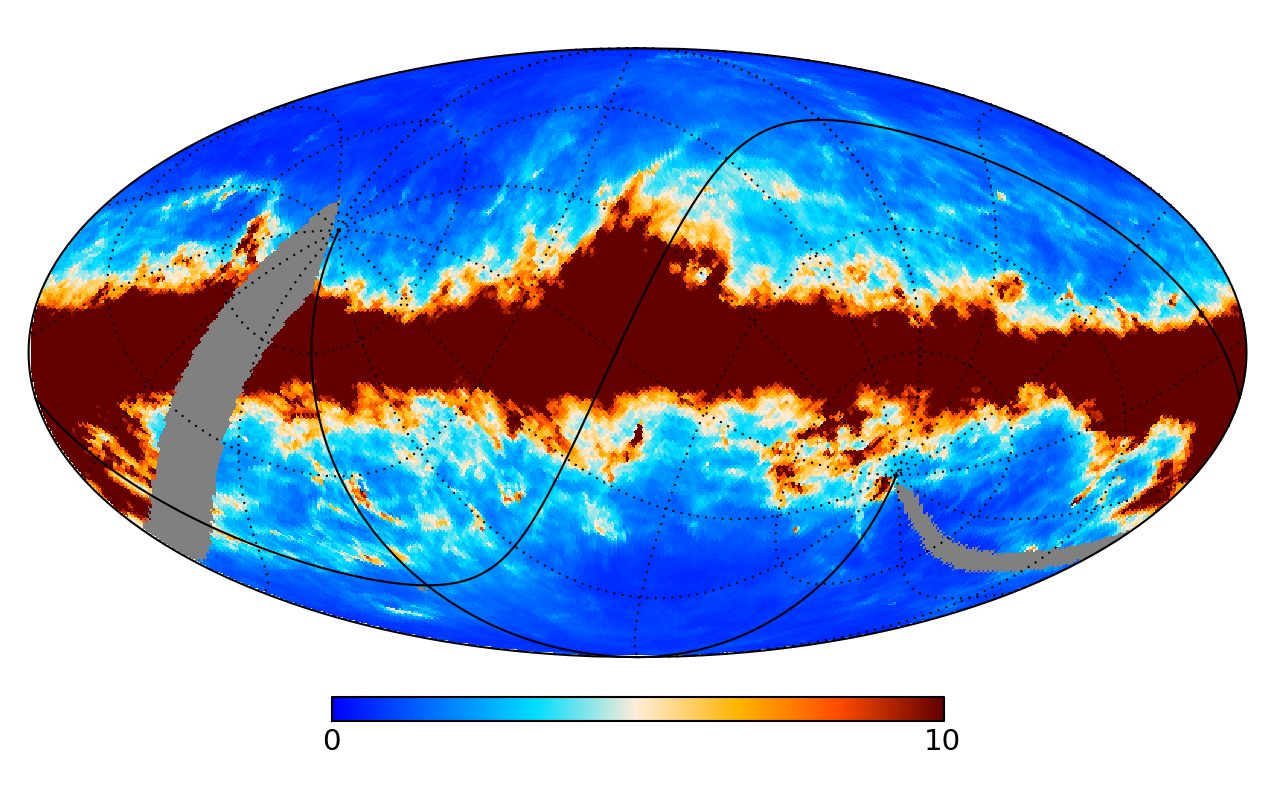} \\
   \includegraphics[width=80mm]{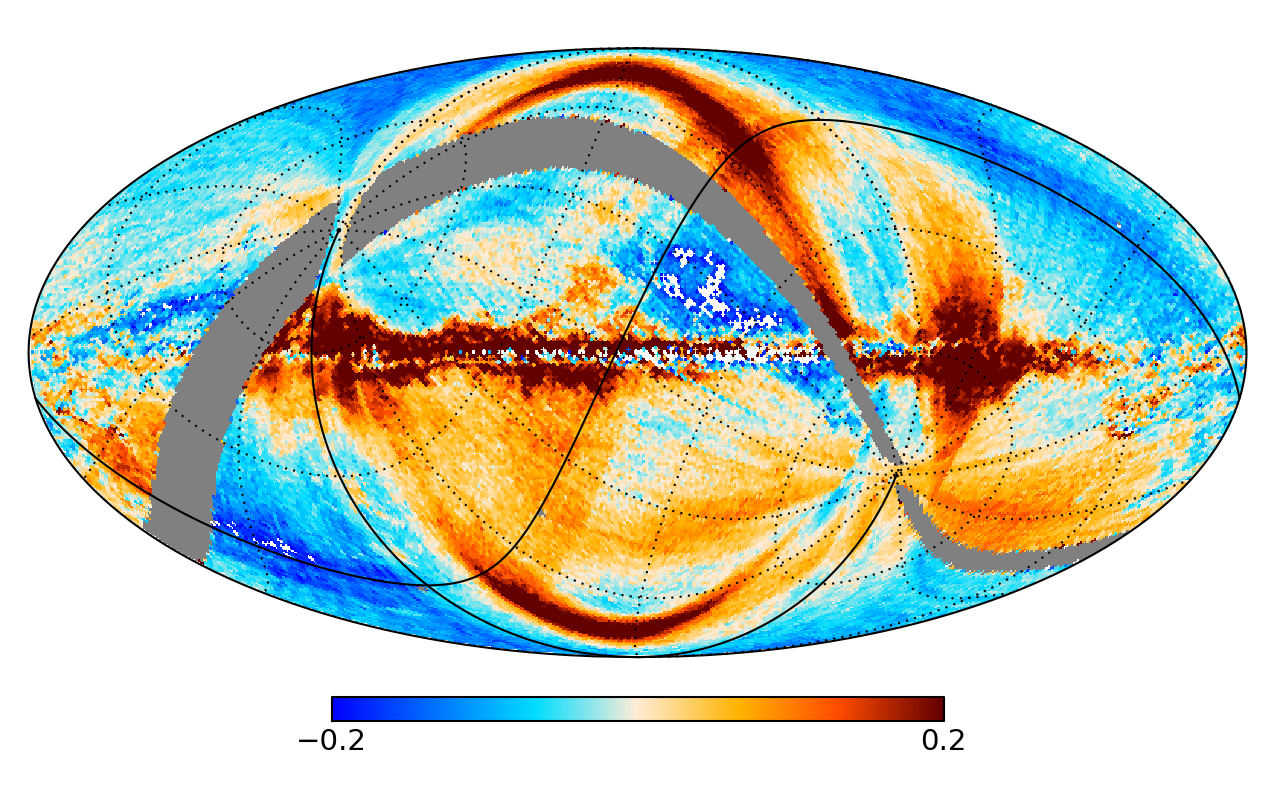}
   \caption{Single-Survey maps in Galactic coordinates for the 857-1
     detector. \emph{Top:} Survey 1 map. \emph{Middle:} Survey 2
     map. \emph{Bottom:} Survey 2 minus Survey 1
     difference map.
     This bottom image shows the zodiacal emission and 
     the residual Galactic emission effects that are discussed in the text.
     The units are MJy\,sr$^{-1}$, assuming a spectrum
     inversely proportional to frequency. Undefined pixels are
     shown in grey. 
     These occur in pixels that either have not
     been observed during the Survey, were observed during the passage
     of a planet, or for a small number of other events. In the top two plots,
     pixels that were observed over periods longer than a week were
     not masked, and thus the masked regions are smaller in the top two images than that
     in the difference (bottom) panel. The units are MJy/sr.}
   \label{fig:surveymaps}
  \end{figure}

  The existence of the zodiacal emission in the \Planck\ maps is
  straightforward to demonstrate by exploiting the fact, noted above, 
  that different \Planck\ Surveys sample different columns of IPD while
  observing the same location on the distant celestial sphere.  
  Figure~\ref{fig:surveymaps} shows Survey~1 and Survey~2 maps for the 857-1 
  detector, and their difference. (See Figs.~\ref{fig:beforeAndAfterJackknives} and~\ref{fig:beforeAndAfterJackknivesCMB}
  for similar figures for all HFI frequencies.)  Three features 
  stand out in the map difference: (1)~The scale has been reduced immensely, but 
  the Galactic plane is still visible; (2)~the 
  ``arcs'' at the top and the bottom of the difference map are the images of the 
  Galactic centre as seen through the instruments' far sidelobes (FSLs); 
  and (3)~the zodiacal emission can be seen as the variations following
  the ecliptic plane. 
    
  In the difference map, zodiacal emission is seen in the ecliptic plane as both positive and
  negative, depending upon the relative geometry of the IPD and the
  \Planck\ satellite when a given location was observed in the two
  Surveys.  While taking the Survey difference reduces the amplitude of the zodiacal emission signal
  \citep[this will be described in Sect.~\ref{sec:model},
  and in particular Fig.~\ref{fig:zodiMaps}; see also][]{maris2006}, it has the advantage
  that most of the Galactic and extra-galactic signals, the main
  contaminants in the analysis at high frequencies, are removed as well.
  What remains of the Galactic signal arises
  from effects such as beam asymmetries and imperfections in the 
  transfer function removal~\citep{planck2013-p03c}. 
  The conclusions of this analysis remain the same for Galactic cuts of anywhere between 5\deg\ and 20\deg. We generally do not use data within 10\deg\ of the Galactic plane, in order to avoid the most visible of the Galactic contamination visible in the bottom panel of Fig.~\ref{fig:surveymaps}. We emphasize that
  analysing difference maps makes this work less sensitive to 
  Galactic contamination than some other \Planck\ analyses. 
  
  It should be noted that the zodiacal emission is much dimmer than many other background
  components in the \Planck\ data. Whereas the zodiacal emission dominates the sky in some
  \textit{IRAS} and \textit{COBE} bands, this is never the case for \Planck, where the cosmic
  infrared background and Galactic
  dust dominate at high frequencies, and the CMB itself dominates at lower frequencies.
  This makes our differencing scheme appealing, but also restricts the analysis in some ways. 
  It is, for example, difficult to look at individual scans, or slices of the sky, as has
  been done successfully with \textit{IRAS}. We are obligated to use almost the entire
  sky, and to use a model of the zodiacal emission to interpret variations, rather than being
  able to directly interpret the total zodiacal emission on the sky.

 \section{Model}\label{sec:model}
 
  This section describes the creation of the zodiacal and far sidelobe 
  templates that we fit to the Survey difference map in Fig.~\ref{fig:surveymaps}. 
   
  \subsection{Zodiacal Components}
  
  The \textit{COBE}/DIRBE zodiacal emission model is described in depth in~K98, but we review
  the salient parts here. 

  \subsubsection{Diffuse Cloud}\label{sec:cloudmath}
  
  The density of the diffuse IPD cloud,
  having both radial and vertical dependence, 
  is taken to be of the form
  \begin{equation}
   n_\mathrm{c}\left(\vec{R}\right)
   =
   n_0
   R_\mathrm{c}^{-\alpha}
   \left\{
    \begin{array}{cl}
     e^{-\beta\left(\zeta^2/2\mu\right)^\gamma} 
      & \mathrm{if}\ \zeta <  \mu \\
     e^{-\beta\left(\zeta -\mu/2\right)^\gamma} 
      & \mathrm{if}\ \zeta\geq\mu \\
    \end{array}
   \right.,
  \end{equation}
  where 
  \begin{equation}
  	R_\mathrm{c} 
  	= 
  	\sqrt{\left(x-x_0\right)^2 + 
  	      \left(y-y_0\right)^2 +
  	      \left(z-z_0\right)^2},
  \end{equation}
  \begin{equation}
   \zeta \equiv \left|Z_\mathrm{c}\right|/R_\mathrm{c},
  \end{equation}
  \begin{eqnarray}
    Z_\mathrm{c}
    & = & 
    \left(x-x_0\right)
    \sin\left(\Omega_R\right)
    \sin\left(i_R\right) 
    \nonumber
    \\
    & - &
    \left(y-y_0\right)
    \cos\left(\Omega_R\right)
    \sin\left(i_R\right) 
    + \left(z-z_0\right)\cos(i_R),
  \end{eqnarray}
  and $\alpha$, $\beta$, $\gamma$, $\mu$, $x_0$, $y_0$, $z_0$, 
  $\Omega_R$, and $i_R$
  are parameters describing the location and shape of the cloud. 
  This form (and others used elsewhere that are similar) is based on an
  approximation of a model of particles orbiting the Sun, accounting for drag from 
  the Poynting-Robertson effect, and with a modified fan distribution used
  to describe the changes in density above and below the plane of the 
  ecliptic. See K98 for details and references. 

\begin{figure*}[htbp]
  \centering
  $\begin{array}{ccc}
     \includegraphics[width=60mm]{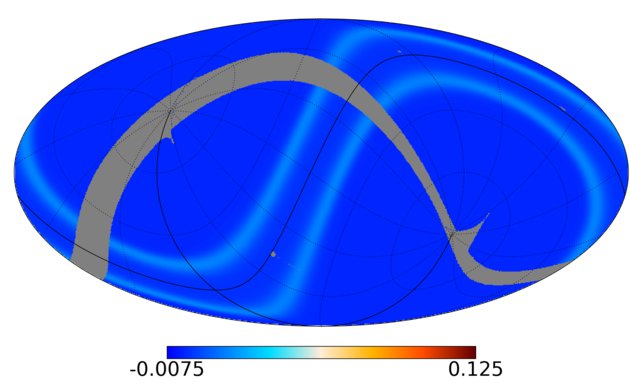}
   & \includegraphics[width=60mm]{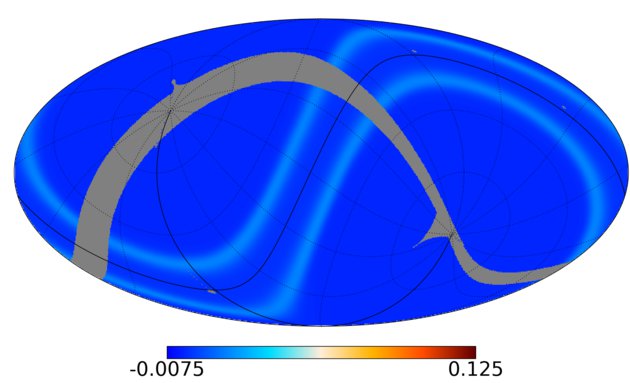}
   & \includegraphics[width=60mm]{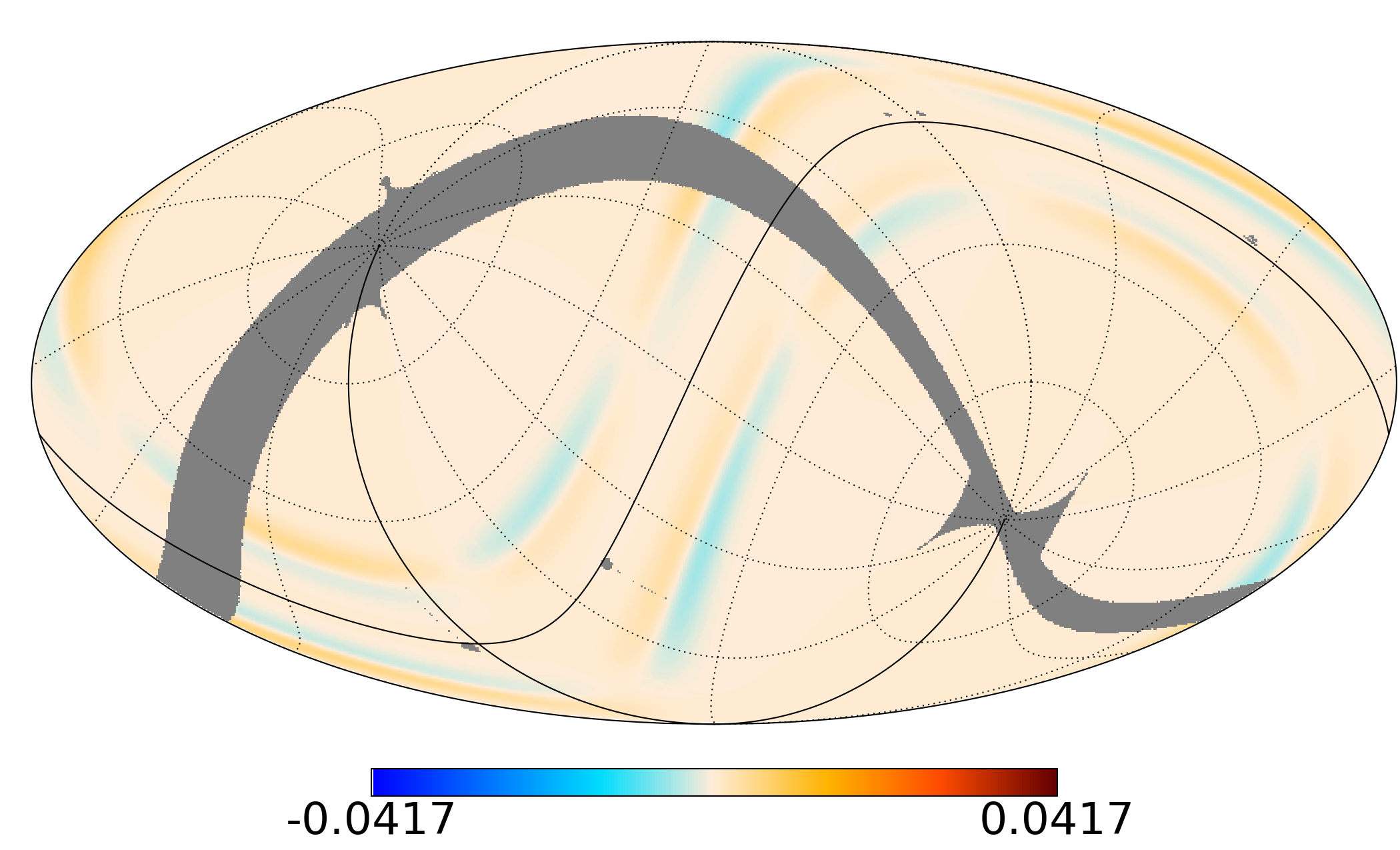}
   \\
     \includegraphics[width=60mm]{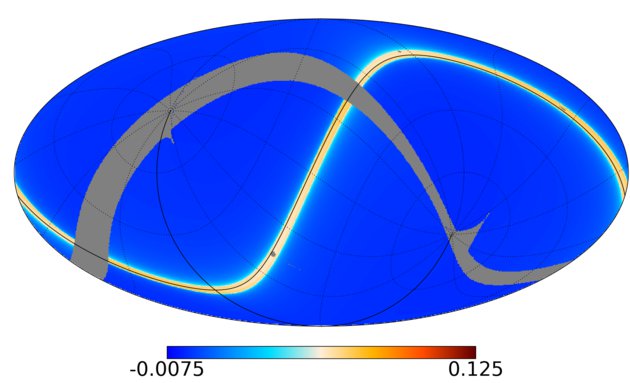}
   & \includegraphics[width=60mm]{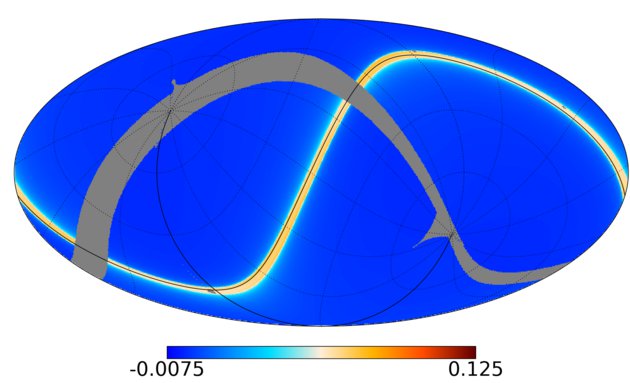}
   & \includegraphics[width=60mm]{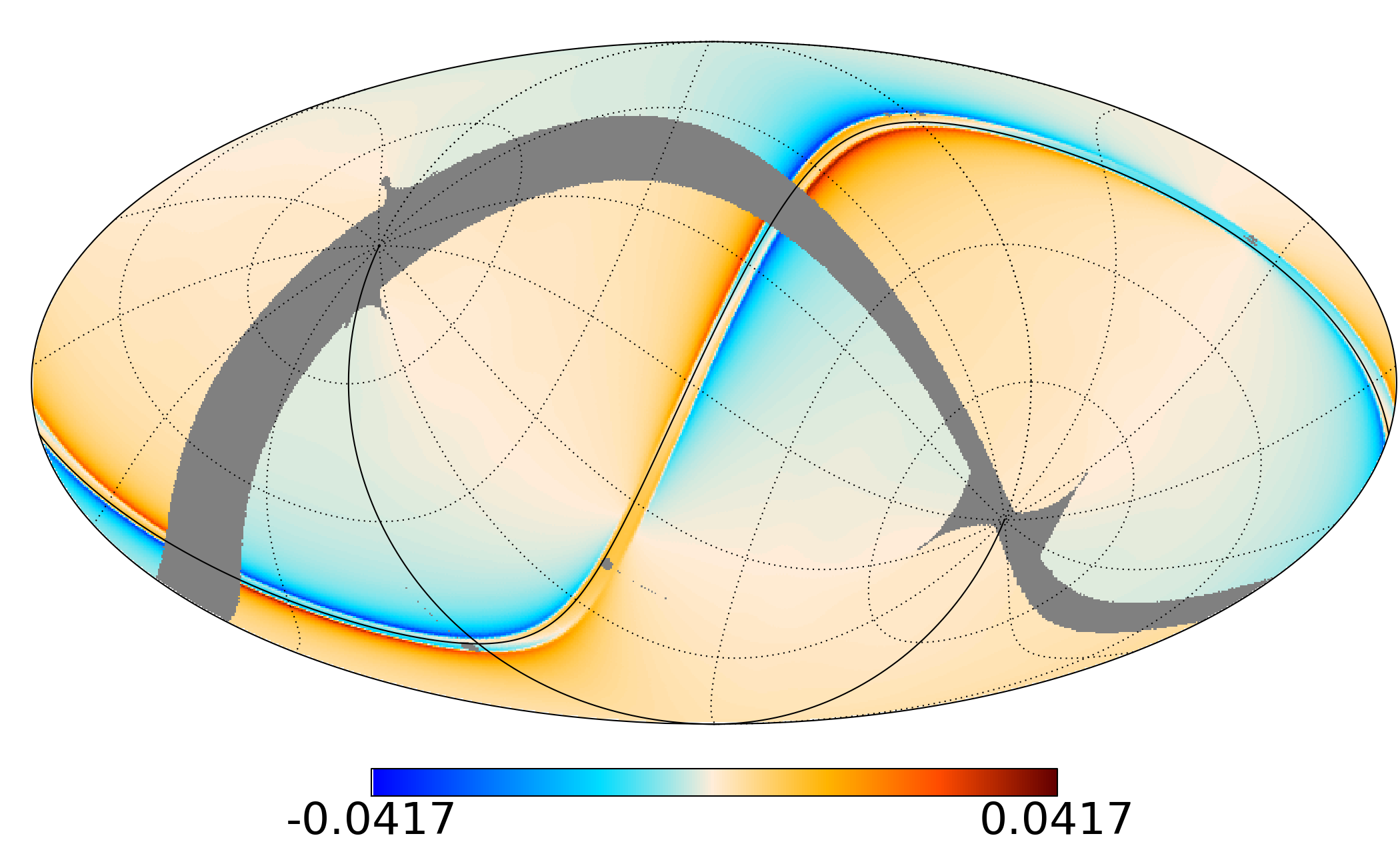}
   \\
     \includegraphics[width=60mm]{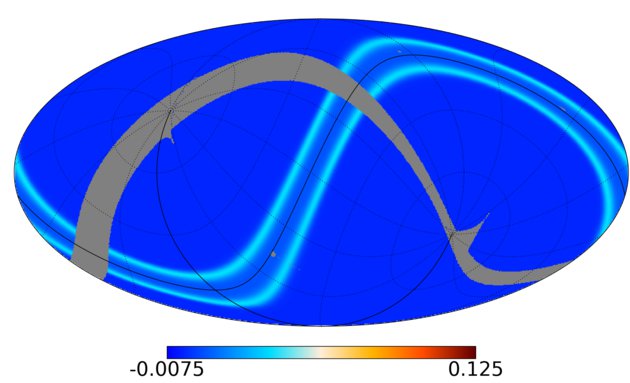}
   & \includegraphics[width=60mm]{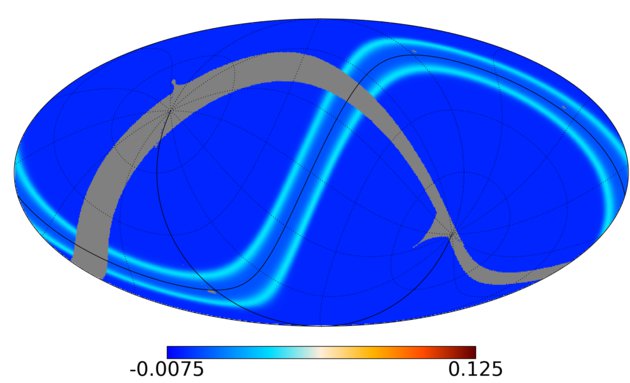}
   & \includegraphics[width=60mm]{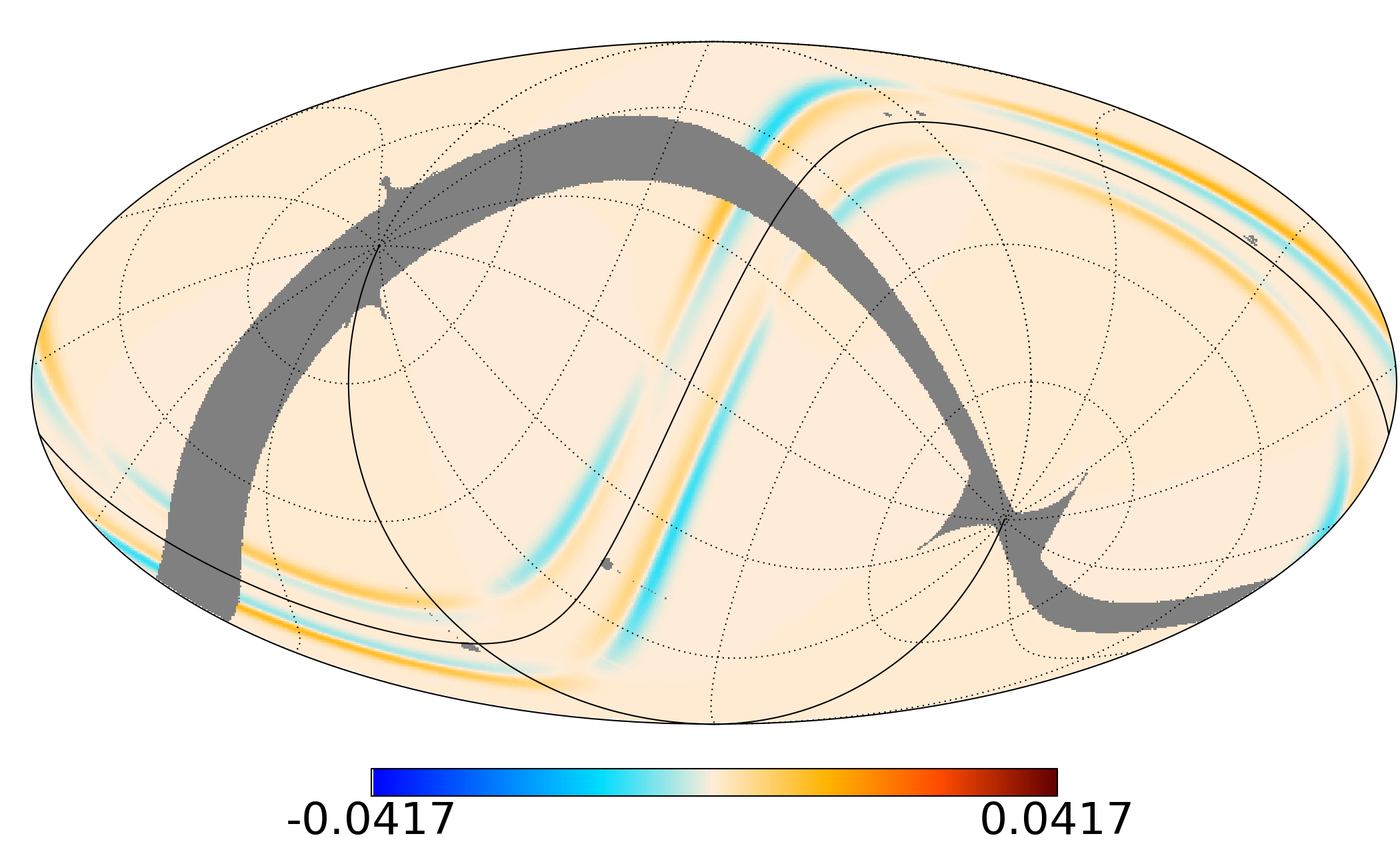}
   \\
     \includegraphics[width=60mm]{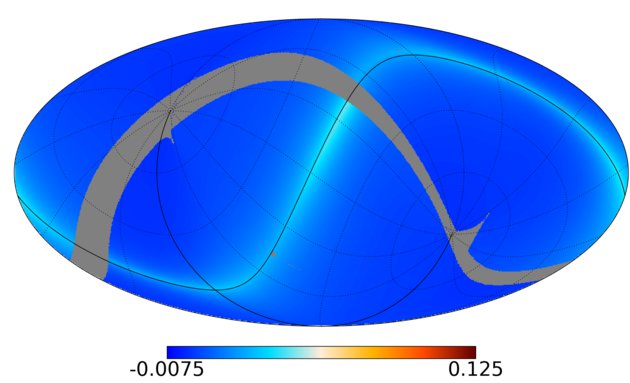}
   & \includegraphics[width=60mm]{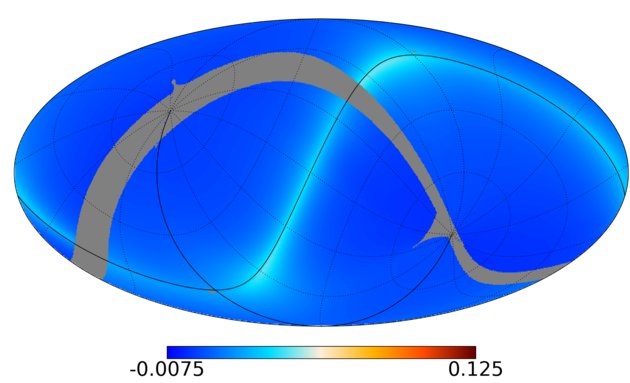}
   & \includegraphics[width=60mm]{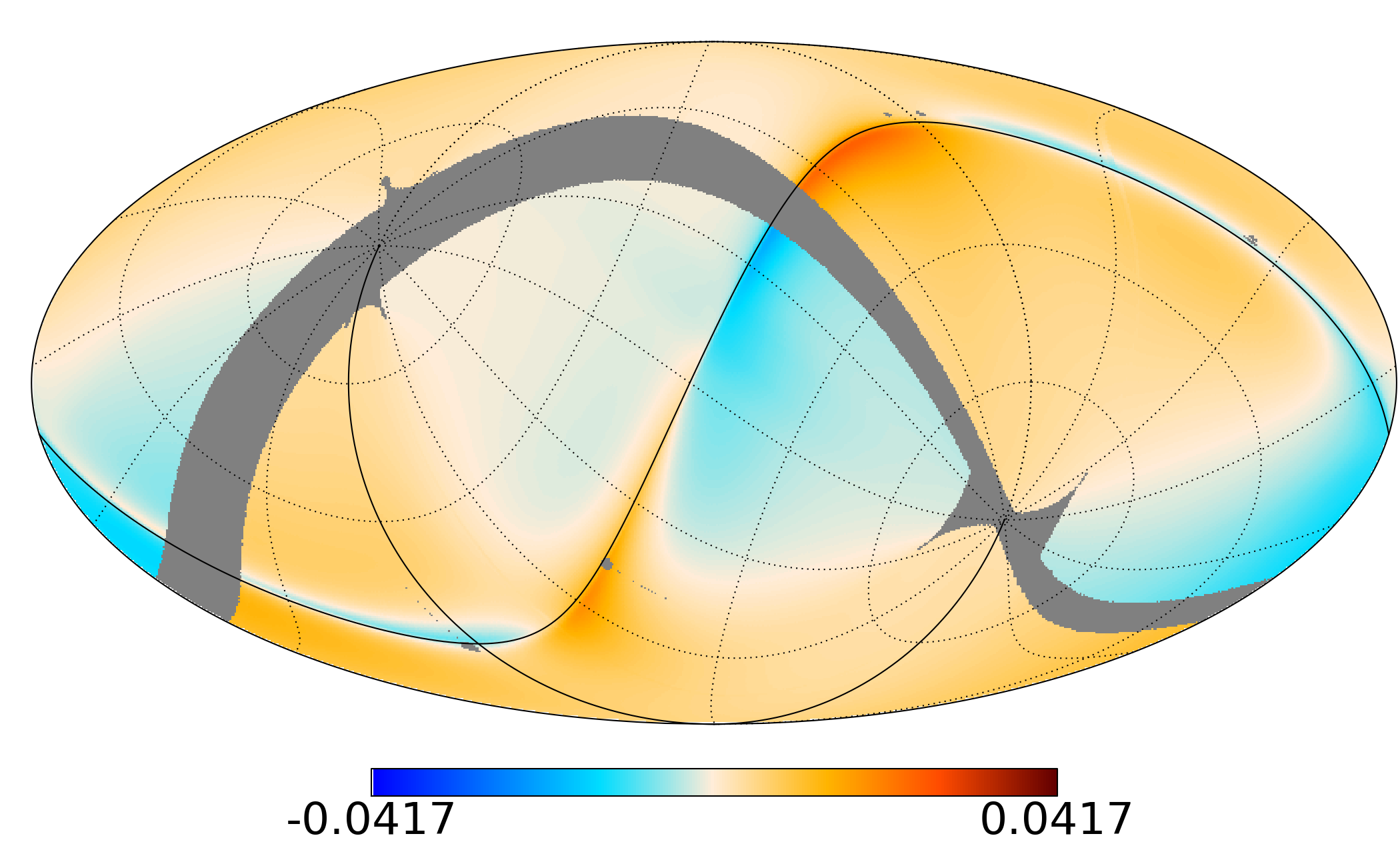}
   \\
     \includegraphics[width=60mm]{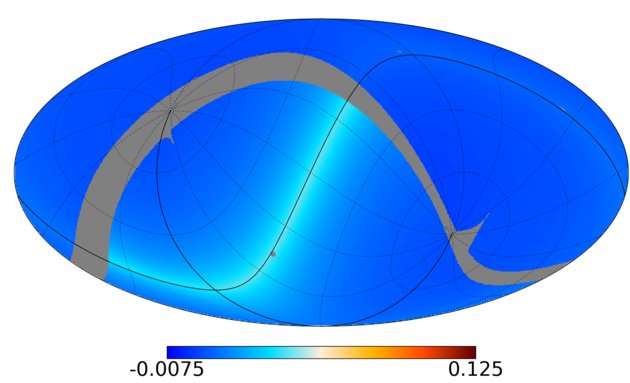}
   & \includegraphics[width=60mm]{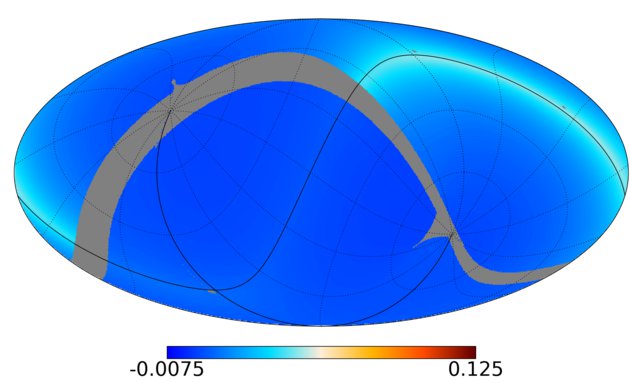}
   & \includegraphics[width=60mm]{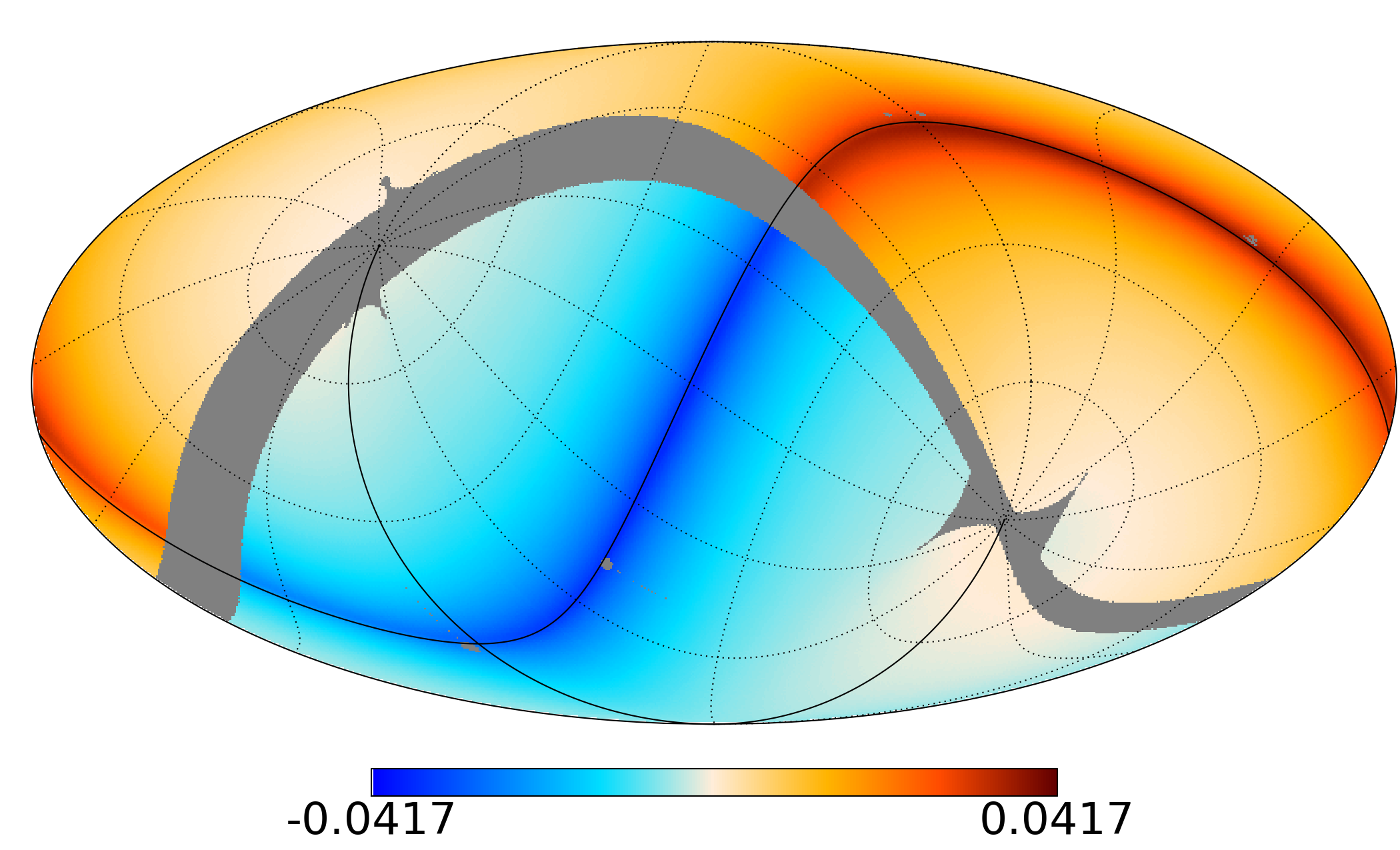}
   \\
     \includegraphics[width=60mm]{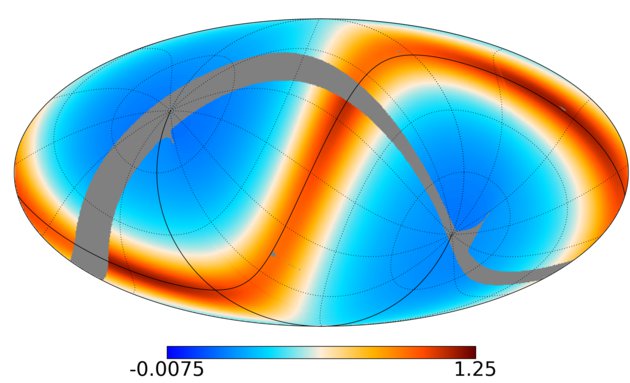}
   & \includegraphics[width=60mm]{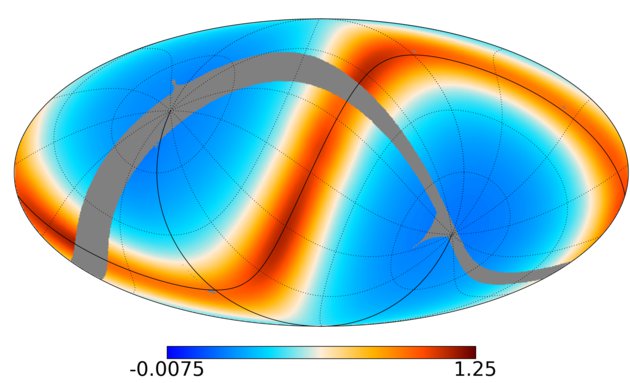}
   & \includegraphics[width=60mm]{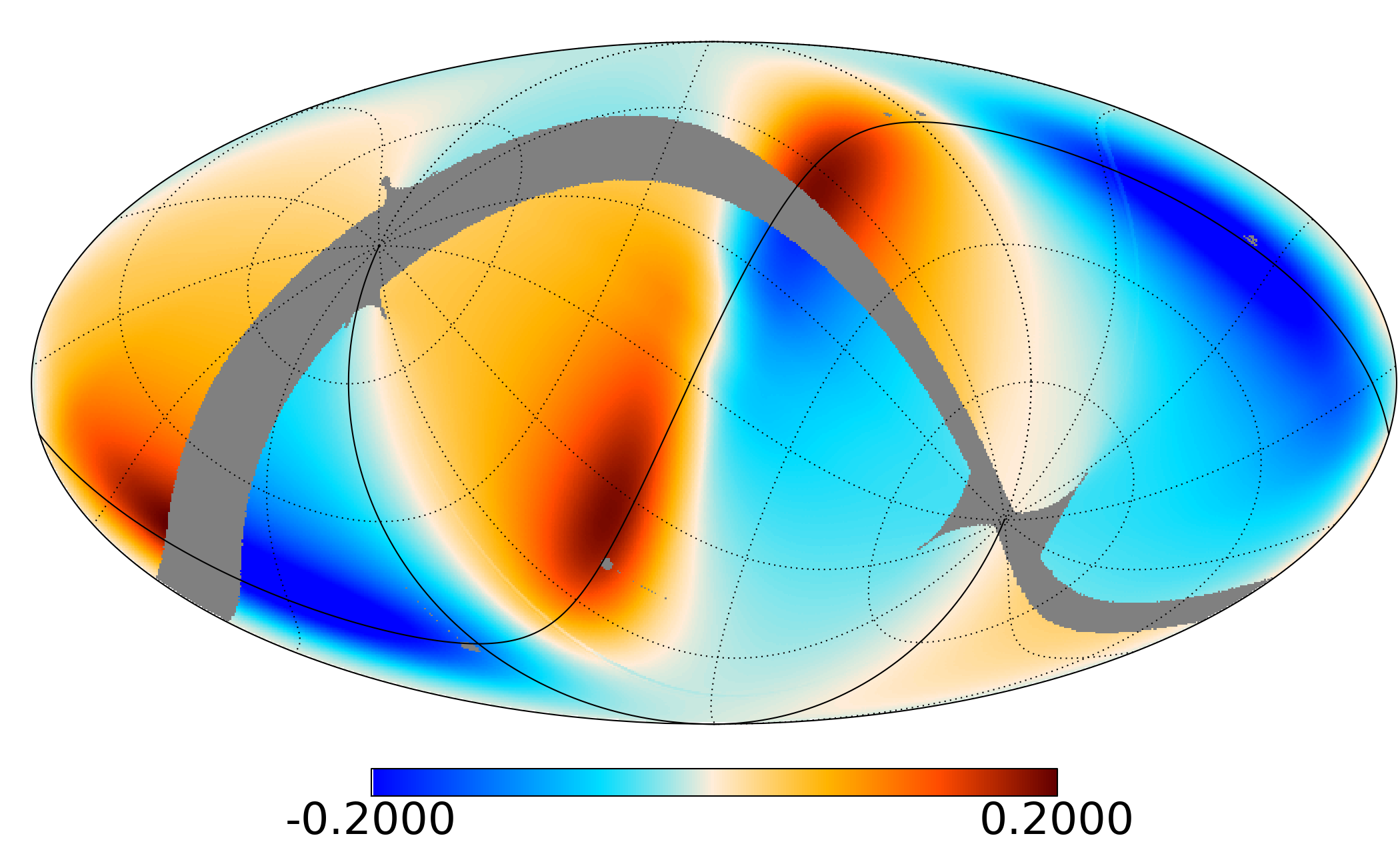}
  \end{array}$
  \caption{Estimated 857\,GHz zodiacal emission templates, in MJy/sr,
   based on the model of~K98, in Galactic coordinates, for Survey~1 (left), 
   Survey~2 (center), and Survey~2 minus Survey~1 (right). From top to bottom we show Dust Band~1,
   Dust Band~2, Dust Band~3, the Circumsolar Ring, the Earth-Trailing Feature, and finally
   the Diffuse Cloud. Note that the scales for the right-hand
   column are different from those of the left-hand and centre columns, and that
   the scale for the bottom row is different from that of the others.
  }
  \label{fig:zodiMaps}
  \end{figure*}

  The numerical values for the parameters can be found in~K98, 
  or from the LAMBDA 
  website\footnote{Legacy Archive for Microwave Background Data Analysis 
  -- NASA (\url{http://lambda.gsfc.nasa.gov/}).}. 
  This is shown, for both Survey 1 and Survey 2, as well as their difference, 
  in the bottom row of Fig.~\ref{fig:zodiMaps}. These plots were made assuming that 
  the particles emit as blackbodies, which is only an
  approximation, as we will see in Sec.~\ref{sec:fit}.
  
  \subsubsection{Dust Bands}
   \label{sec:bandmath} 
  
   The zodiacal dust bands were first seen by \textit{IRAS}~\citep{Low1984}, and 
   appear as pairs of bright, parallel bands equally spaced above and below
   the ecliptic plane. They were quickly associated with asteroid 
   families, and then understood 
   to be the relics of asteroid collisions or collapses~\citep{Dermott1984,sykes86}. 
   \cite{Reach1997} study them in detail. 
   
   The K98 model contains three bands\footnote{While three bands 
   are described in the text of K98,
   there are actually four in the code. There is, however, no
   ambiguity, as the density of the fourth band is set to zero in the
   code at the LAMBDA site.} called Bands 1, 2, and~3. 
   They appear at ecliptic latitudes $\pm 1\pdeg4$, $\pm 10\deg$, and $\pm 15\deg$.
   \textit{IRAS}, having higher angular resolution than DIRBE, found many bands, called 
   $\alpha$, $\beta$, $\gamma$, E/F, G/H, J/K, and M/N~\citep{Sykes1990}, 
   though some are more firmly detected than others.
   
   The K98 Band~1 (= \textit{IRAS}\ Band $\gamma$), was originally 
   associated with the Eos family of
   asteroids~\citep{Dermott1984}, although \cite{Grogan2001} 
   called this into question and \citet[hereafter N03]{Nesvorny2003} 
   found better correspondence with the Veritas family of asteroids.
  
   K98 note that their Band~2 is a blend of \textit{IRAS}\ Bands $\alpha$ and~$\beta$~\citep{Sykes1990}. \cite{sykes86} tentatively 
   associated the $\alpha$ band with the Themis family of asteroids. 
   \citet[][hereafter N08]{Nesvorny2008} has 
   narrowed this association to a cluster within this family 
   associated with the Beagle asteroid. The $\beta$
   band was associated by \cite{sykes86} 
   with the Koronis family of asteroids, and
   N03 narrowed this to the Karin 
   cluster within the Koronis family. Anticipating the discussion
   in Sect.~\ref{subsec:Bands}, we note also that 
   the $\beta$ band appears brighter than the $\alpha$
   band~\citep{Sykes1988,Reach1997,Nesvorny2008}.
   
   K98 states that their Band~3, that furthest from the ecliptic plane, 
   has been associated with both the Io and Maria families of
   asteroids~\citep{Sykes1988,Reach1997}, 
   corresponding to \textit{IRAS}\ Bands J/K and M/N. N03
   have more recently noted, however, that 4652 Iannini and/or 845 Naema may be
   better asteroid associations for the J/K band-pair, and 
   that 1521 Seinajoki may work better for the M/N pair.
   We summarize these associations in Table~\ref{tab:bandInfo}.

\begin{table*}
\begingroup
\caption{Zodiacal bands identified by {\it COBE} and {\it IRAS}, and information about the asteroid families with which they have been associated.}
\label{tab:bandInfo}
\nointerlineskip
\vskip -3mm
\footnotesize
\setbox\tablebox=\vbox{
 \newdimen\digitwidth 
 \setbox0=\hbox{\rm 0} 
 \digitwidth=\wd0 
 \catcode`*=\active 
 \def*{\kern\digitwidth}
 \newdimen\signwidth 
 \setbox0=\hbox{+} 
 \signwidth=\wd0 
 \catcode`!=\active 
 \def!{\kern\signwidth}
\halign{\hfil#\hfil\tabskip 1em&
    \hfil#\hfil\tabskip=3.0em&
    \hfil#\hfil\tabskip=1.0em&
    \hfil#\hfil\tabskip=3.0em&
    #\hfil\tabskip=1.0em&
    \hfil#\hfil&
    \hfil#\hfil&
    \hfil#\hfil&
    \hfil#\hfil\tabskip=0pt\cr 
\noalign{\doubleline}
\multispan2\hfil\sc{\it COBE\/}\hfil&
\multispan2\hfil\sc{\it IRAS\/}\hfil&\multispan5\hfil\sc Associated Asteroid Family\hfil\cr
\noalign{\vskip -3pt}
\multispan2\hrulefill&\multispan2\hrulefill&\multispan5\hrulefill\cr
\omit\hfil Name$^{\rm a}$\hfil&$\delta_\zeta$$^{\rm b}$&Name$^{\rm c}$&$i_p$$^{\rm d}$&
     \omit\hfil Name$^{\rm e}$\hfil&$i_p$$^{\rm f}$&Type$^{\rm g}$&Age$^{\rm h}$&a$^{\rm i}$\cr
\noalign{\vskip 3pt\hrule\vskip 5pt}     
1&*8.78&$\gamma$&*9.35&490 Veritas&*9.26&  C/Ch/\dots&        8.3&3.169\cr
\noalign{\vskip 7pt}
2&*1.99&$\alpha$&*1.34&656 Beagle&*1.34&  \dots/\dots/C& $\lsim$10&3.157\cr
\omit&   &$\beta$&*2.11&832 Karin&*2.11&\dots/\dots/S&         5.8&2.866\cr
\noalign{\vskip 7pt}
3&15.0*&J/K&     12.0*&845 Naema&11.96&\dots/C/\dots&  $\gsim$20& \dots\cr
&&&&                   4562 Iannini&12.17&\dots/\dots/S&  $\lsim$5&2.644\cr
\noalign{\vskip 4pt}
&&      M/N&     15.0*&1521 Seinajoki&15.02&\dots/\dots/\dots&\dots&2.852\cr
\noalign{\vskip 5pt\hrule\vskip 3pt}}}
\endPlancktablewide
\tablenote {{\rm a}} K98 band designation.\par
\tablenote {{\rm b}} $\delta_\zeta$ parameter of the K98 zodiacal emission model. This parameter determines roughly at which ecliptic latitudes, in degrees, the band appears.\par
\tablenote {{\rm c}} \textit{IRAS}\ bands associated with the given \textit{COBE}\ band.\par
\tablenote {{\rm d}} Modelled proper inclination of the given \textit{IRAS}\ band.  The first three are from~\cite{Grogan2001}, while the last two are from~\cite{Sykes1988}.\par
\tablenote {{\rm e}} Asteroid family associated with the given \textit{IRAS}\ band.  These all come from N03, except that for 832 Karin, which comes from~N08.\par
\tablenote {{\rm f}} Average proper inclination for the associated asteroid family.  These all come from N03, except that for 832 Karin, which comes from~N08.\par
\tablenote {{\rm g}} Spectral type/classifications.  The first two entries  in the triplets correspond to the Tholen and SMASSII classes~\cite{Bus2002}, while the third corresponds to the SDSS-based classification~\citep{Carvano2010}. Ellipses are used to
      indicate that the given classification was not found.\par
\tablenote {{\rm h}} Time since the asteroid disruption that created the associated asteroidal family, in Myr. These come from N03, except for 832 Karin, which comes from~N08.\par
\tablenote {{\rm i}} Proper semi-major axis of the asteroidal family associated with the band, in AU.  These come from N03, except for 832 Karin, which comes from~N08.\par
\endgroup
\end{table*}

   For each of the three dust bands in the \textit{COBE}\ model, the density is given by
   \begin{equation}
    n_{B}\left(\vec{R}\right)
    =
    \frac{3N_0}{R}
    e^{-\left(\zeta/\delta_\zeta\right)^6}
    \left(1 + \frac{(\zeta/\delta_\zeta)^{p}}{v_B}\right)
    (1-e^{-(R/\delta_R)^{20}}),
    \label{eq:bandDensity}
   \end{equation} 
   where $B$ denotes the band, and we have used a simplified notation based on that
   of K98, where one can also find the numerical values for 
   the parameters. Note that Eq.~\ref{eq:bandDensity}
   matches the code used for the zodiacal model 
   (which can be found on the LAMBDA website), but that there is a factor 
   of $1/v_B$ difference between Eqn.~\ref{eq:bandDensity} and Eqn.~8 of K98
   (note that $v_B$ is a shape parameter, not a frequency).
   Also, K98 assumed that the emissivities of the three sets of 
   bands were all equal. We relax
   this assumption below and allow the emissivities of 
   each of the sets of bands to be different.  
   The bands are shown, assuming unit emissivity,  
   as the first, second, and third rows in Fig.~\ref{fig:zodiMaps}.
   
  \subsubsection{Circumsolar Ring and Earth-Trailing Feature}
   \label{sec:ringmath} 
   
   IPD particles drifting towards the Sun in their orbits can become trapped in
   orbital resonances near the Earth's orbit, thus creating a ring of enhanced density 
   in these regions~\citep{Dermott1984}.
   The functional form for the density of this Circumsolar Ring is taken to be
   \begin{equation} 
    n_{B}\left(\vec{R}\right)
    =
    n_\mathrm{SR}\cdot e^{-\left(R-R_\mathrm{SR}\right)^2/\sigma_\mathrm{rSR}^2-\left|Z_R\right|/\sigma_\mathrm{zSR}}.
   \end{equation}
   Similar to the treatment of the bands, K98 assumed that the 
   emissivity of the 
   Circumsolar Ring was the same as that of the Earth-Trailing Feature, below.
   We also relax this assumption, and allow them to be different. 
   The shape of the expected signal from the Circumsolar Ring is 
   shown in the fourth row of Fig.~\ref{fig:zodiMaps}.
   
  The density of the Earth-Trailing Feature is given by
  \begin{equation} 
   n_{B}\left(\vec{R}\right)
    = 
   n_\mathrm{TB}\cdot 
   e^{-\left(R-R_\mathrm{TB}\right)^2/\sigma_\mathrm{rTB}^2-\left|Z_R\right|/\sigma_\mathrm{zTB}
                  -\left(\theta-\theta_\mathrm{TB}\right)^2/\sigma_\mathrm{\theta TB}^2}. 
  \end{equation}
  We note that for both the Circumsolar Ring and the Earth-Trailing Feature,
  there is an error in the text of K98 -- a factor of~2 in the denominator of 
  the first and third terms in the exponential has been added in the text,
  compared to what is in the code. We follow the code. 
  The expected signal from the Earth-Trailing Feature is shown in
  the fifth row of Fig.~\ref{fig:zodiMaps}.

  \subsubsection{Integrated Emission}
  
  The total zodiacal emission is calculated as
  \begin{equation}
   I_x\left(\nu\right) 
   = 
   \epsilon_x\int d\vec{R}
    \cdot 
   n_x\left(\vec{R}\right)
    \cdot 
   B\left(\nu, T\left(\vec{R}\right)\right),
  \end{equation}
  where $x$ is the zodiacal component, $\nu$ is the frequency, $\vec{R}$
  is a location in the Solar System, and 
  $B\left(\nu, T\left(\vec{R}\right)\right)$ is the Planck
  function for the frequency and temperature at the specific location, 
  given by $T_0/R^\delta$, with $T_0$ and $\delta$ being parameters. 
  $\epsilon$ is the emissivity for the given component, which we will be
  finding with our fit, and $n_x$ is the density for the given
  component, described above. The integral is carried out along the line of sight, from the 
  location of the satellite to 5.2\,AU.
  
 \subsection{Galactic Emission Seen Through Sidelobes}
  \label{subsec:galFSL}
 
\begin{figure}[htbp]
   \centering 
   \includegraphics[width=88mm]{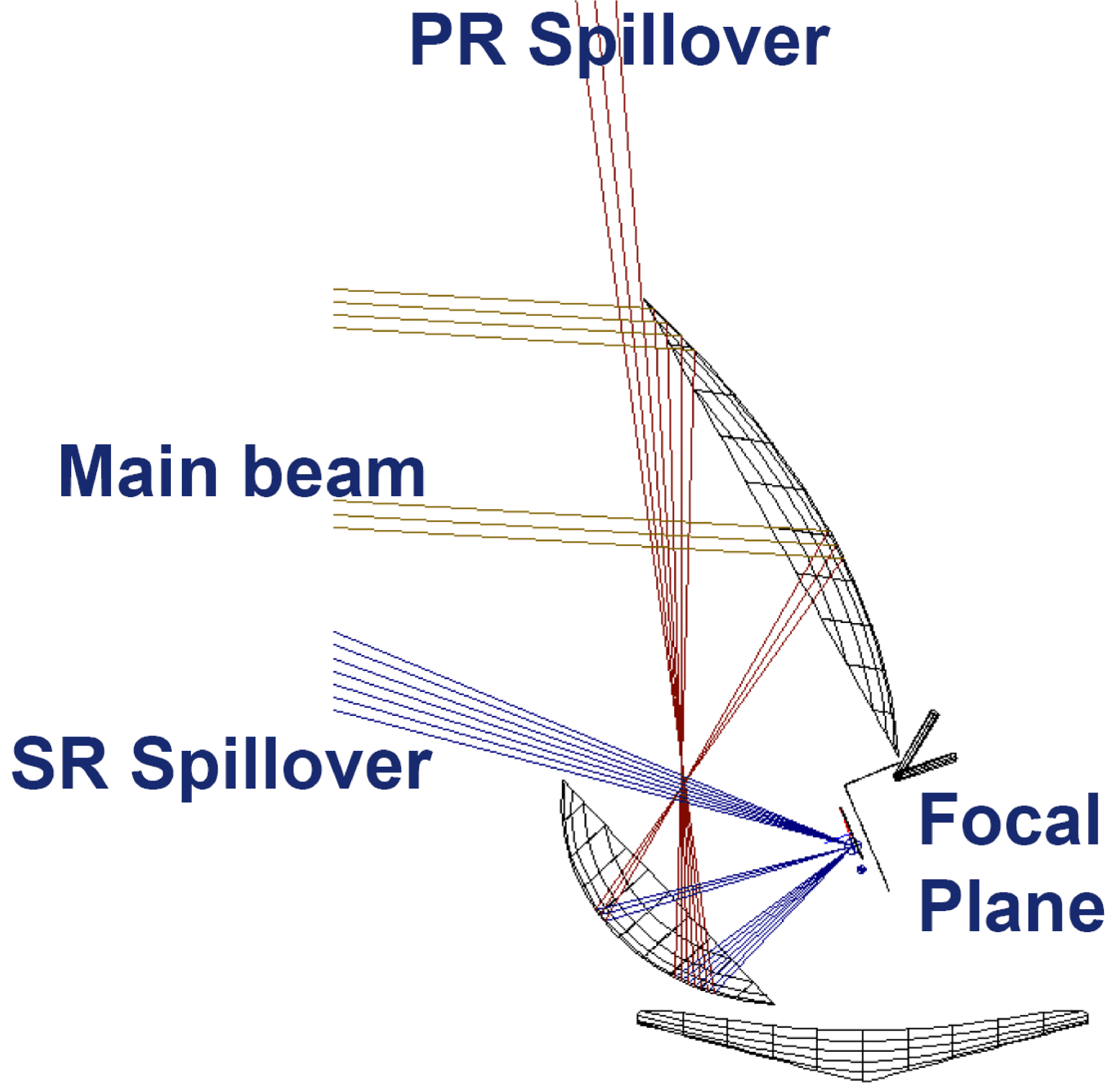}
   \caption{Origin of far sidelobes. 
    The ``SR Spillover'' (for Secondary Reflector Spillover; 
    the lowest set of rays on
    the left of the figure) arrives at the focal plane from outside
    the secondary mirror, directly from the sky. 
    The ``PR Spillover'' (for Primary Reflector Spillover), 
    arrives at the focal plane from above the
    primary mirror and reflects off the secondary to arrive
    at the focal plane. The set of rays between these two contributions 
    represents the main beam. The ``baffle'' contribution, light reaching the 
    focal plane after reflecting from the inner sides of the baffles, 
    is not shown here. It is often included as part of the SR Spillover. 
    Adopted from~\cite{tauber2010b}.
   } 
   \label{fig:sidelobeDiagram}
  \end{figure}

  For the \Planck\ telescope, we estimate a contribution to the total solid 
  angle from far sidelobes -- defined as beam response more than 5\deg\ from 
  the centre of each beam (\citealt{planck2013-p02d}; \citealt{planck2013-p03c}) -- of a fraction of a percent.
  Figure~\ref{fig:sidelobeDiagram} shows the main far sidelobe paths.
  
  The secondary reflector (SR) spillover arises from radiation that reaches the focal
  plane without reflecting off the primary reflector. As such, 
  a major component is radiation from the general direction of the telescope
  boresight, though well outside the main beam. 
  The ``baffle'' contribution to the SR spillover results from radiation that reflects 
  off the baffles to arrive at the focal plane.   
  The primary reflector (PR) spillover arises from radiation that comes to 
  the satellite from just above the primary mirror, 
  reflects off the secondary mirror and arrives
  at the detectors. 
  
  At the highest \Planck\ frequencies, the Galactic centre is bright enough to
  be seen through the far sidelobes, even if faintly. 
  Since the orientations of these
  sidelobes change as the instrument scans, the Survey differences done
  to detect the zodiacal emission are also sensitive tests of the far side lobes. 
  Though the Galactic emission mechanism and amplitude is different, 
  analogous effects are discussed for the 
  LFI in~\cite{planck2013-p02d}.  
    
  To this point, resource constraints have limited this study to a single far sidelobe calculation for all detectors. We use a {\tt GRASP}\footnote{\url{http://www.ticra.com/products/software/grasp}} calculation of the far sidelobes for the 353-1 horn~\citep[see Fig.~9 of][]{planck2013-p03c}. The calculation is based on the multi-reflector, geometrical theory of diffraction and used backward ray-tracing. We do not attempt to correct for differences in either frequency or location for other horns. While this is not optimal, and will be improved in later releases, the primary, large-scale features of the far sidelobes are defined by the telescope, rather than the horns or their placement, so first-order effects should be captured. Some of the limitations imposed by this assumption are discussed in Sect.~\ref{sec:farSidelobeResults}. 
  
\begin{figure*}[htbp]
  \centering
  $\begin{array}{ccc}
     \includegraphics[width=60mm]{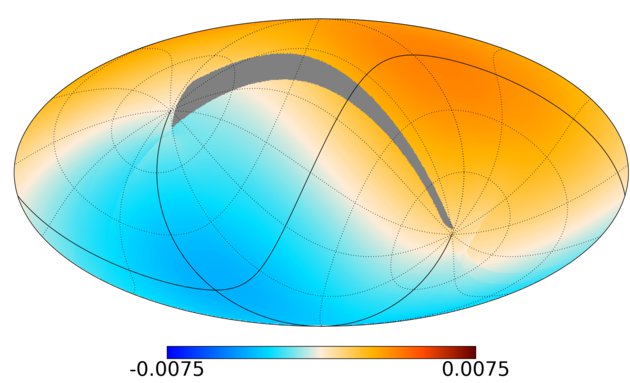}
   & \includegraphics[width=60mm]{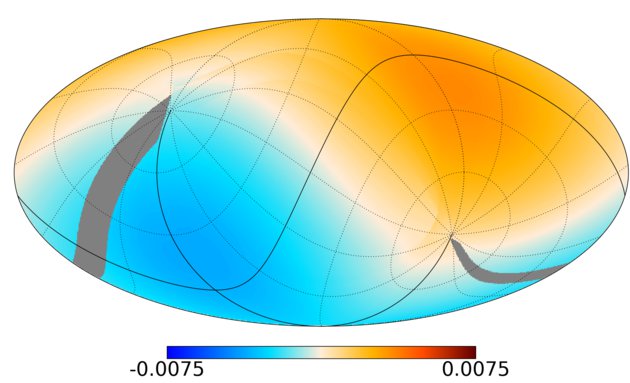}
   & \includegraphics[width=60mm]{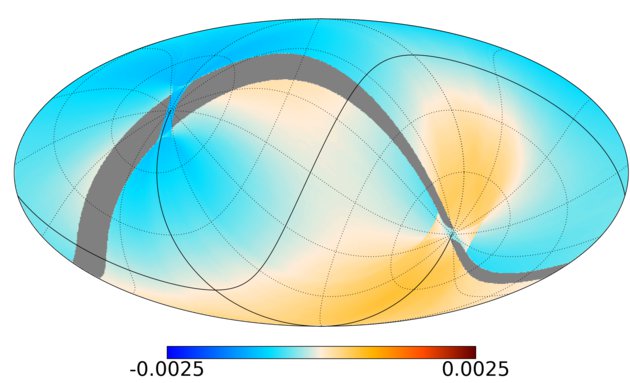}
   \\
     \includegraphics[width=60mm]{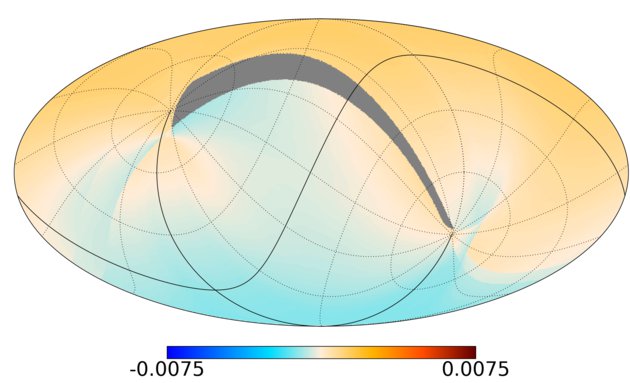}
   & \includegraphics[width=60mm]{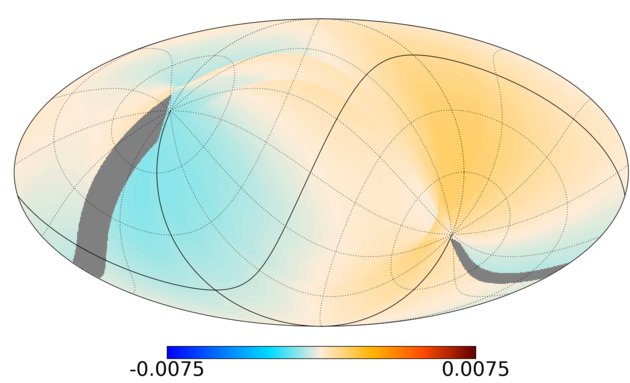}
   & \includegraphics[width=60mm]{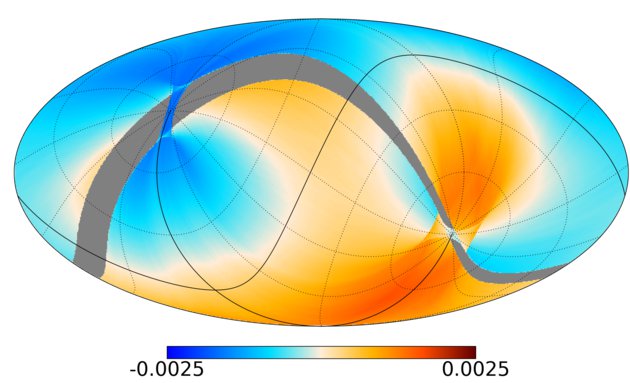}
   \\
     \includegraphics[width=60mm]{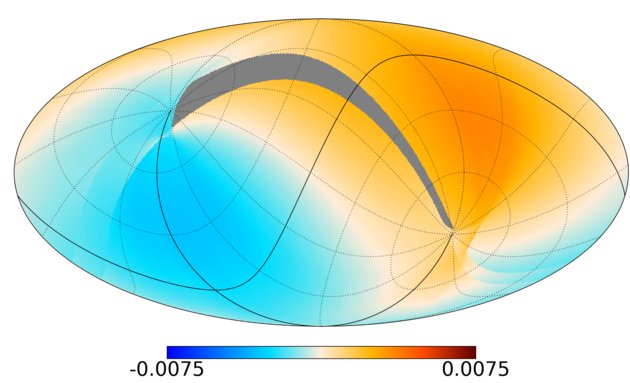}
   & \includegraphics[width=60mm]{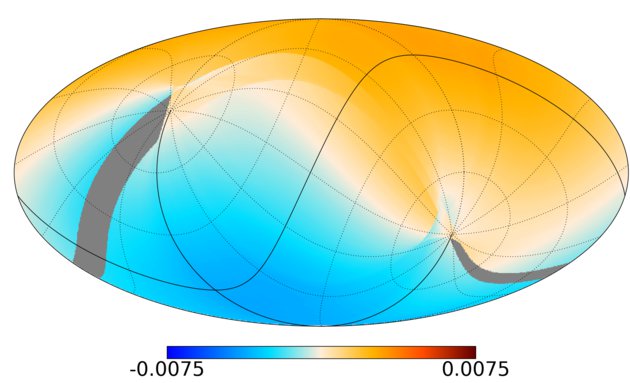}
   & \includegraphics[width=60mm]{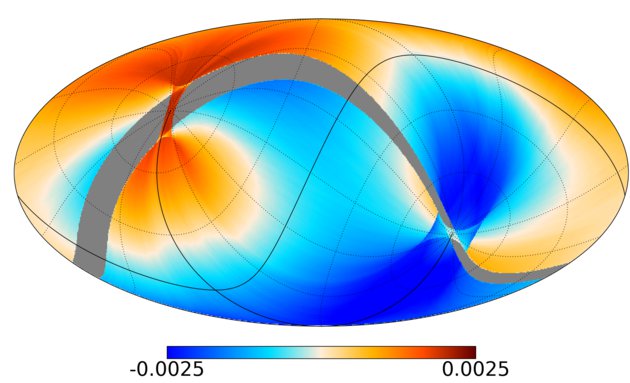}
   \\
     \includegraphics[width=60mm]{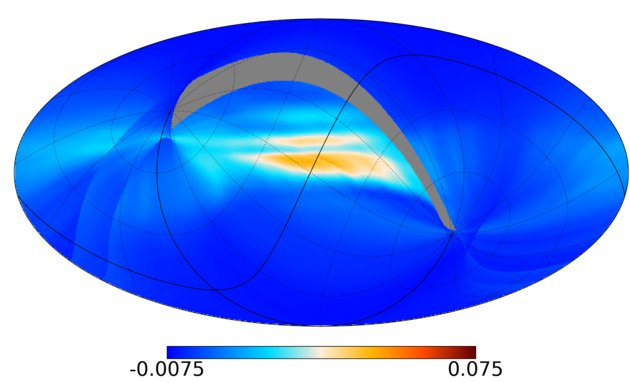}
   & \includegraphics[width=60mm]{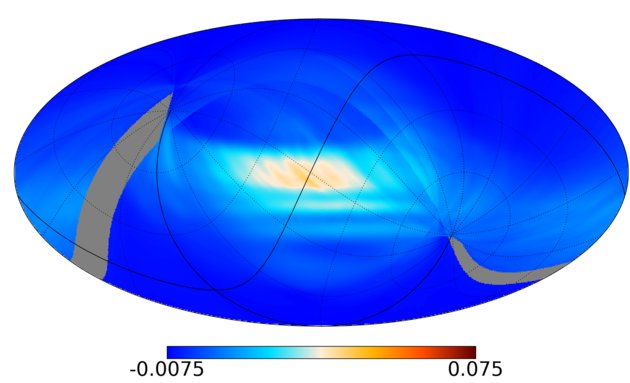}
   & \includegraphics[width=60mm]{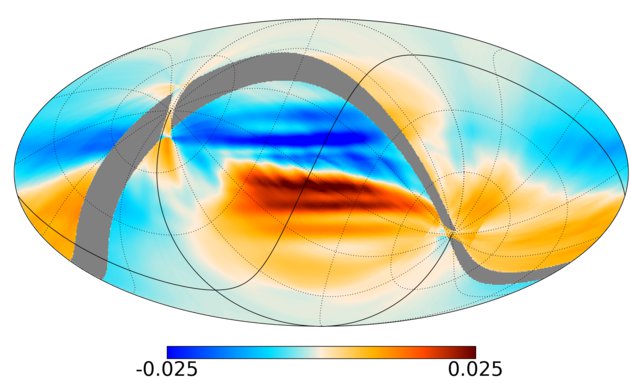}
   \\
     \includegraphics[width=60mm]{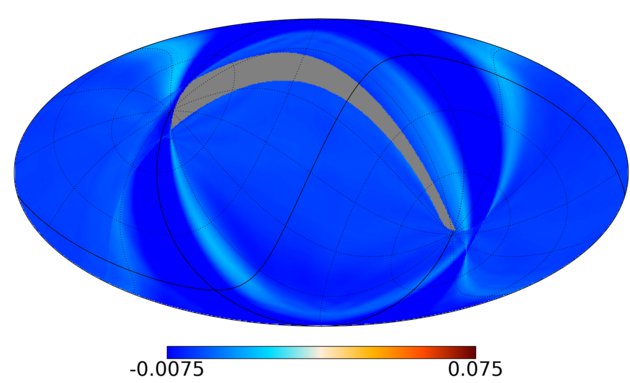}
   & \includegraphics[width=60mm]{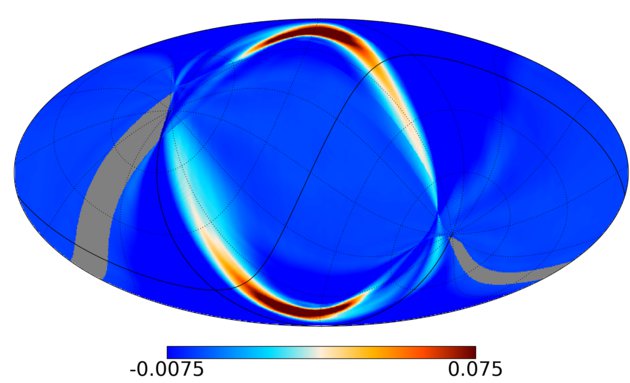}
   & \includegraphics[width=60mm]{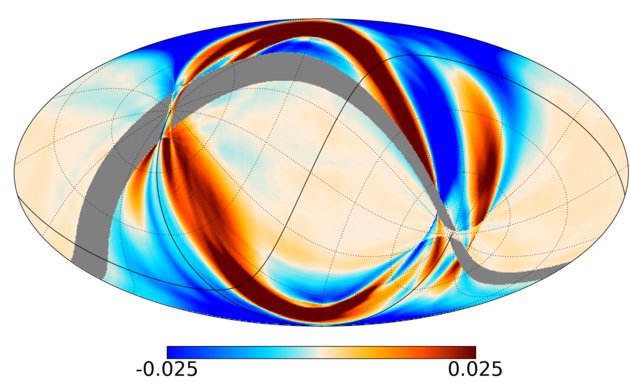}
   \\
     \includegraphics[width=60mm]{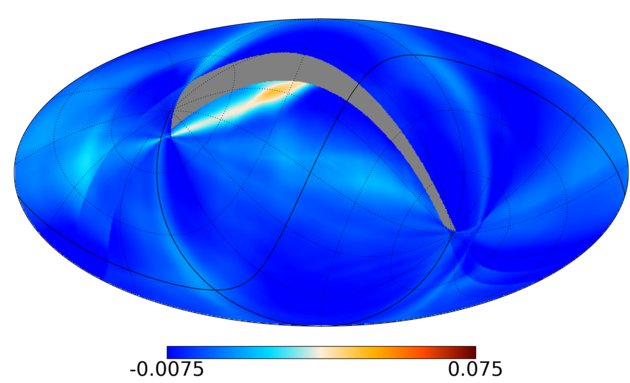}
   & \includegraphics[width=60mm]{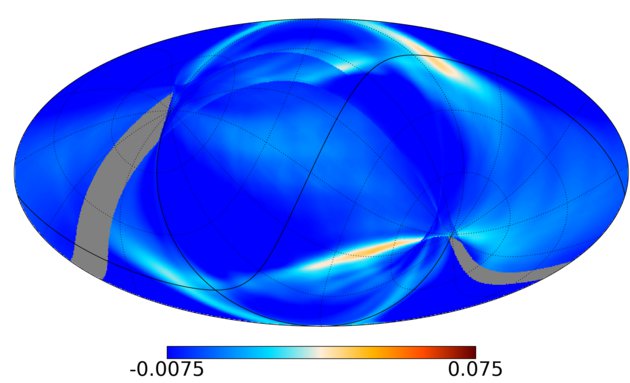}
   & \includegraphics[width=60mm]{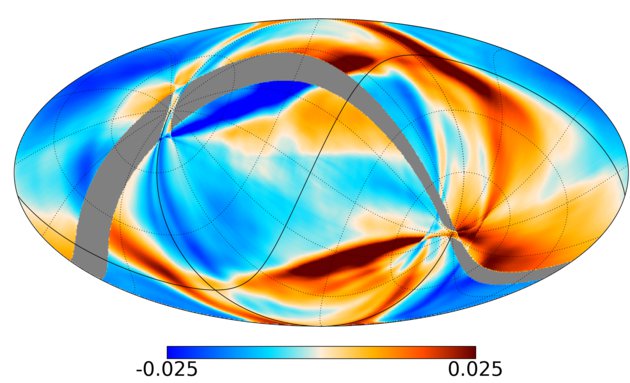}
  \end{array}$
   \caption{Templates of the dipole and the Galaxy, in MJy/sr, seen through our 
    far sidelobes for Surveys 1 (left) and 2 (centre), 
    and for the difference in these two (right). 
    Row~1: dipole seen through the direct SR contribution.
    Row~2: dipole seen through the PR contribution.
    Row~3: dipole seen through the baffle SR contribution.
    Row~4: Galaxy seen through the direct SR contribution.
    Row~5: Galaxy seen through the PR contribution.
    Row~6: Galaxy seen through the baffle SR contribution.
    The simulations in Rows~4--6 are made using 857\,GHz data passed through
    the far sidelobe calculation described in Sect.~\ref{subsec:galFSL}. 
    The scales are different for the top and bottom three rows, and 
    for the first two and the last columns. 
   }
   \label{fig:FSLs}
  \end{figure*}

  To make templates of what we might see from the Galactic centre through
  the far sidelobes, we use the \Planck\ simulation software described in
  \cite{reinecke2006}, with the {\tt GRASP}-calculated 353-1 
  far sidelobe pattern and the actual calibrated \Planck\ frequency maps of the sky 
  as inputs. 
  The far sidelobe templates are made at the timeline level
  and run through the relevant parts of the pipeline software. In particular, 
  the offset removal, or ``destriping'' must be done on the templates before fitting 
  in order to obtain reasonable results. 
  The templates made using these FSL calculations with a \Planck\ 857\,GHz sky
  as input are shown in the bottom three rows of Fig.~\ref{fig:FSLs}.   
  
  One factor for which we do
  not account with these templates is the difference in spillover between 
  the different frequencies~\citep{Lamarre2010,tauber2010b}. Since
  we illuminate more of the telescope at lower frequencies than at higher
  frequencies, if our templates simply scale with the spillover, we would
  expect different fit values for our templates at different frequencies. 
 
\begin{figure*}[htbp]
  \centering
  $\begin{array}{ccc}
     \includegraphics[width=60mm]{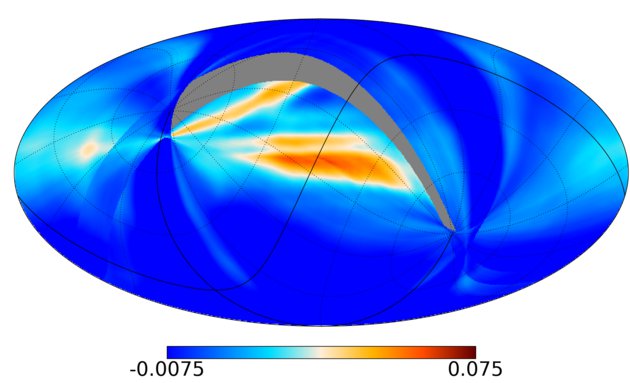}
   & \includegraphics[width=60mm]{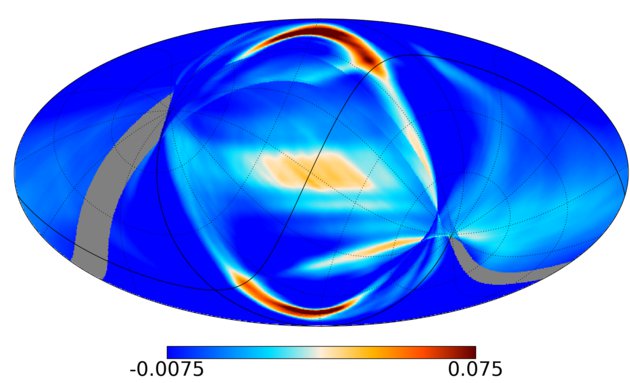}
   & \includegraphics[width=60mm]{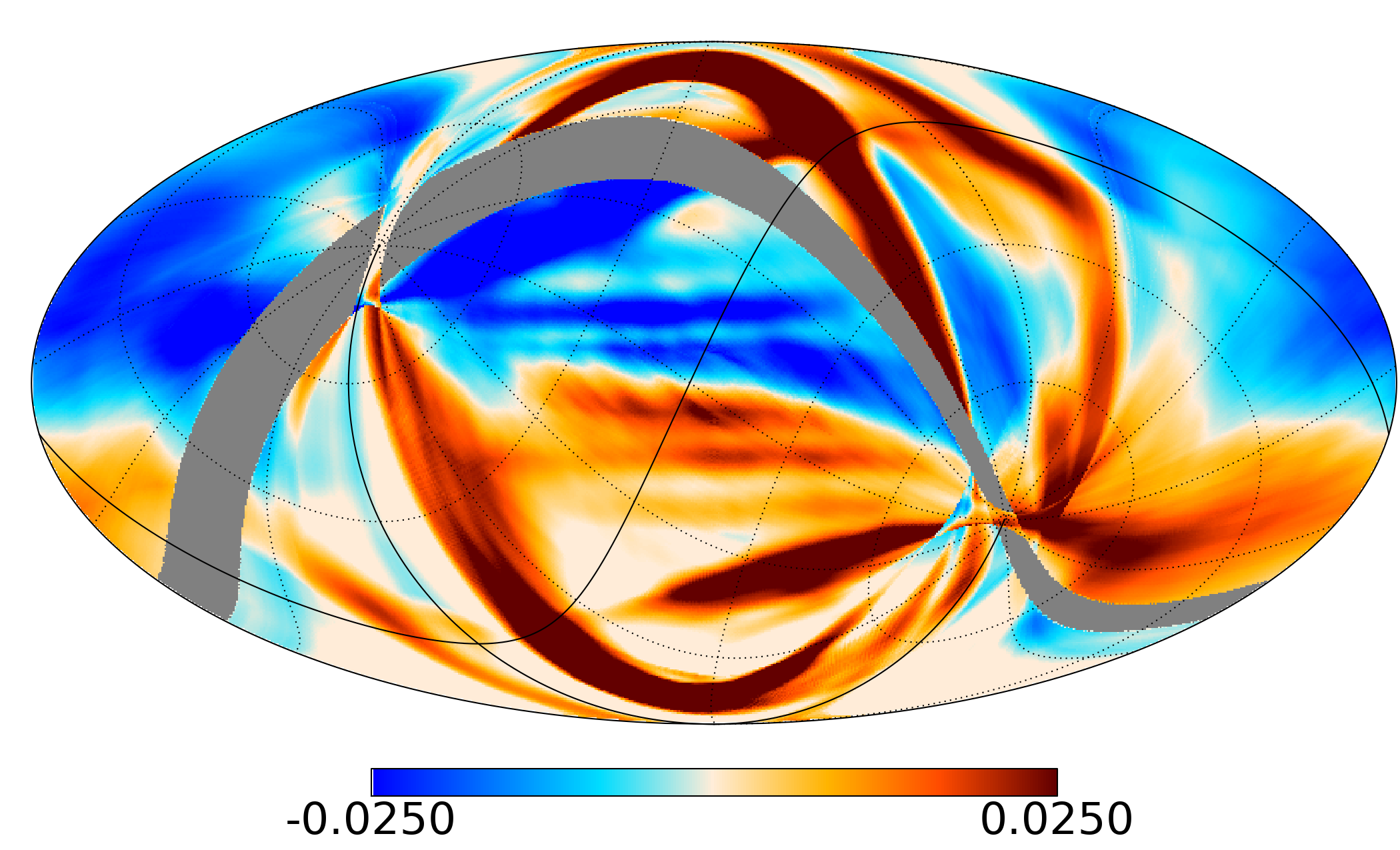}
   \\
     \includegraphics[width=60mm]{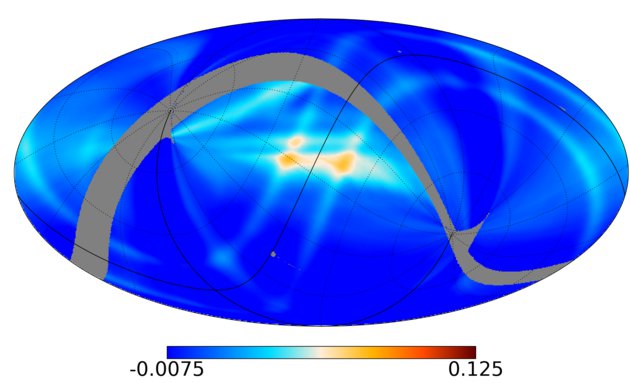}
   & \includegraphics[width=60mm]{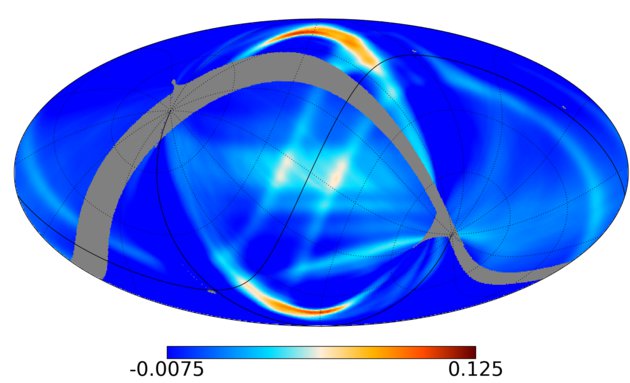}
   & \includegraphics[width=60mm]{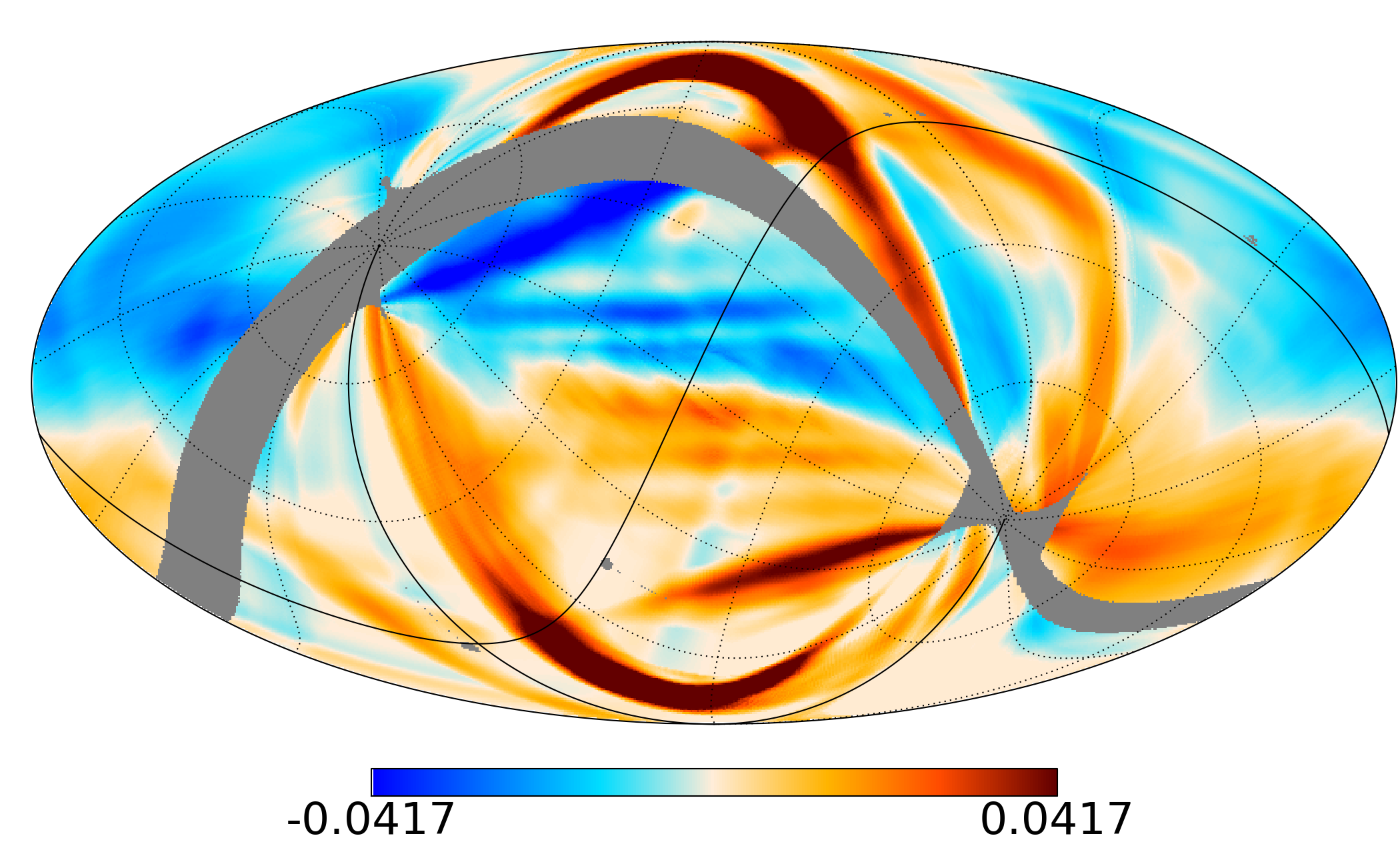}
   \\
     \includegraphics[width=60mm]{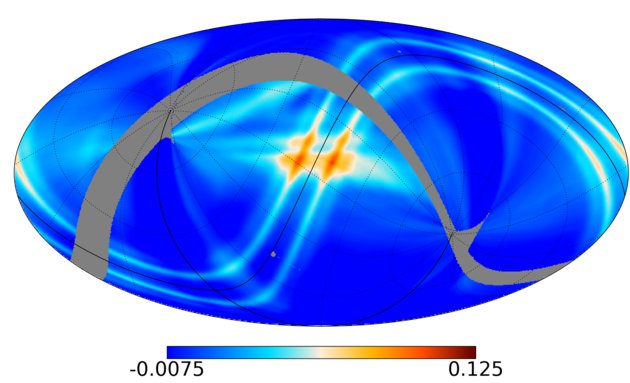}
   & \includegraphics[width=60mm]{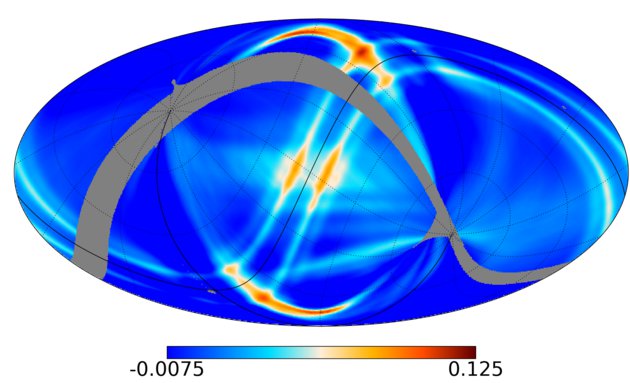}
   & \includegraphics[width=60mm]{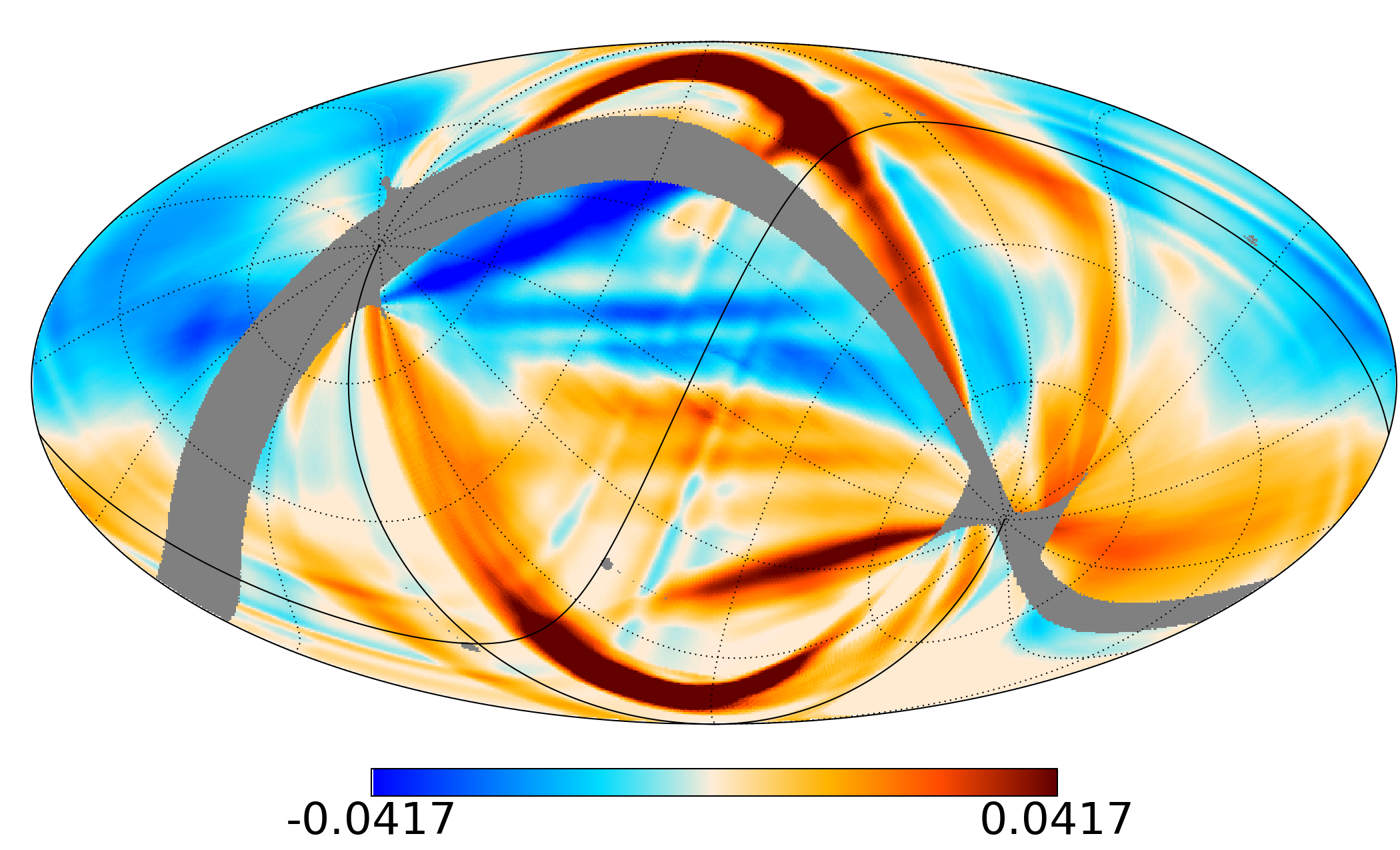}
   \\
     \includegraphics[width=60mm]{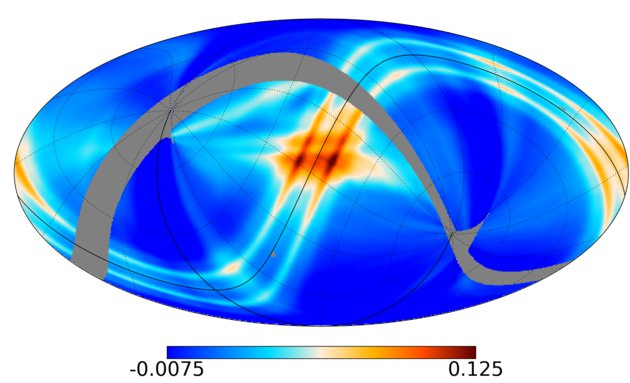}
   & \includegraphics[width=60mm]{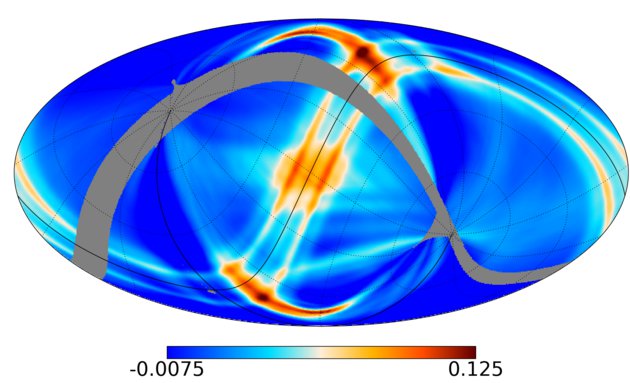}
   & \includegraphics[width=60mm]{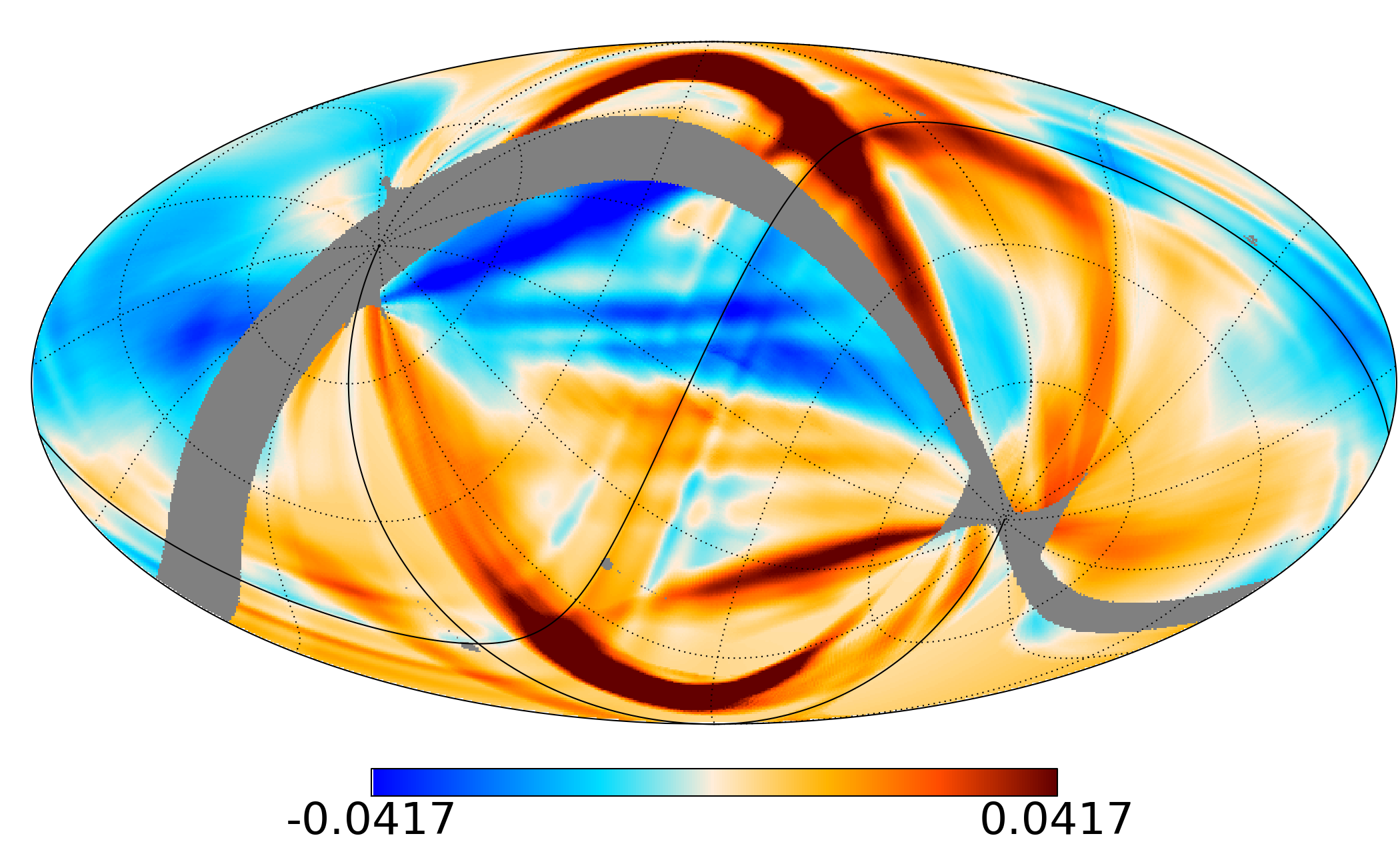}
   \\
     \includegraphics[width=60mm]{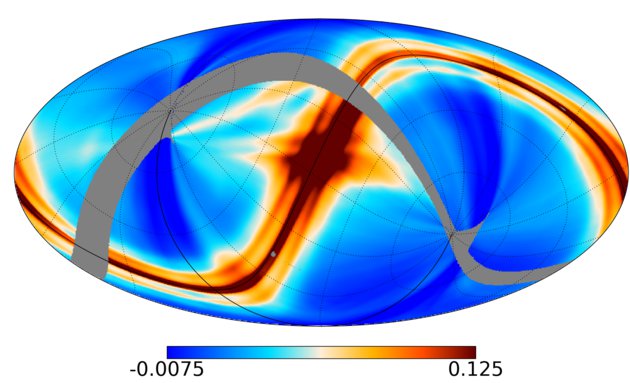}
   & \includegraphics[width=60mm]{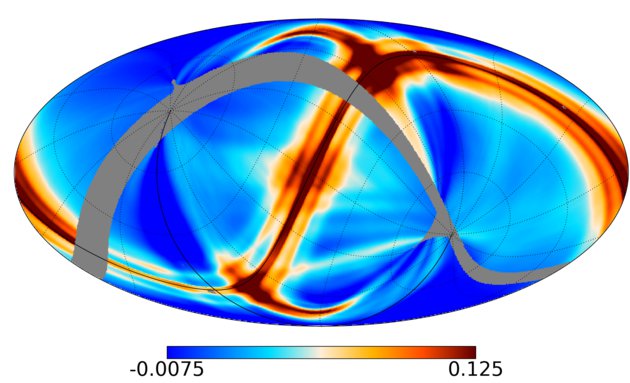}
   & \includegraphics[width=60mm]{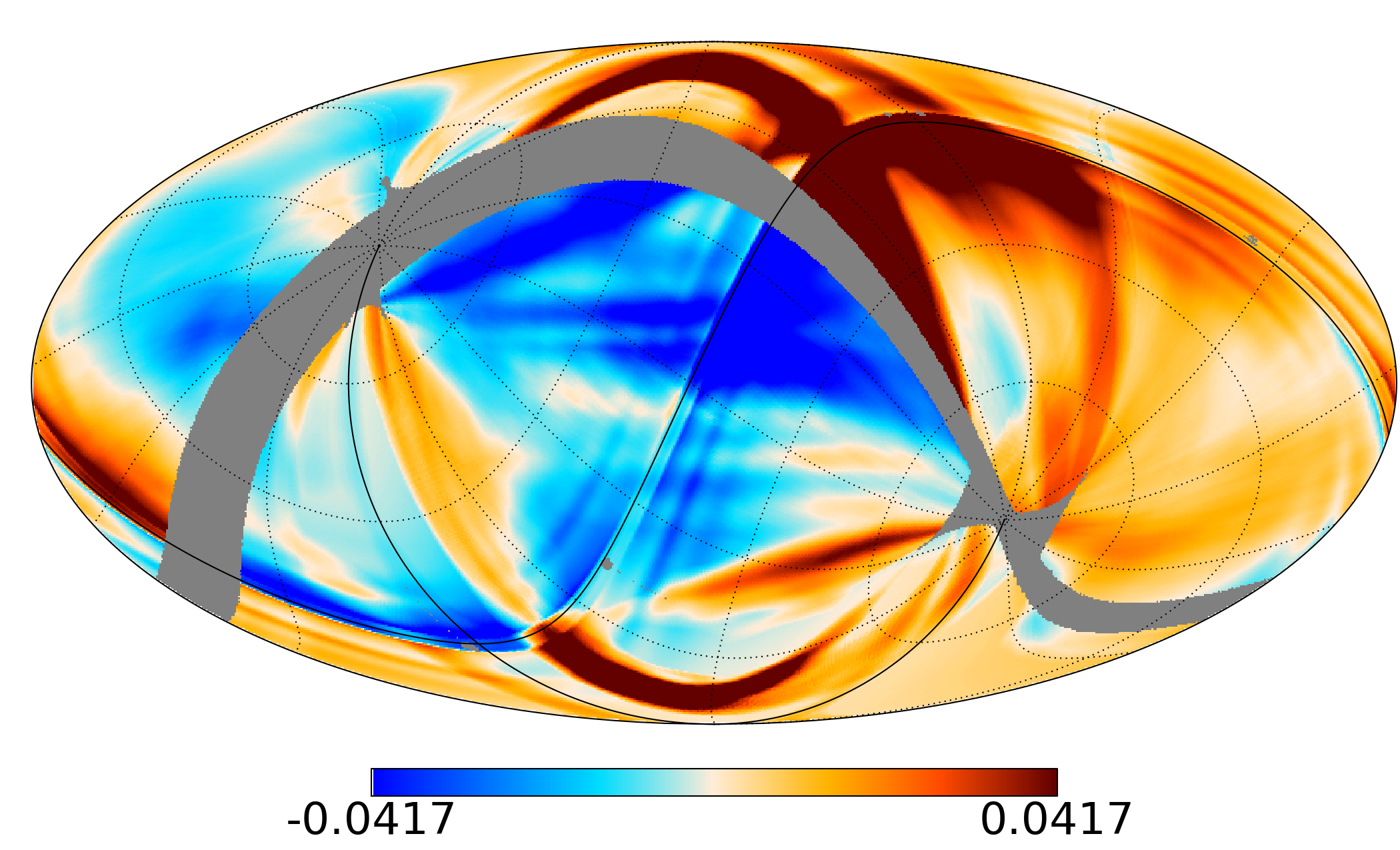}
   \\
     \includegraphics[width=60mm]{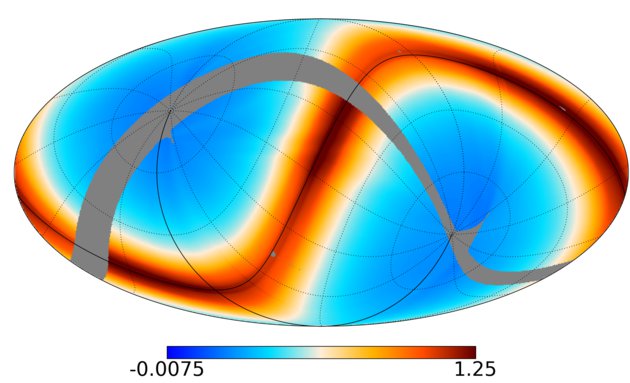}
   & \includegraphics[width=60mm]{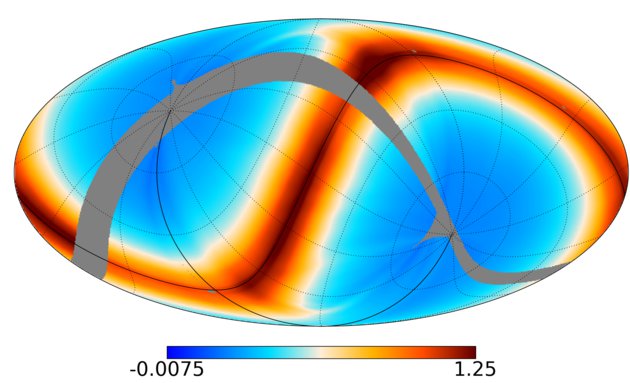}
   & \includegraphics[width=60mm]{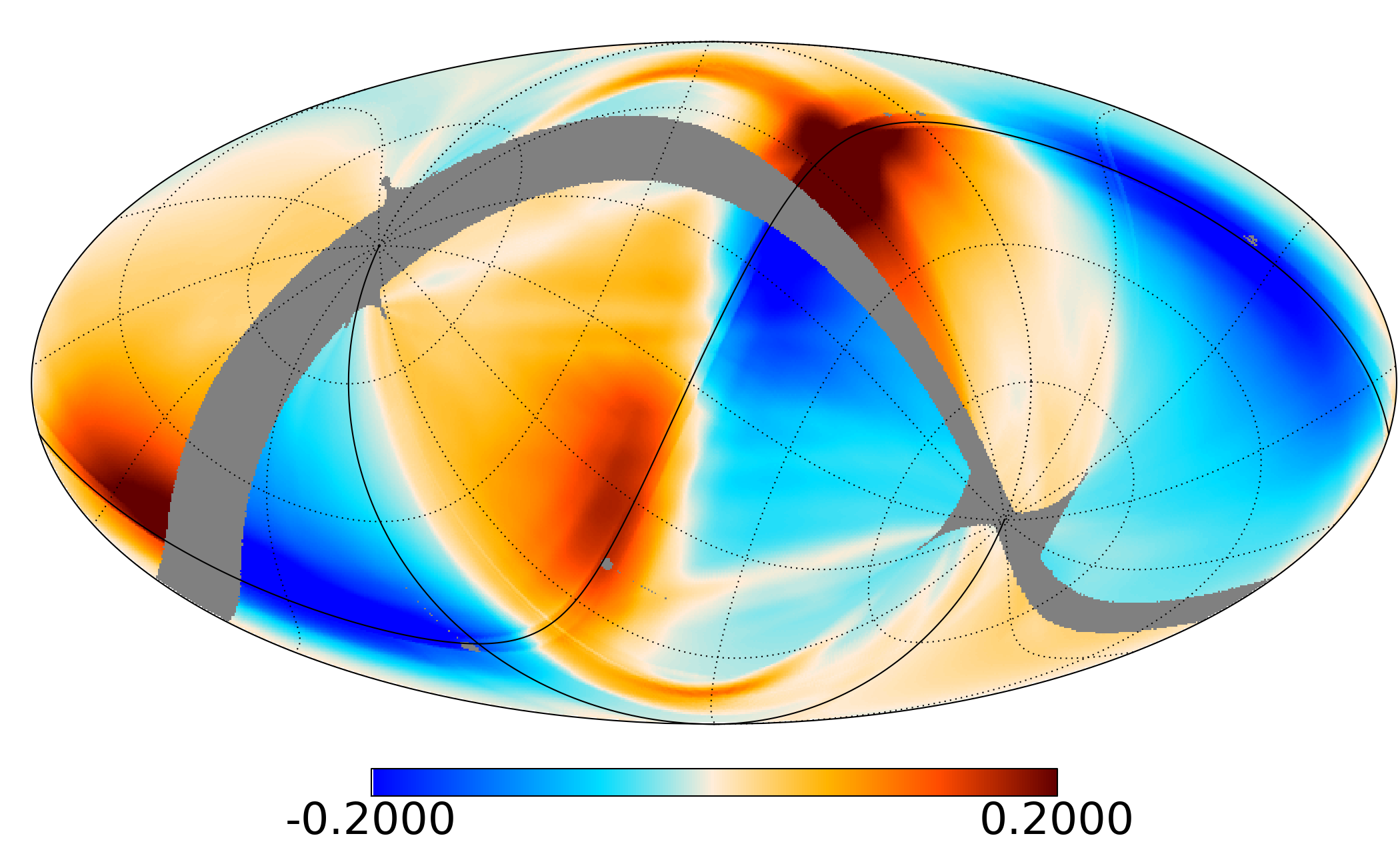}
  \end{array}$
   \caption{
    Sequence of maps in MJy/sr, each building upon that above, designed to show
    the relative contributions of the various templates in
    Figs.~\ref{fig:zodiMaps} and~\ref{fig:FSLs}. 
    Row 1: the sum of the last three templates in Fig.~\ref{fig:FSLs}.
    Row 2: Row 1 plus Dust band~1.
    Row 3: Row 2 plus Dust band~2.
    Row 4: Row 3 plus Dust band~3.
    Row 5: Row 4 plus the Circumsolar Ring and Earth-Trailing Feature.
    Row 6: The far sidelobes and all zodiacal components.
    \emph{Left}: Survey 1. \emph{Centre}: Survey 2. \emph{Right}: Survey 2 minus Survey 1. Note that
    the amplitudes used are only approximate, the figure being for illustrative
    purposes only. The scales change between the first and second rows, and between 
    the fifth and sixth rows, as well as between the second and third column. 
    }
   \label{fig:evolution}
  \end{figure*}

  To illustrate the relative contributions of these various templates, 
  Fig.~\ref{fig:evolution} shows
  a series of maps. Maps in each row are similar to those in the previous row, 
  except that one more template or group of templates has been added to form the new row. 
  For all rows, the first column corresponds to data from 
  Survey~1, the second column corresponds to Survey~2,
  and the third column is Survey~2 minus Survey~1.   
  The first row shows the sum of all far sidelobes. The second row
  shows the result when we add Dust Band~1 to the far sidelobes -- note that
  the scales change from the first to the second row. The third row shows
  the sum of the far sidelobes and Dust Bands~1 and 2. The fourth row shows
  the sum of the far sidelobes and all dust bands. The fifth row shows
  the fourth row plus the Circumsolar Ring and Earth-trailing
  feature. Finally, the bottom row shows the sum of the far sidelobes and
  all zodiacal templates (the scales have again changed). 
  As this is simply illustrative, we have assumed unit emissivities for the zodiacal components, and multiplied the far sidelobe components by a factor of 15 (which we will see in Sect.~\ref{sec:fit} is representative of the most extreme case). With these caveats in mind, the Survey difference of the sums of all the components, the lower right image, can be compared to the bottom image in Fig.~\ref{fig:surveymaps}. The zodiacal and far sidelobe structures can be seen in both figures. In addition to these, the largest differences are associated with the Galactic plane, where asymmetries in the main lobes plus beam orientation changes from one survey to the next will cause such signals.
   
 \section{Spectrum}\label{sec:fit} 

 We fit the data shown in the bottom panel of
 Fig.~\ref{fig:surveymaps}, as well as the analogous data at other
 frequencies and for the second year of observations, to a constant
 plus combinations of the templates shown in 
 Figs.~\ref{fig:zodiMaps} and~\ref{fig:FSLs}.

 For each fit, we assume that the Survey difference map
 at sky pixel $p$, called $D_{p}$, can be modelled
 as
 \begin{equation}
  D_{p} 
  = 
  \sum_t 
  \left(
   \epsilon_{2,t} T_{2,t,p}-\epsilon_{1,t} T_{1,t,p}
  \right)
  + \mathrm{constant},
 \end{equation}
 where $\epsilon_{\left[1|2\right],t}$ is the emissivity fit for
 template $t$ during Survey 1 or 2 at
 the given frequency and $T_{\left[1|2\right],t,p}$ is the value 
 of the $t^\mathrm{th}$
 template at pixel $p$ for Survey 1 or 2, calculated as described in
 Sect.~\ref{sec:model}. For example, for our ``basic'' fit, 
 we will have 19 templates -- one each
 for the Diffuse Cloud, Circumsolar Ring, Earth-Trailing Feature, and
 each of the three dust bands, as well as one each for the Galactic
 far sidelobes. All of these are repeated twice, once
 for each Survey in a yearly difference map. Finally 
 we also fit to an overall constant, to which \Planck\ is not sensitive. 
 We then minimize 
 \begin{equation}
  \chi^2 = \sum_p \left(\Delta_{p} - D_{p}\right)^2.
 \end{equation}
 
 Separating the templates into Surveys has the disadvantage of
 increasing the number of parameters in our fits. While we do not 
 expect either the emissivity of the zodiacal emission or the 
 far sidelobe calculations to change from one survey to the next, we separate
 them in this way for two reasons. The first is simply as a basic reality
 check -- if we see significant differences between fits to two 
 different Surveys, we should be sceptical. Beam asymmetries or
 transfer function effects, for example, 
 might cause differences from Survey to Survey, as might imperfections
 in the model itself.  
 
 The second reason is to calculate uncertainties. As just noted, we are often 
 as concerned by systemic effects as much as by ``random'' noise. By separating
 the data by Survey, we may calculate uncertainties using the standard error of the
 successive measurements as a proxy for the uncertainties, rather than propagating white noise 
 estimates. This should provide us with a more conservative estimate of our uncertainties, 
 one that accounts for model deficiencies or low levels of systematics that change by Survey
 (e.g., the aforementioned beam asymmetries and transfer functions).
 
 The bulk of the zodiacal emission is, of course, in the ecliptic
 plane. \Planck, on the other hand, has more statistical weight, 
 pixel by pixel, at the ecliptic poles. We therefore use uniform
 weights over pixels, rather than statistical weights, since this would
 down-weight specifically the regions with our signal. 
  
\begin{figure*}[htbp]
  \centering
  \includegraphics[width=180mm,type=pdf,ext=.pdf,read=.pdf]{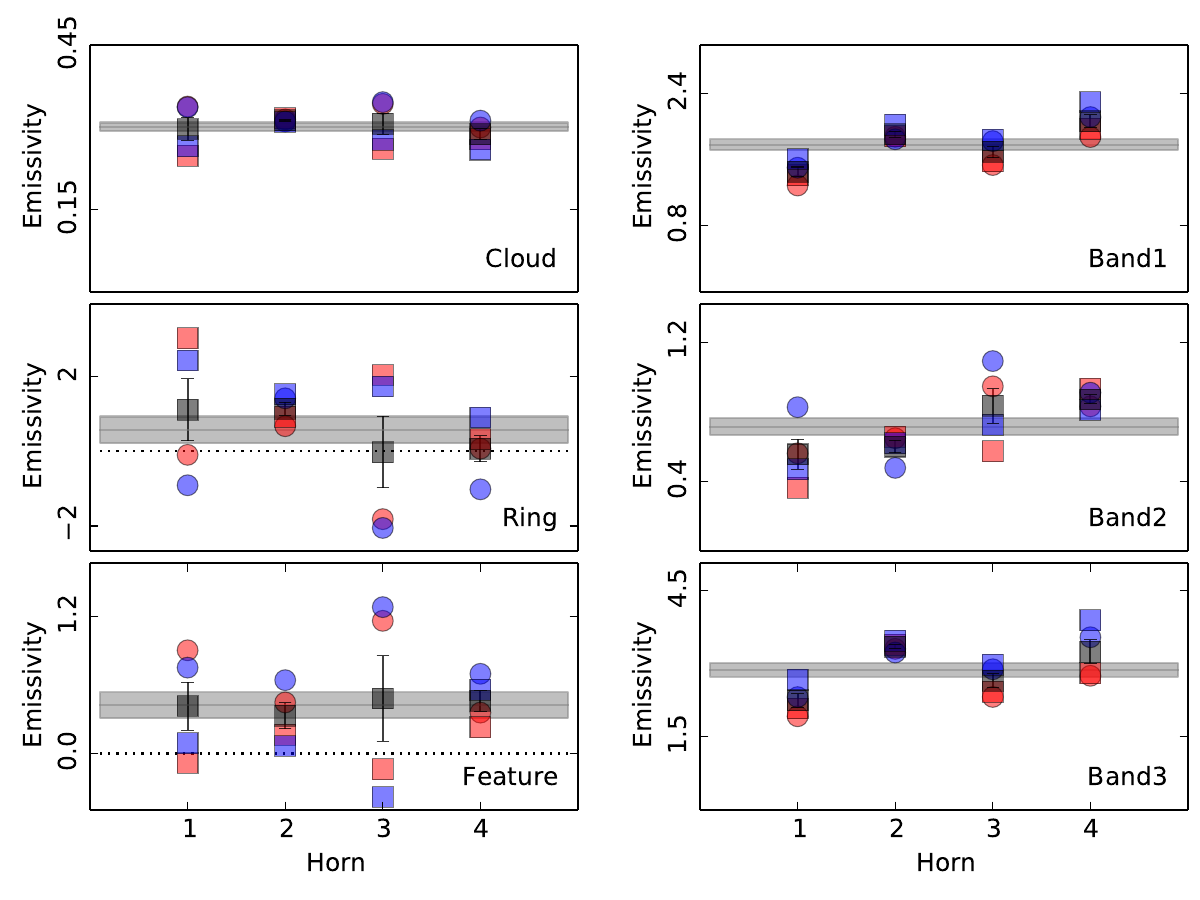}
  \caption{Emissivities of components from fits at 857\,GHz.  Fits to the first year of \Planck\ observations are in red, divided into Survey~1    
  (circles) and Survey~2 (squares).  Fits to the second year are in blue, divided into Survey~3 (circles) and Survey~4 (squares). 
  Absent time variability of the zodiacal emission, little difference would be expected between corresponding red and blue
  symbols.  Agreement or disagreement between squares and circles gives some indication of systematic errors in the data and the correctness of 
  the templates. 
  The average of all measurements for each horn is shown as a black square, and the average and standard errors for the entire frequency, is 
  given by the horizontal grey band. 
  Dotted lines mark zero levels where appropriate. 
  Similar plots for all HFI frequencies can be found in the \Planck\ Explanatory Supplement \citep{planck2013-p28}. }
  \label{fig:fit857}
 \end{figure*}

\begin{figure*}[htbp]
   \centering
   \includegraphics[width=180mm,type=pdf,ext=.pdf,read=.pdf]
   {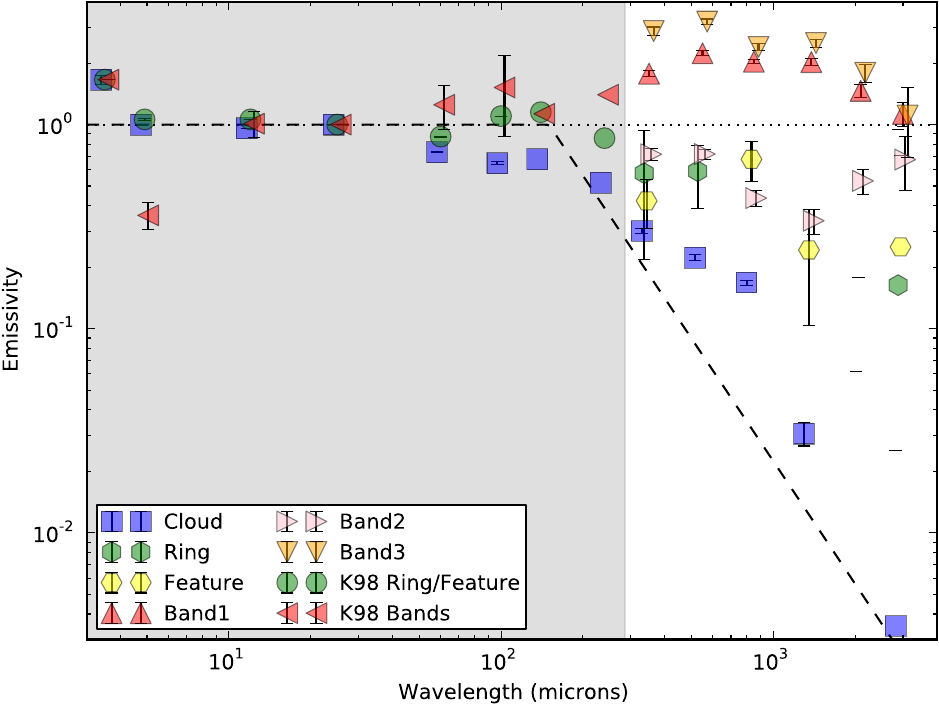}
    \caption{
     Emissivities of components of the K98 zodiacal emission 
     model obtained from \Planck/HFI ($\lambda > 250\micron$) and \textit{COBE}/DIRBE ($\lambda \le 250\micron$, grey shading; K98). 
     The Diffuse Cloud is shown as blue squares. 
     For DIRBE, all dust bands (red, left-pointing triangles) were assumed have the same emissivity.
     For HFI, the bands (red, up-pointing triangles for Dust Band 1; pink, right-pointing triangles for Dust Band 2; and orange, down-pointing  
     triangles for Dust Band 3) were allowed to have different emissivities. 
     Similarly, K98 assumed that the Circumsolar Ring and  Earth-Trailing Feature (green circles) had the same emissivities.  
     For \Planck\ the Circumsolar Ring (green hexagons) and Earth-Trailing Feature (yellow hexagons) were allowed to be different. 
     \Planck\ values were obtained by fitting an amplitude to each component, as well as the Galaxy seen through the sidelobes. 
     All other parameters in the model were fixed at their K98 values. 
     Each point is the average of the corresponding values obtained for all individual horns and Surveys at the given frequency,
     over the first two years of HFI data (grey-shaded regions in each panel of Fig.~\ref{fig:fit857}.  
     Error bars give the standard errors of these different measures.  
     Numerical values are given in Table~\ref{tab:fullFit}. 
     Note that a few Cloud, Circumsolar Ring, and Earth-Trailing Feature values are negative, and so do not appear in this log--log plot. 
     In such cases, the upper limit will appear as a short horizontal line. 
     The dotted line indicates an emissivity of unity at all wavelengths, and the dashed line indicates an emissivity that is unity at
     wavelengths below 150\microns\ and proportional to $\lambda^{-2}$ at longer wavelengths.}
    \label{fig:noDipoleSidelobesFit}
  \end{figure*}

 As mentioned above, we fit each of the Survey difference maps
 to a cloud, Circumsolar Ring, Earth-Trailing Feature, three bands and
 three far sidelobe templates, plus a constant. The results for 
 the four 857\,GHz horns are shown in Fig.~\ref{fig:fit857}. 
 Averaging over horns and Surveys at all six HFI frequencies 
 yields Fig.~\ref{fig:noDipoleSidelobesFit}. Numerical
 values are given in Tables~\ref{tab:fullFit}
 and~\ref{tab:galacticSidelobeFit}.

\begin{table*}[htbp] 
\begingroup
\caption{Emissivities of the zodiacal dust components from the fit result averages.}
\label{tab:fullFit}
\nointerlineskip
\vskip -3mm
\footnotesize
\setbox\tablebox=\vbox{
 \newdimen\digitwidth 
 \setbox0=\hbox{\rm 0} 
 \digitwidth=\wd0 
 \catcode`*=\active 
 \def*{\kern\digitwidth}
 \newdimen\signwidth 
 \setbox0=\hbox{+} 
 \signwidth=\wd0 
 \catcode`!=\active 
 \def!{\kern\signwidth}
\halign{\hbox to 1in{#\leaderfil}\tabskip 2em&
    \hfil$#$\hfil\tabskip=1em&
    \hfil$#$\hfil&
    \hfil$#$\hfil&
    \hfil$#$\hfil&
    \hfil$#$\hfil&
    \hfil$#$\hfil\tabskip=0pt\cr 
\noalign{\doubleline}
\omit&\multispan6\hfil\sc Emissivity\hfil\cr
\noalign{\vskip -3pt}
\omit\hfil\sc Frequency\hfil&\multispan6\hrulefill\cr
\omit\hfil [GHz]\hfil&\omit\hfil Cloud\hfil&\omit\hfil Ring\hfil&\omit\hfil Feature\hfil&\omit\hfil Band 1\hfil&
   \omit\hfil Band 2\hfil&\omit\hfil Band 3\hfil\cr
\noalign{\vskip 3pt\hrule\vskip 5pt}
857& !0.301\pm0.008& !0.578\pm0.359& !0.423\pm0.114& 1.777\pm0.066& 0.716\pm0.049& 2.870\pm0.137\cr
545& !0.223\pm0.007& !0.591\pm0.203& -0.182\pm0.061& 2.235\pm0.059& 0.718\pm0.041& 3.193\pm0.097\cr
353& !0.168\pm0.005& -0.211\pm0.085& !0.676\pm0.149& 2.035\pm0.053& 0.436\pm0.041& 2.400\pm0.100\cr
217& !0.031\pm0.004& -0.185\pm0.143& !0.243\pm0.139& 2.024\pm0.072& 0.338\pm0.047& 2.507\pm0.109\cr
143& -0.014\pm0.010& -0.252\pm0.314& -0.002\pm0.180& 1.463\pm0.103& 0.530\pm0.073& 1.794\pm0.184\cr
100& !0.003\pm0.022& !0.163\pm0.784& !0.252\pm0.455& 1.129\pm0.154& 0.674\pm0.197& 1.106\pm0.413\cr
\noalign{\vskip 5pt\hrule\vskip 3pt}}}
\endPlancktablewide
\endgroup
\end{table*}
 
      
\begin{table*}
\begingroup
\caption{Fit coefficients for the Galaxy seen through the far sidelobes.}
\label{tab:galacticSidelobeFit}
\nointerlineskip
\vskip -3mm
\footnotesize
\setbox\tablebox=\vbox{
 \newdimen\digitwidth 
 \setbox0=\hbox{\rm 0} 
 \digitwidth=\wd0 
 \catcode`*=\active 
 \def*{\kern\digitwidth}
 \newdimen\signwidth 
 \setbox0=\hbox{+} 
 \signwidth=\wd0 
 \catcode`!=\active 
 \def!{\kern\signwidth}
\halign{\hbox to 1in{#\leaderfil}\tabskip=3em&
    \hfil$#$\hfil\tabskip=1em&
    \hfil$#$\hfil\tabskip=4em&
    \hfil$#$\hfil\tabskip=1em&
    \hfil$#$\hfil&
    \hfil$#$\hfil\tabskip=0pt\cr 
\noalign{\doubleline}
\omit&\multispan2\hfil\sc Primary Spillover\hfil&\multispan3\hfil\sc Secondary Spillover\hfil\cr
\noalign{\vskip -3pt}
\omit\hfil\sc Frequency\hfil&\multispan2\hrulefill&\multispan3\hrulefill\cr
\omit\hfil [GHz]\hfil&\omit\hfil Fit\hfil&\omit\hfil Prediction$^{\rm a}$\hfil&\omit\hfil Direct\hfil&\omit\hfil Baffle\hfil&
    \omit\hfil Prediction$^{\rm b}$\hfil\cr
\noalign{\vskip 3pt\hrule\vskip 5pt}
100& -25.8\pm5.7& *7&26.3\pm15.7&        56.7\pm5.7& 10\cr
143& *-9.1\pm4.1& *6&13.0\pm*4.8&        23.8\pm5.4& 10\cr
217& *!0.6\pm1.1& *5&*6.3\pm*2.1&        *6.5\pm1.3& *6\cr
353& *-1.2\pm0.5& *1&*\llap{-}4.3\pm*2.1&*3.6\pm0.7& *1\cr
\noalign{\vskip 5pt}
545& *!7.7\pm1.7& 15&*8.8\pm*3.1&        *7.9\pm1.0& *1\cr
857& !17.1\pm3.4&1.5&23.9\pm*4.2&        16.7\pm3.1& 0.005\cr
\noalign{\vskip 5pt\hrule\vskip 3pt}}}
\endPlancktablewide
\tablenote {{\rm a}} Unitless value we would expect for the fit to the primary spillover sidelobe contribution. It is the ratio of the primary spillover at the given frequency to the spillover at 353\,GHz, as calculated in Table~2 of \cite{tauber2010b}. If all our data and predictions were perfect, this value would match the corresponding value in the column labelled ``Fit.''\par
\tablenote {{\rm b}} Unitless value we would expect for the fit to the secondary spillover sidelobe contribution. It is the ratio of the secondary spillover at the given frequency to the spillover at 353\,GHz, as calculated in Table~2 of \cite{tauber2010b}. 
If all our data and predictions were perfect, this value would match the corresponding fit values in the columns labelled ``Direct'' and ``Baffle'', which would also be equal to each other.\par
\endgroup
\end{table*}    

 \begin{figure*}[htbp]
  \centering
  $\begin{array}{cc}
   \includegraphics[width=90mm]{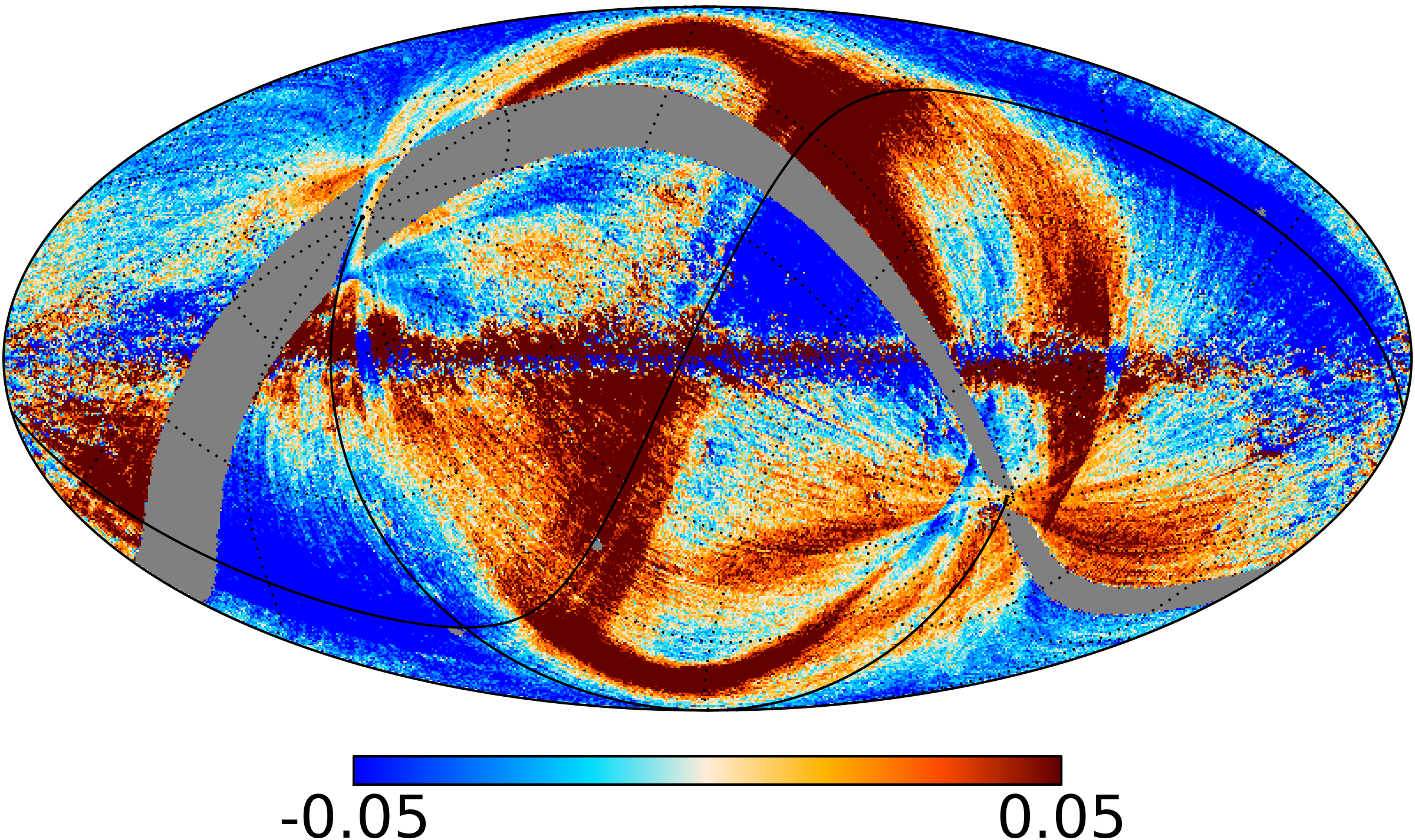} &
   \includegraphics[width=90mm]{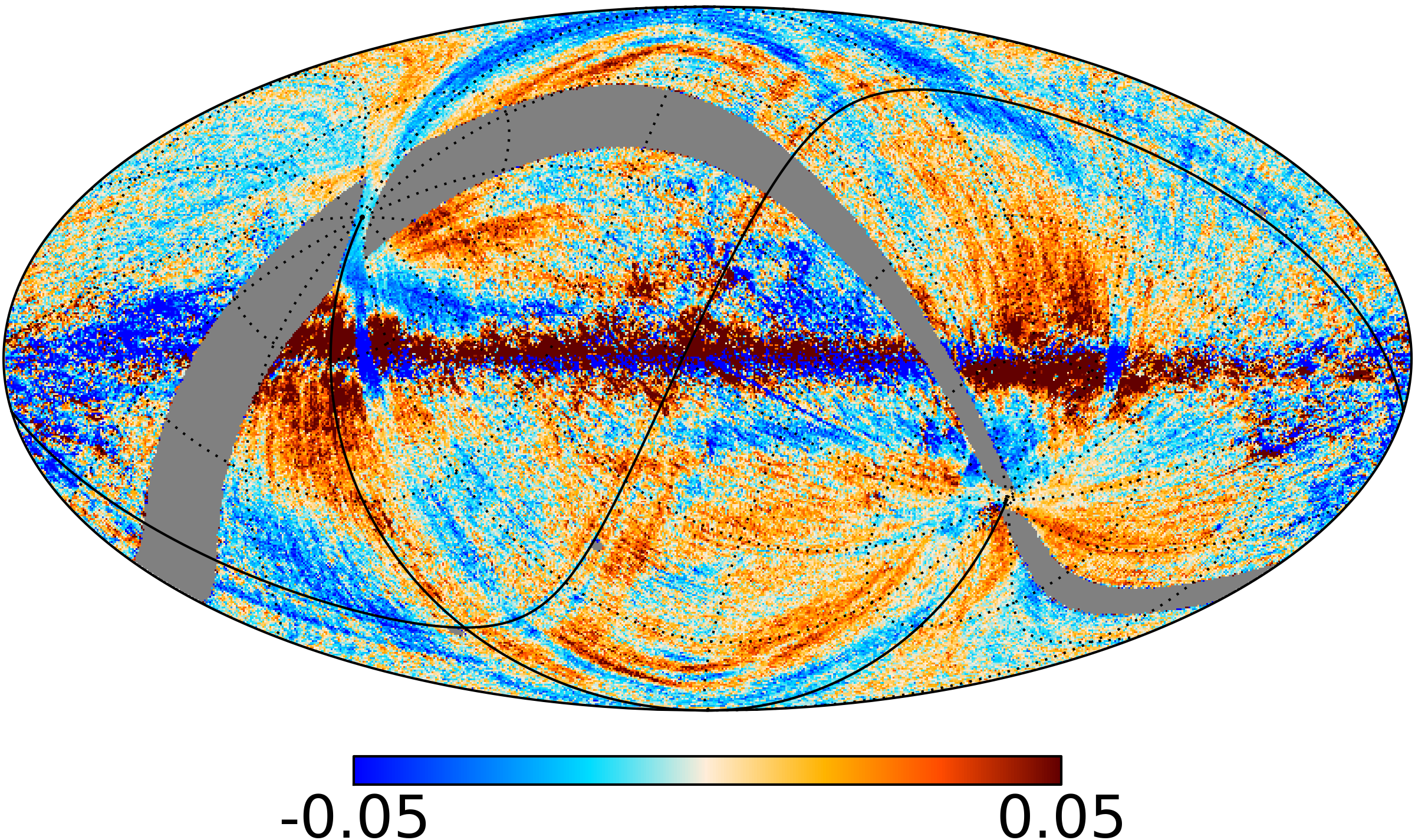} \\
   \includegraphics[width=90mm]{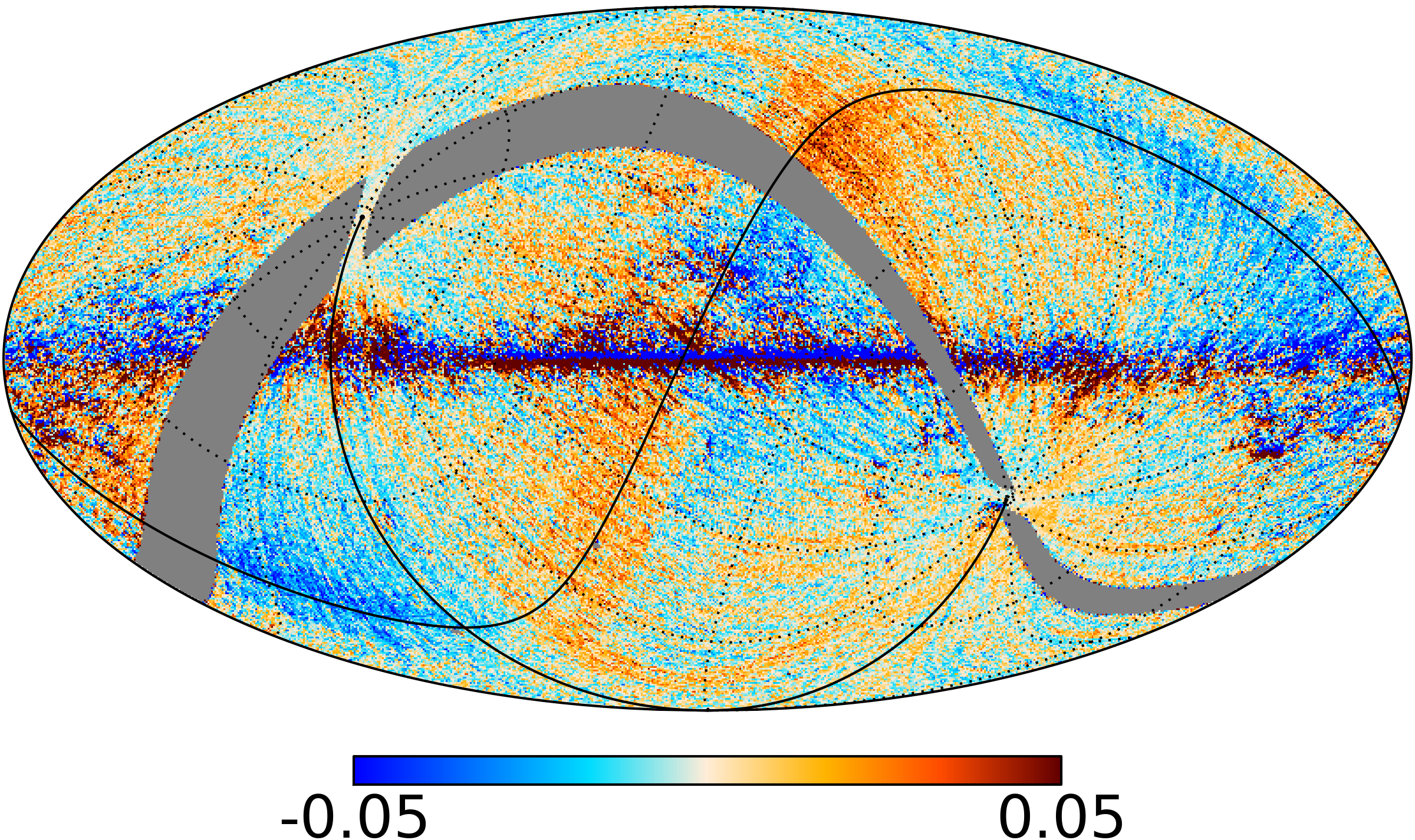} &
   \includegraphics[width=90mm]{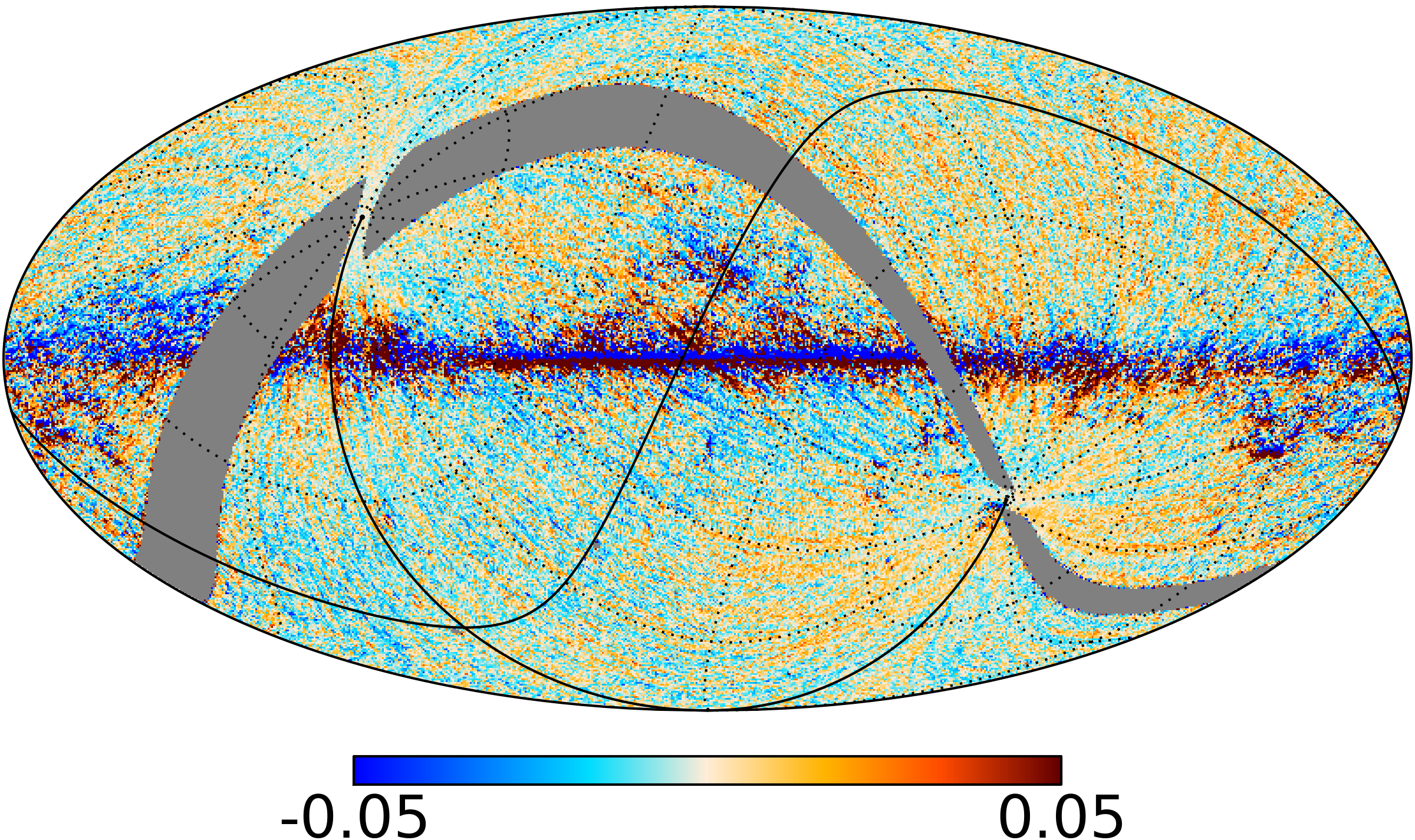} \\
   \includegraphics[width=90mm]{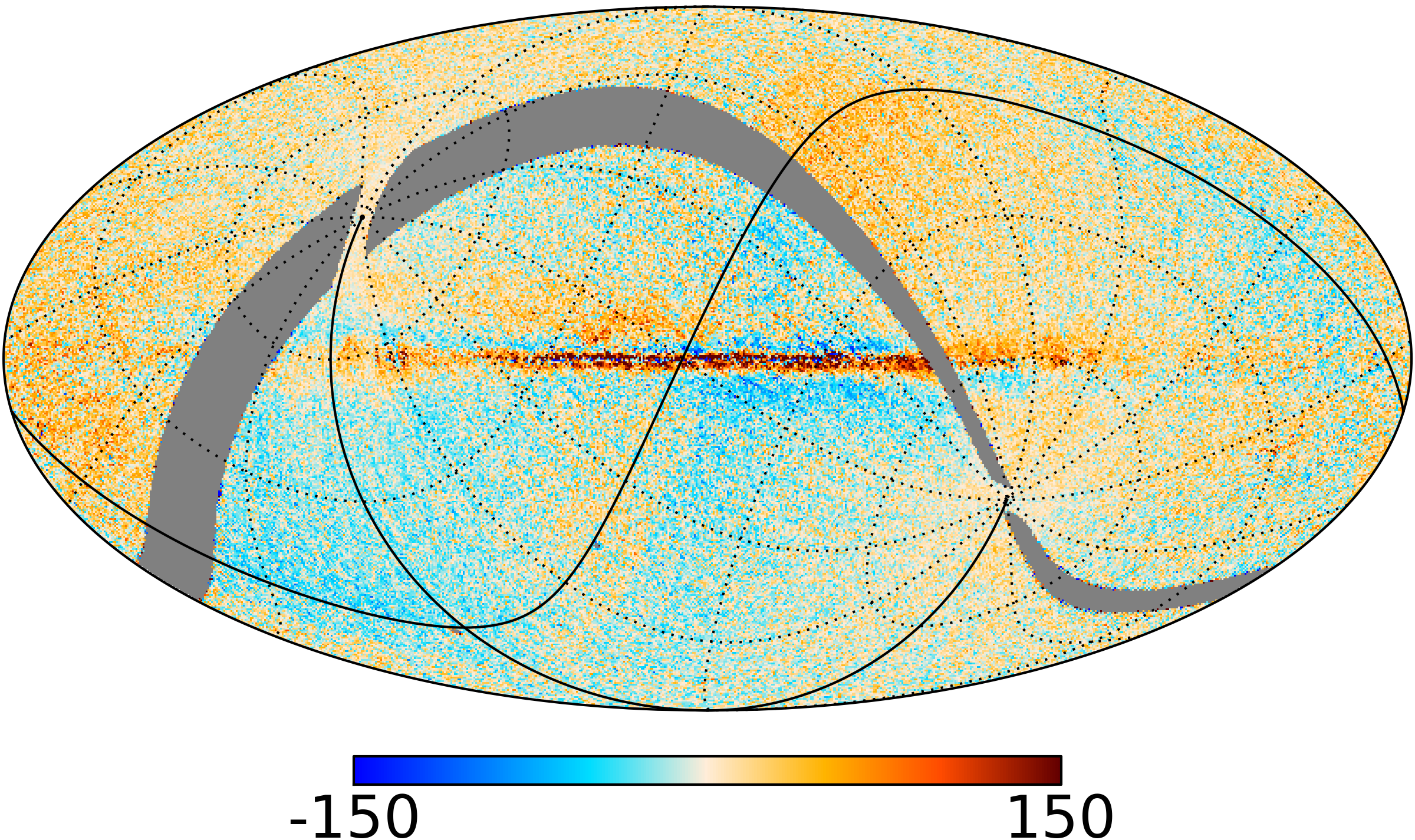} &
   \includegraphics[width=90mm]{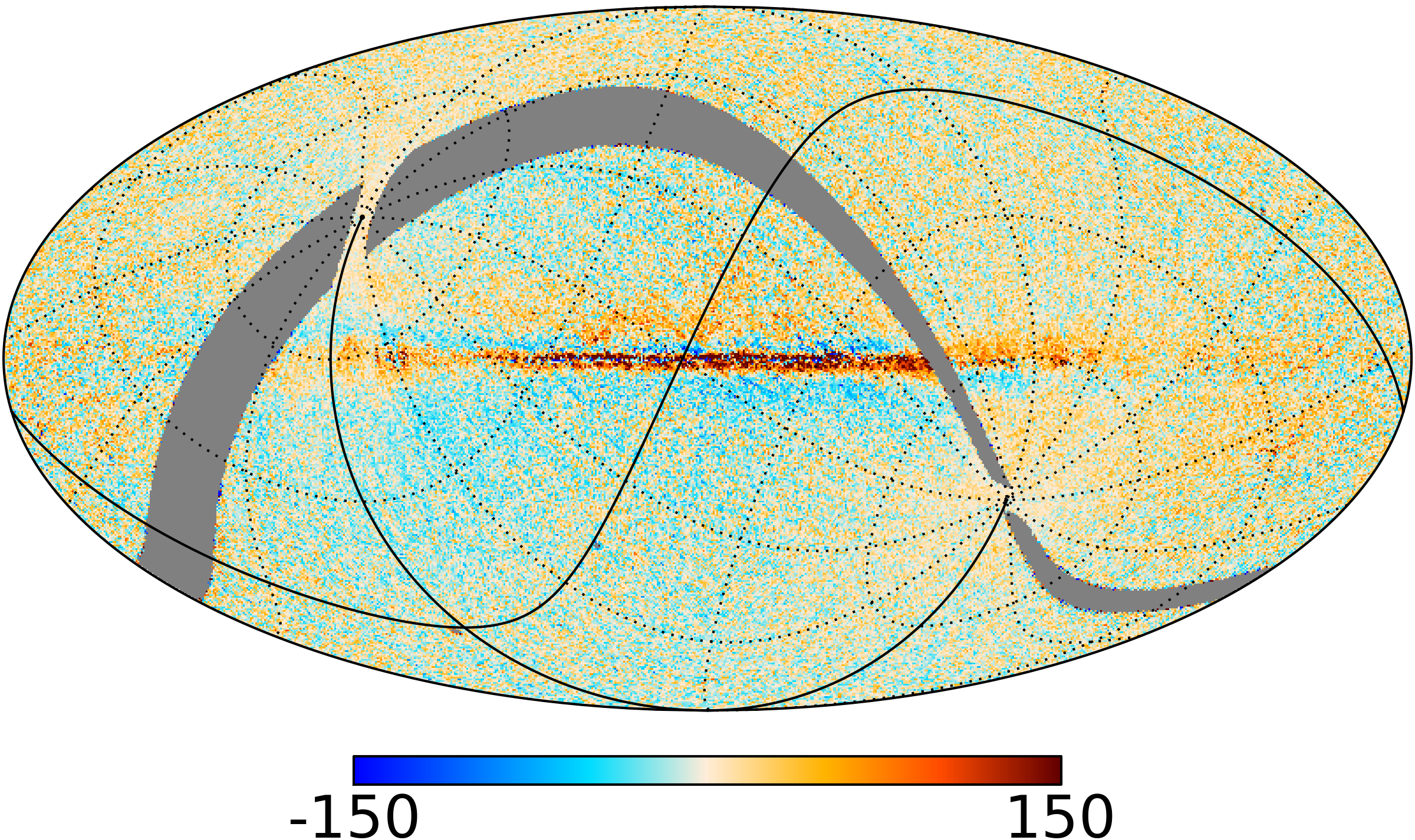} 
   \end{array}$
  \caption{Survey 2 minus Survey 1 difference maps, before (left) 
   and after (right) zodiacal emission removal. 
   The top two rows, for 857 and 545\,GHz, are in units MJy\,sr$^{-1}$, while the
   bottom row, for 353\,GHz, is in units of $\mu\mathrm{K}_\mathrm{CMB}$.}
  \label{fig:beforeAndAfterJackknives}
 \end{figure*} 
 
 \begin{figure*}[htbp]
  \centering
  $\begin{array}{cc}
   \includegraphics[width=90mm]{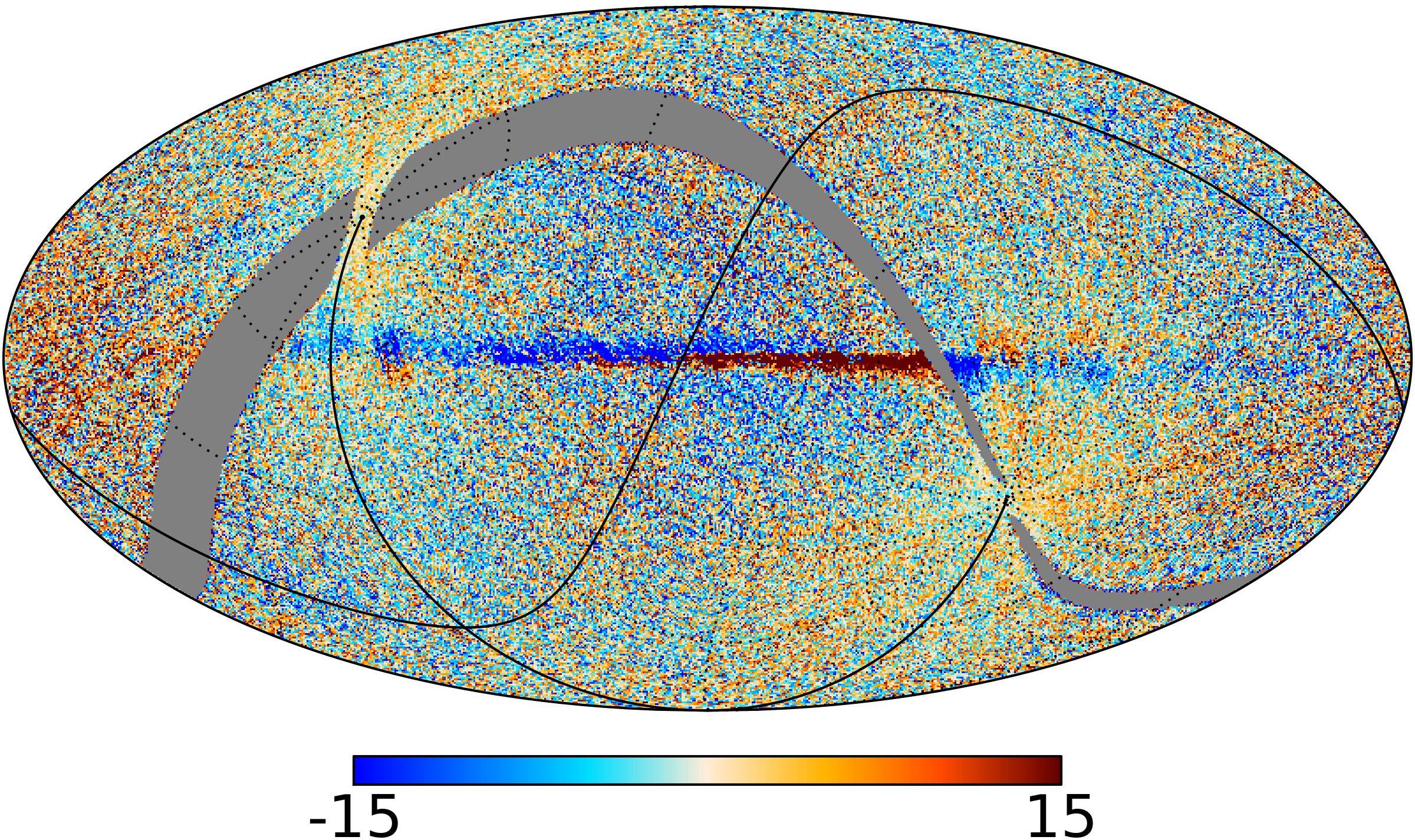} &
   \includegraphics[width=90mm]{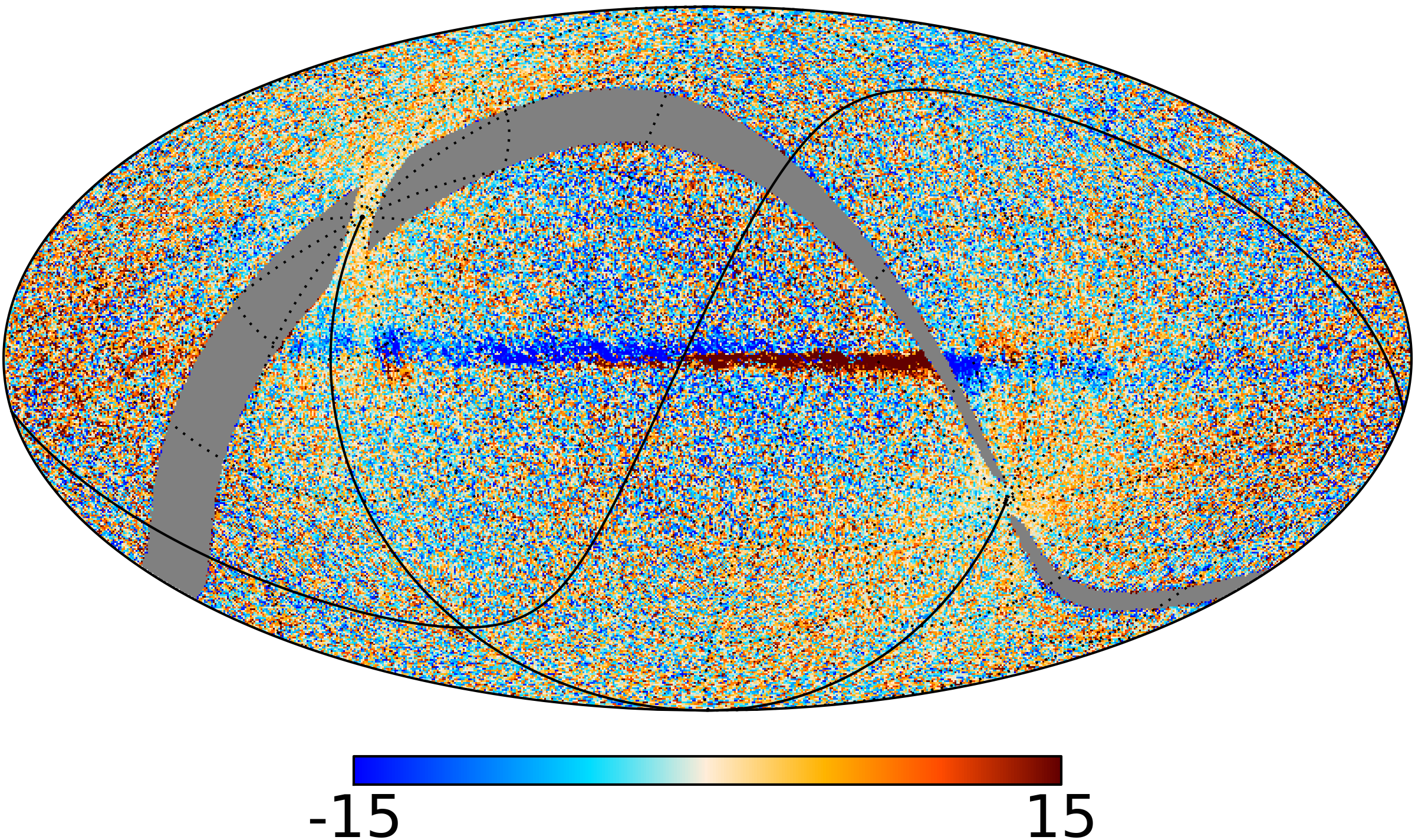} \\
   \includegraphics[width=90mm]{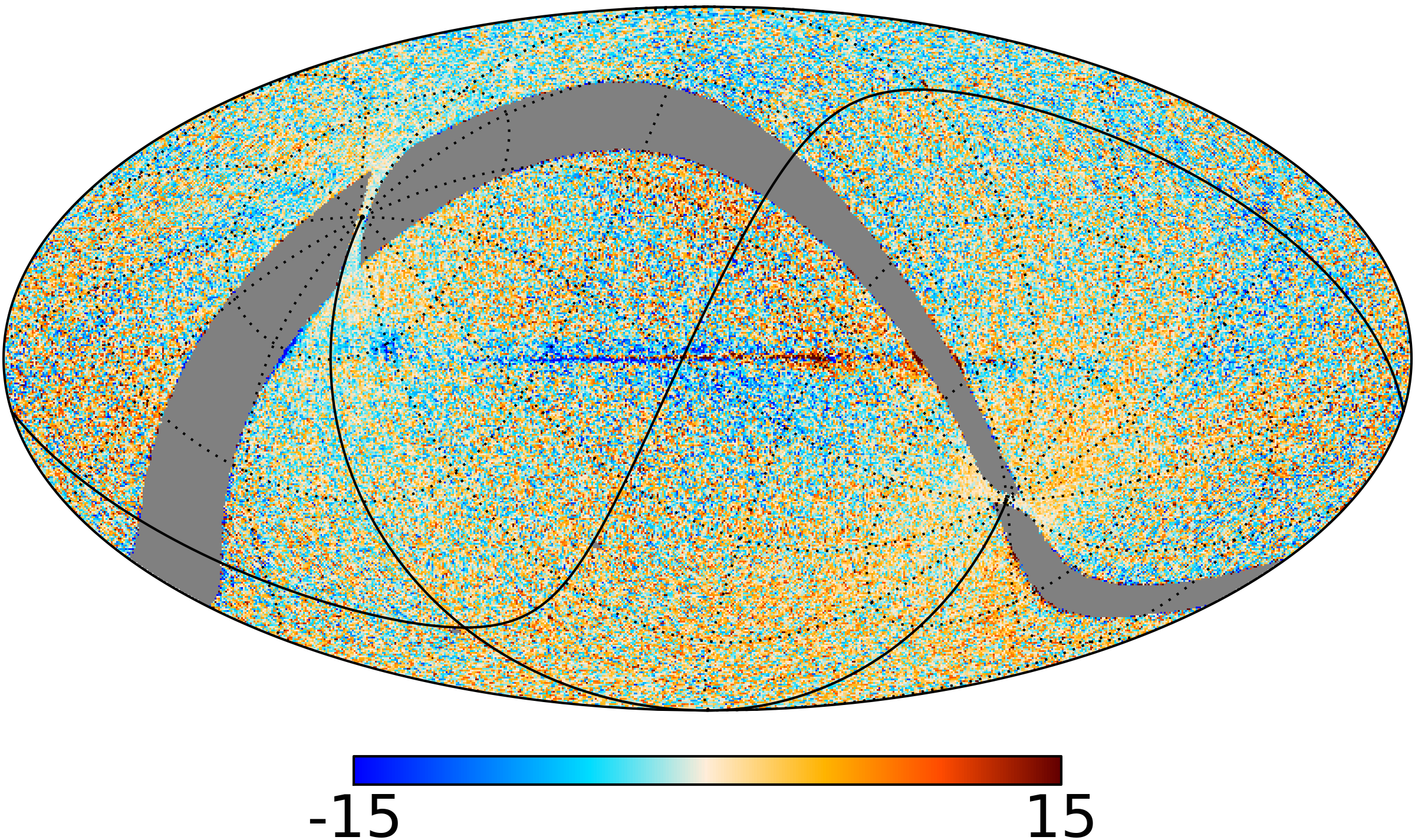} &
   \includegraphics[width=90mm]{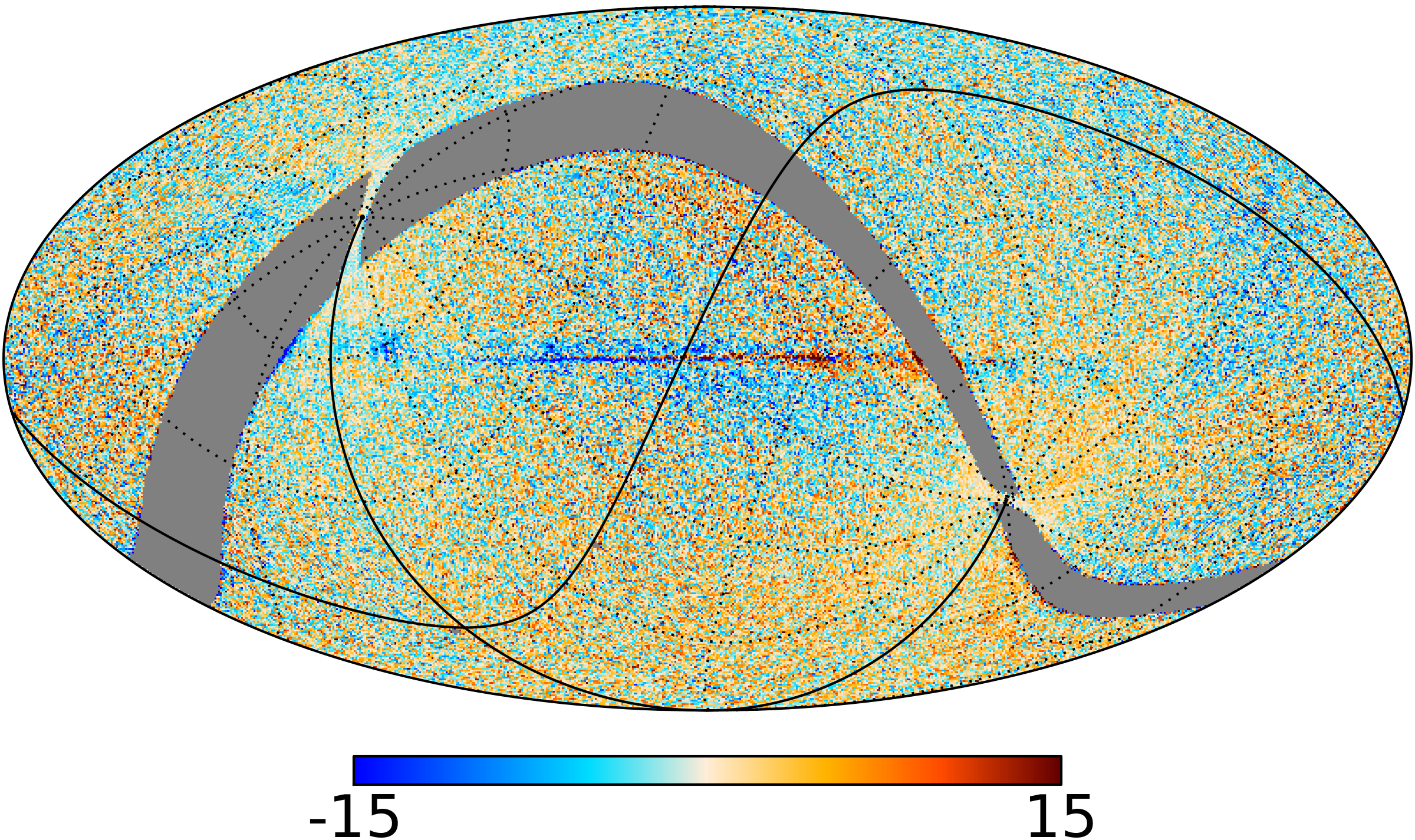} \\
   \includegraphics[width=90mm]{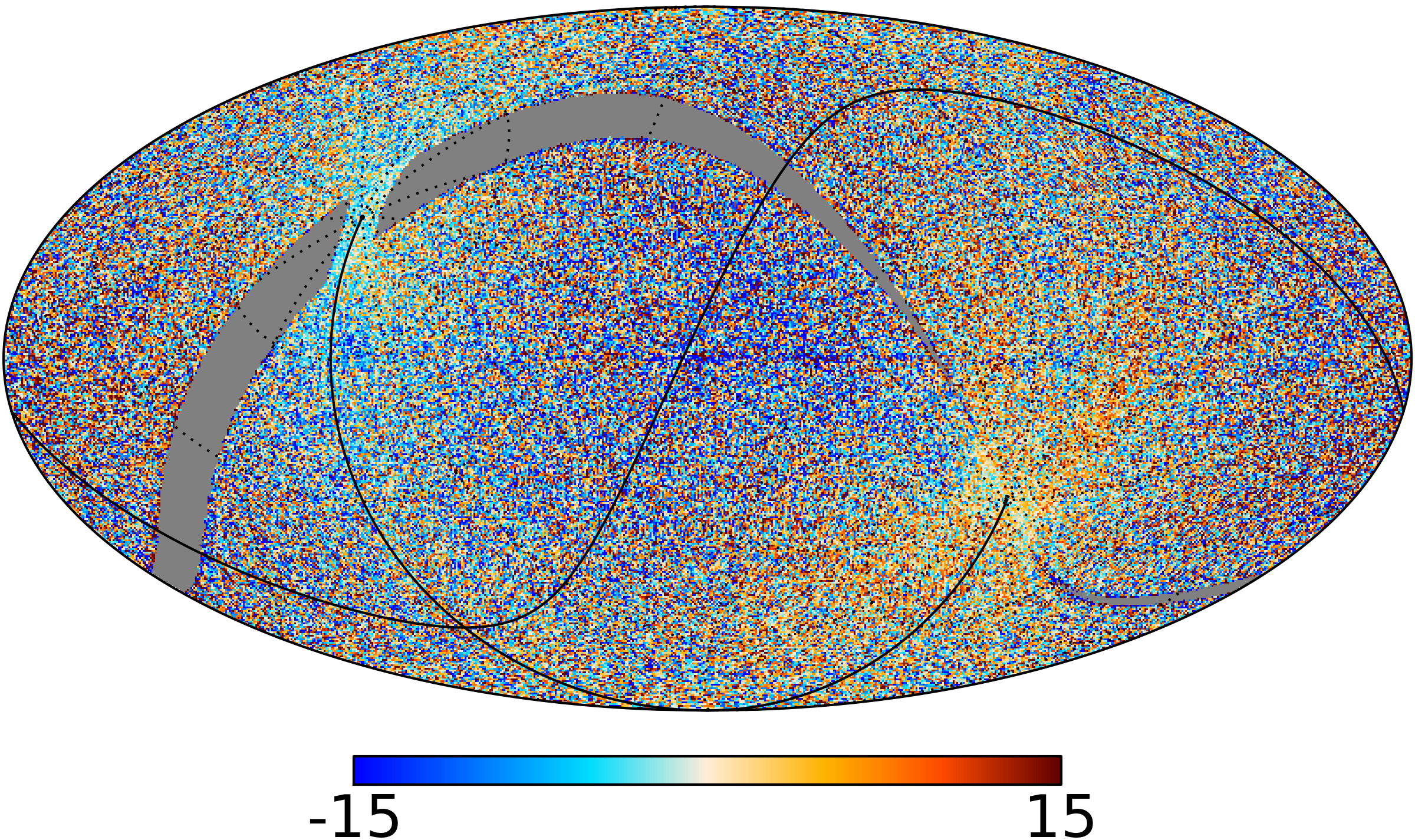} &
   \includegraphics[width=90mm]{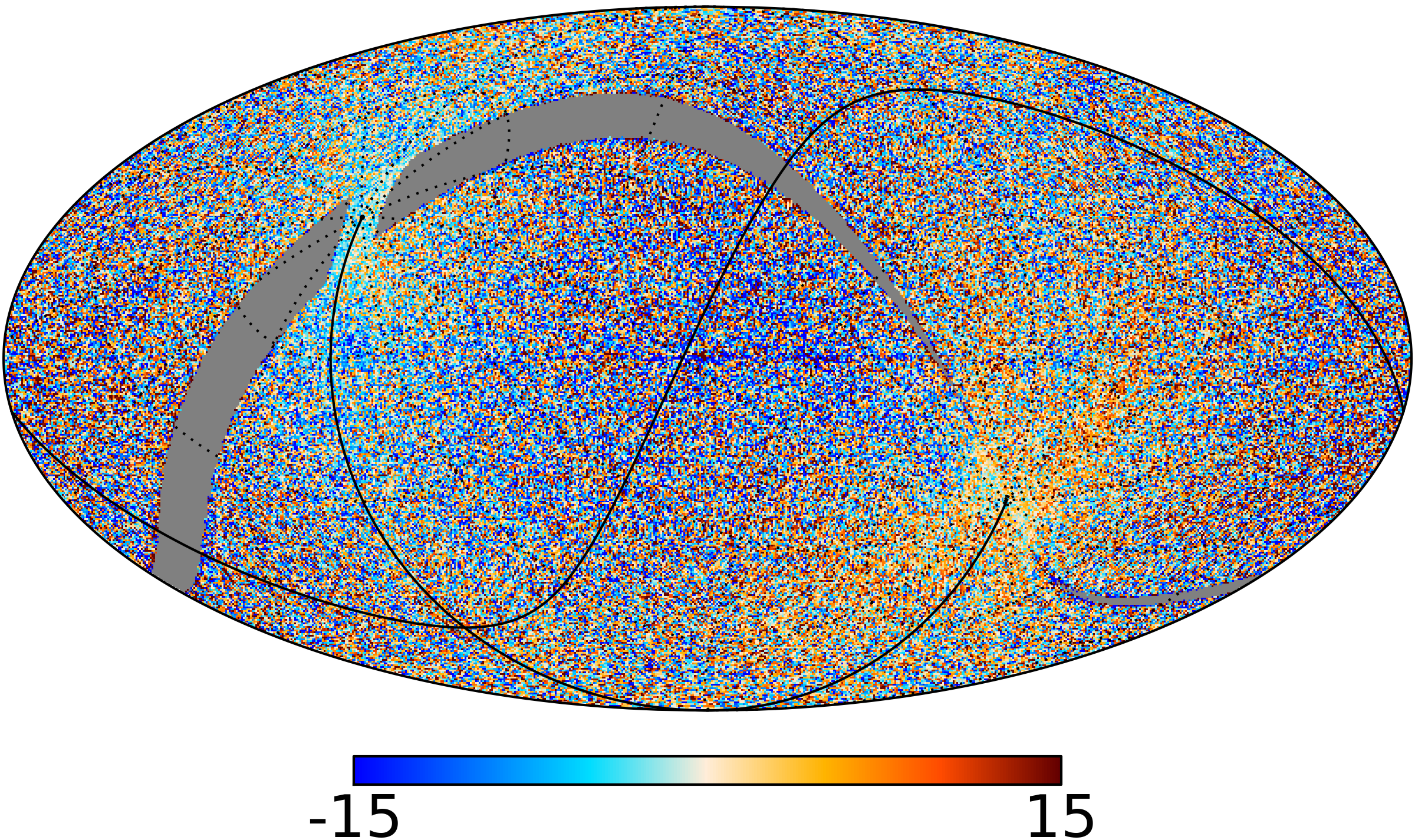} 
   \end{array}$
  \caption{Survey 2 minus Survey 1 difference maps before (left) 
   and after (right) zodiacal emission removal. 
   The rows, from top to bottom, are for 217, 143, and 100\,GHz.
   All maps are in units of $\mu\mathrm{K}_\mathrm{CMB}$.}
  \label{fig:beforeAndAfterJackknivesCMB}
 \end{figure*} 
 
 As discussed in the HFI processing paper~\citep{planck2013-p03} 
 and in the \Planck\ explanatory supplement~\citep{planck2013-p28}, 
 these fits are used to create the implied zodiacal
 and sidelobe emission in each HFI observation, 
 which can then be removed from the data before the maps are recreated.
 Note that the fitted values are strictly used for the removal -- even, for example, 
 when they are negative.
 The Survey~2 minus Survey~1 difference maps for the 
 857-1 horn both with and without zodiacal and far sidelobe removal are shown 
 in Figs.~\ref{fig:beforeAndAfterJackknives} and~\ref{fig:beforeAndAfterJackknivesCMB}.
 
 Inspection of the right-hand columns of 
 Figs.~\ref{fig:beforeAndAfterJackknives} and~\ref{fig:beforeAndAfterJackknivesCMB}
 shows some artefacts of the aforementioned systematics, especially at submillimetre frequencies.
 In Table~\ref{tab:syst}, we show the differences between emissivities measured from Surveys~2
 and~4 and emissivities measured from Surveys~1 and~3, divided by the statistical uncertainty. 
 To be more concrete, the first six entries in the 857\,GHz (top) row compare the values of
 the circles with those of the squares in Fig.~\ref{fig:fit857}. 
 Were there no systematic effects at all, we would expect these values to be of order 1. 
 A larger number here is an indication there are either effects that need to be accounted for, 
 or that the zodiacal or far sidelobe model may need improvement.

\begin{table*}
\begingroup
\caption{Survey-to-survey consistency of the measurements.  For each component and frequency, the quantity $\sum_{d,y}\left(s_{d,2y}-s_{d,2y-1}\right)\left(e_{d,2y}^{-2}+e_{d,2y-1}^{-2}\right)^{-\nicefrac{1}{2}}$ is given, where the sum is over years $y=1,2$ and all detectors $d$ in a given frequency, and where $s_{d,2y}$ and $e_{d,2y}$ are the values and errors for emissivity measurements from even surveys. Quantities involving $2y-1$ are the corresponding values for odd surveys.  Thus, the entries in the
tables are essentially the differences in fitted emissivities between even and odd surveys divided by the expected dispersion.}
\label{tab:syst}
\nointerlineskip
\vskip -3mm
\footnotesize
\setbox\tablebox=\vbox{
 \newdimen\digitwidth 
 \setbox0=\hbox{\rm 0} 
 \digitwidth=\wd0 
 \catcode`*=\active 
 \def*{\kern\digitwidth}
 \newdimen\signwidth 
 \setbox0=\hbox{+} 
 \signwidth=\wd0 
 \catcode`!=\active 
 \def!{\kern\signwidth}
\halign{\hbox to 1in{#\leaderfil}\tabskip 2em&
    \hfil$#$\hfil\tabskip=1em&
    \hfil$#$\hfil&
    \hfil$#$\hfil&
    \hfil$#$\hfil&
    \hfil$#$\hfil&
    \hfil$#$\hfil&
    \hfil$#$\hfil&
    \hfil$#$\hfil&
    \hfil$#$\hfil\tabskip=0pt\cr 
\noalign{\doubleline}
\omit&\multispan9\hfil$\sum_{d,y}\left(s_{d,2y}-s_{d,2y-1}\right)\left(e_{d,2y}^{-2}+e_{d,2y-1}^{-2}\right)^{-\nicefrac{1}{2}}$\hfil\cr
\noalign{\vskip -3pt}
\omit\hfil\sc Frequency\hfil&\multispan9\hrulefill\cr
\omit\hfil[GHz]\hfil&\omit\hfil Cloud\hfil&\omit\hfil Ring\hfil&\omit\hfil Feature\hfil&\omit\hfil Band 1\hfil&\omit\hfil Band 2\hfil&
    \omit\hfil Band 3\hfil&\omit\hfil PR\hfil&\omit\hfil SR Direct\hfil&\omit\hfil SR Baffle\hfil\cr
\noalign{\vskip 3pt\hrule\vskip 5pt}
857 &  -4.8 &  !4.2 & *-5.1 &   0.7 &  -1.6 &  !0.7 &  !0.4 &  -0.3 &  -0.9\cr
545 &  -3.1 &  !2.7 & *!1.1 &   1.2 &  -0.2 &  !1.3 &  !3.0 &  -1.0 &  -1.4\cr
353 &  !2.2 &  -0.5 & *-8.4 &   0.7 &  !2.2 &  -0.1 &  -1.6 &  !3.2 &  !2.0\cr
217 &  -1.4 &  !5.4 & -10.4 &   0.4 &  -0.2 &  !0.4 &  -3.1 &  !1.4 &  -0.5\cr
143 &  -4.3 &  !5.8 & *-1.1 &   0.4 &  -1.7 &  !0.5 &  -2.1 &  !1.1 &  !0.2\cr
100 &  -1.2 &  !2.7 & *-3.2 &   0.2 &  -0.9 &  !0.2 &  -2.2 &  !2.4 &  !0.4\cr
\noalign{\vskip 3pt\hrule\vskip 5pt}}}
\endPlancktablewide
\endgroup 
\end{table*}

 \section{Discussion}\label{sec:discussion}

  Here we discuss the fit implications for both the HFI instrument and the zodiacal cloud.

  \subsection{Far Sidelobes}\label{sec:farSidelobeResults}
    
   In addition to fit values obtained for the Galaxy seen in 
   each component of the far sidelobes, the columns labelled ``Prediction'' in
   Table~\ref{tab:galacticSidelobeFit} show the expected
   values of the spillover, normalized to that of the 353\,GHz
   channel, from \cite{tauber2010b}. These are the ratios of the expected
   spillover in each frequency, compared to that at 353\,GHz, the frequency
   for which the sidelobe calculations were done. 
   Since the fit values account for the changes in Galactic
   emission with frequency, if our predictions and data were perfect, the
   fit values would match those of the predictions. 
   
   The FSL signature is clearly visible at 857\,GHz in the bottom panel of Fig.~\ref{fig:surveymaps}, and quantified in Table~\ref{tab:galacticSidelobeFit}. As the 857 and 545\,GHz channels are multi-moded, the differences are not that surprising; unlike single-moded horns, multi-moded horns allow propagation of multiple, interacting electromagnetic modes.  It is difficult to perform the calculations necessary for the prediction, since each mode must be accounted for, and each mode interacts with the others~\citep{Murphy2010}. In addition, the specifications for the horn fabrication were quite demanding, and small variations could give large variations in the amount of spillover.

   For the lower frequency, single-moded channels, however, the 
   situation is different. There is no clear detection of PR
   spillover. While the significant negative values may indicate some
   low-level, large-scale systematic, there seems to be nothing with
   the distinctive signature of primary spillover at frequencies between
   100 and 353\,GHz. 
      
   For the direct contribution of the secondary SR spillover, the situation is similar at 
   353\,GHz, but at 217 and 143\,GHz we find a 3\,$\sigma$ detection at about 
   the level expected, while at 100\,GHz the value is about 2.5
   times higher than expected, though the signal-to-noise ratio of the detection is 
   less than 2\,$\sigma$. The baffle contribution to the SR spillover
   seems to be in accord with expectations at 353 and 217\,GHz, and
   higher than what is predicted at 100\,GHz. 
   
   The values for the PR spillover, which is the most distinctive of 
   the far sidelobe patterns and therefore presumably the easiest to disentangle 
   from other effects, suggest that the PR spillover values
   in Table~2 of \cite{tauber2010b} may be slightly overestimated. The values
   for the direct contribution of the SR spillover roughly confirm 
   the far sidelobe calculations. The baffle contribution to the SR spillover
   seems a bit high. We take the ensemble of these numbers as rough confirmation
   that our beam calculations are not drastically incorrect, but do not use
   the specific numbers in either \cite{planck2013-p03c} or \cite{planck2013-p03f}. 
   Similar conclusions are drawn for the LFI in \cite{planck2013-p02d}.
   
The pre-launch optical system measurement campaign reinforces this conclusion.  It found no significant SR baffle spillover excesses, or anything indicating a problem with the primary spillover calculations that might lead to significant, negative spillover values~\citep{tauber2010b}. We attribute these numbers to systematics, perhaps linked with long time constants~\citep{planck2013-p03c}, which might move these large-scale features on the sky, or to the fact our sidelobe model has not accounted for the offset in the focal plane between the horn used to make the model and the 100\,GHz horns.
     
   We have also included a 
   template of the dipole as seen through the far sidelobes
   in some fits to check if they are detected. As expected, they are not.
   The results quoted above are from fits that do not include these dipole templates.

  \subsection{Diffuse Cloud}
  
   Figure~\ref{fig:noDipoleSidelobesFit} shows the emissivity of the 
   Diffuse Cloud falling off with increasing wavelength, as would be expected for particles 
   with characteristic sizes of order 30\micron. The dashed line shows a 
   flat emissivity to 150\micron,
   with values proportional to the frequency squared at longer wavelengths.
   This is to aid comparison with Fig.~2 of~\cite{fixsendwek2002},
   who used FIRAS data to investigate the 
   far-infrared/submillimetre behaviour of the zodiacal cloud. 
   Our results are consistent with their conclusions\footnote{Care 
   must be taken with direct comparisons between the emissivities quoted
   in~\cite{fixsendwek2002}, who quote emissivities relative to a 
   245\,K cloud, and those in this work, 
   which follow K98 and assume a cloud of 
   temperature 286\,K at 1\,AU from the Sun.}. 
 
   As the Diffuse Cloud is so much brighter in the mid-infrared than
   in the submillimetre, its relatively low level at \Planck\ wavelengths has
   been exploited in~\cite{planck2013-p03d} to set in-flight limits on 
   any possible out-of-band leaks in the instrument's spectral transmission.
   
  \subsection{Circumsolar Ring and Earth-Trailing Feature}
 
   We draw no conclusions about the Circumsolar Ring or the 
   Earth-Trailing Feature. The fit values obtained for their
   emissivities are inconsistent from frequency to frequency, 
   and often negative. This remains true for \Planck\ data even when 
   the two components are required to have the same emissivities in the fit, 
   as was done in K98.
   
   This is not necessarily surprising, as the angle between the satellite
   spin axis and the direction of observation is less than 90\deg. This, in turn, 
   means that in addition to observing deep in the Rayleigh-Jeans region of the dust
   spectrum, \Planck\ rarely observes the centre of the Circumsolar Ring or Earth-Trailing Feature, 
   which are nominally $\approx$90\deg\ from the Sun-Earth line, in contrast to
   \textit{IRAS} and DIRBE, both of which often scanned through the regions of maximum
   density enhancement of the Ring and Feature.
   
   Inspection of the middle- and lower-left panels of Fig.~\ref{fig:fit857}
   shows systematic differences between results from even- and odd-numbered
   Surveys (that is, the circles and squares seem to be systematically different,
   regardless of whether they are blue or red). 
   \Planck's observing pattern was different for odd- and even-numbered Surveys, but 
   similar for even-numbered or odd-numbered Surveys alone.
   Random noise is not important here, since the measurements with similar observations are repeatable, 
   so this indicates either that the Circumsolar Ring and Earth-trailing model templates
   themselves need improvement at \Planck\ wavelengths, particularly in the outer regions of
   the features sampled by Planck, or that systematic errors are affecting these specific components.
      
   As the results for these components in particular are difficult to interpret,
   we have checked that the conclusions presented elsewhere in this
   work remain essentially the same whether or not we include the 
   Circumsolar Ring and Earth-Trailing Feature. 
      
  \subsection{Bands}\label{subsec:Bands}

   An interesting feature of Fig.~\ref{fig:noDipoleSidelobesFit}, 
   and the primary result of this work, is the difference between the 
   emissivities of the bands and that of the Diffuse Cloud. This indicates 
   that the particles in the bands are larger than those in the Diffuse Cloud.  
   While there may be hints of this in the longest wavelength DIRBE data, the effect
   becomes clear at \Planck\ wavelengths.
   For Bands~1 and~3, the emissivity seems to cut off near $\lambda_\mathbf{cutoff}\simeq 1\,\mathrm{mm}$. 
   Since the cutoff is related to the characteristic particle size, $a$, as
   $\lambda_\mathbf{cutoff} \simeq 2\pi a$, this would indicate a particle size of order 150\micron\
   or greater. This can be compared to an implied characteristic size of around 30\micron\ for the particles
   in the Diffuse Cloud.
   
   This is not unexpected. The composition of the Diffuse Cloud is disputed, 
   but is often claimed to be both asteroidal and 
   cometary~\citep[see, for example,][]{Kortenkamp1998,Nesvorny2010,Tsumura2010}. 
   Since the bands, on the other hand, are understood to be
   asteroidal debris only~\citep{sykes86}, the difference may simply be a
   reflection of these different origins.
   
   The fact that the fitted emissivities of Bands~1 and~3 rise above unity is 
   perplexing. At first glance, one might imagine some new, 
   cold component in the cloud causing an enhancement that might be interpreted as an excess in emissivity in some 
   other component.  However, to peak around 545\,GHz, this component would have 
   to have a temperature of the order of 10\,K, and therefore be much more distant
   than most of the dust usually associated with the zodiacal cloud. 
   It is difficult to understand how such a component could survive the differencing process
   used in this analysis, which reduces signals from distant sources more than 
   those nearby, or how such a component could mimic an excess in two
   dust bands above the ecliptic plane, but not do the same in the other 
   zodiacal cloud components. 
   
   One might worry that covariance between between the various components might be
   causing problems in the fitting procedure. To check this, we have repeated the 
   fit including and omitting various combinations of the the 
   Circumsolar Ring and Earth-Trailing Feature, or both, and assuming their
   emissivities were independent or equal. In no case did the difference
   between the Diffuse Cloud and the dust bands disappear. 
   
   The excess may ultimately be explained by degeneracies in the model for the density of the bands. 
   As presented in Sect.~\ref{sec:bandmath}, the normalization of the density of particles is completely 
   degenerate with the emissivity for each band. In addition, the emission is also roughly proportional to
   the temperature normalization, because we are observing 
   in the Rayleigh-Jeans tail of the zodiacal emission.
   While any overall change in the temperature of the IPD particles
   would scale all components of the zodiacal emission, because temperature is nearly 
   inversely proportional to the square-root of the distance
   from the Sun, the location of the bands is important. 
   While the excess over unity is too large to be explained by errors in 
   distance and thus temperatures alone, one might appeal to
   a change in a combination of distance, particle density normalization, and 
   emissivity of these bands to arrive at mutually 
   consistent results for both \Planck\ and DIRBE. As this will involve 
   a simultaneous study of both \Planck\ and DIRBE data, it is beyond
   the scope of this paper. 
   
   Bands~1 and~3 also seem to show different behaviour than Band~2. Since
   Bands~1 and~3 are both at high ecliptic latitude, while Band~2
   is not, one might again worry that one of the other templates to 
   which we are fitting might have significant overlap with a subset of the 
   bands, which in turn could cause an apparent difference in emissivities. 
   To check this, we have repeated the fits with and without
   various combinations of the cloud, Circumsolar Ring, and Earth-Trailing Feature, 
   as well as the far sidelobes. In all cases, Bands~1 and~3 are always 
   significantly different than Dust Band~2. When the Diffuse Cloud itself is omitted
   from the fit, the emissivity of Dust Band~2 goes up, but is still distinctively different
   from that of Dust Bands~1 and~3. 
   
   As Dust Band~2 is a combination of the \textit{IRAS}\ $\alpha$ and 
   $\beta$ bands, one may also worry that one of these two is more important
   for the shorter \textit{IRAS}\ and \textit{COBE}\ wavelengths, but that the other might 
   be more important for the longer \Planck\ wavelengths. 
   We note that the Band~2 emission is dominated by contributions from the 
   Karin/Koronis family~\citep[see][Fig.~1]{Nesvorny2008}, 
   but have therefore confirmed specifically that varying the $\delta_\zeta$ parameter of the second
   band between values appropriate for either $\alpha$ or $\beta$
   does not remove this difference (see Table~\ref{tab:bandInfo}).

   If the age of Band~2 was significantly different from those of Bands~1 and~3, we might 
   argue that Poynting-Robertson drag had depleted some of the bands of more
   small particles than the others~\citep{Wyatt2011}. N03 and N08, however, 
   have estimated the ages of most of the asteroid families that might be associated with the bands (reproduced in 
   Table~\ref{tab:bandInfo}), and the age of any of the associations with
   Band~2 is between those of any of the possible associations with Bands~1 or~3.
   These same figures tend to rule out modifications of the material properties due to photo-processing
   or solar wind exposure for differing periods. Band~2 also seems to be
   roughly the same distance from the Sun as the other two bands, so it is difficult to appeal
   to differences in environment as the cause. 
   
   We speculate on the following to explain any differences: 
   Veritas, the asteroid family proposed to be associated with Dust Band~1, 
   is classified as carbonaceous~\citep{Bus2002}. 
   As noted above, the \textit{IRAS}\ $\beta$ band, associated with the Karin
   family of asteroids, seems to dominate the emission from Dust Band~2. 
   Karin and its larger sibling, Koronis, are classified as siliceous, or stony, 
   objects~\citep{Bus2002,Carvano2010}. While Dust Band~3 has a 
   number of asteroid families that may be contributing to it
   (see Table~\ref{tab:bandInfo}), we propose that the emission
   is dominated by carbonaceous-based asteroid families 
   (three quarters of the asteroids in the Solar System are carbonaceous), 
   and that the difference
   in emissivity between Dust Band~2 and Dust Bands~1 and~3 arises from 
   this difference in composition. The differing emissivities may be either
   due to this intrinsic composition difference, or the size-frequency
   distribution of particles that results from different kinds of asteroids 
   colliding~\citep[for example]{Grogan2001}. 
   This explanation would not be valid, however, if it were to turn 
   out that Dust Band~3 was dominated by dust associated with the Iannini
   asteroid, for example, since it is siliceous.

 \subsection{Implications for the CMB}
 
\begin{figure*}[htbp]
  \includegraphics[width=180mm]{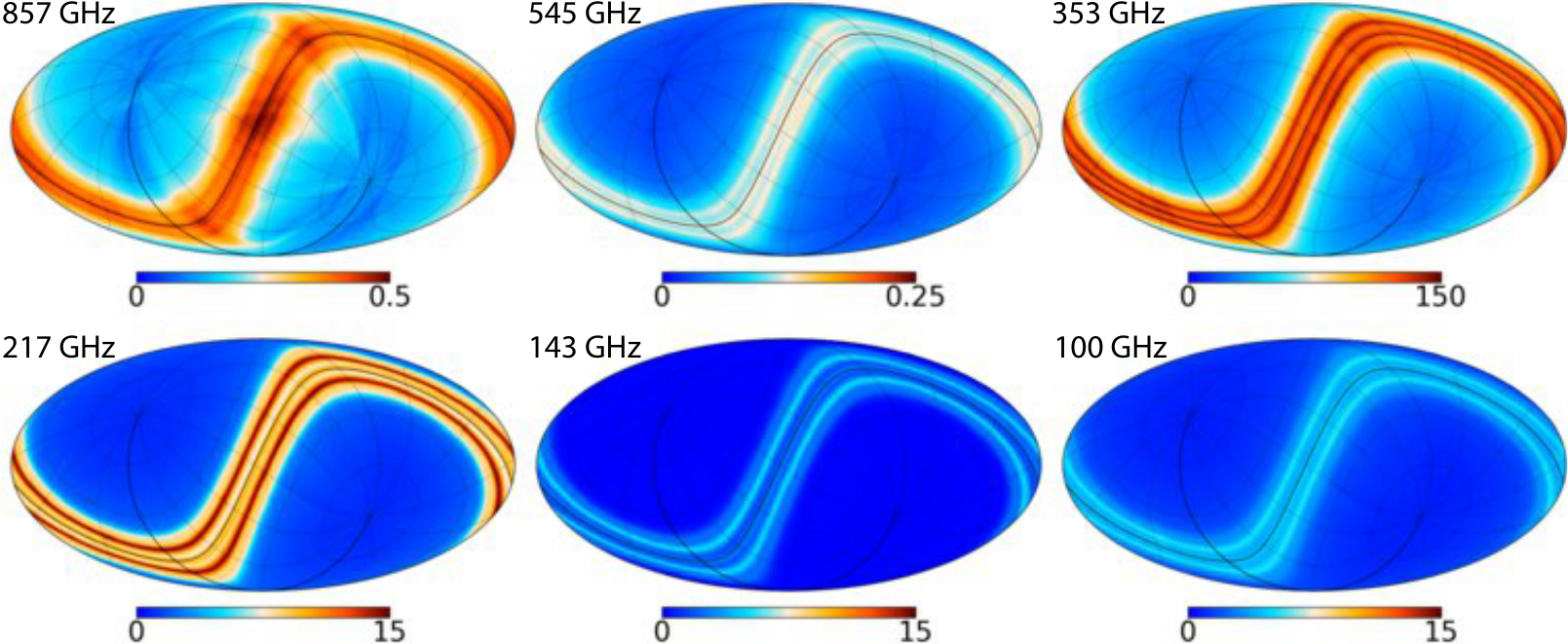}
  \caption{Difference maps (uncorrected for zodiacal emission minus corrected) showing the zodiacal corrections calculated in this paper.   
  Units are MJy/sr for 857 and 545\,GHz, 
  $\mu\mathrm{K}_\mathrm{CMB}$ for the other frequencies.}
  \label{fig:MapCorrections}
 \end{figure*}

  Figure~\ref{fig:MapCorrections}
  shows the zodiacal emission implied by the fits, created by subtracting the maps 
  made after applying the zodiacal emission correction from those that 
  were made without the correction. One can see here the difference in 
  the relative amplitudes of the bands versus the Diffuse Cloud, the
  bands being relatively more important at low than at high frequencies. 
  
\begin{figure}[htbp]
   \centering
   \includegraphics[width=88mm]{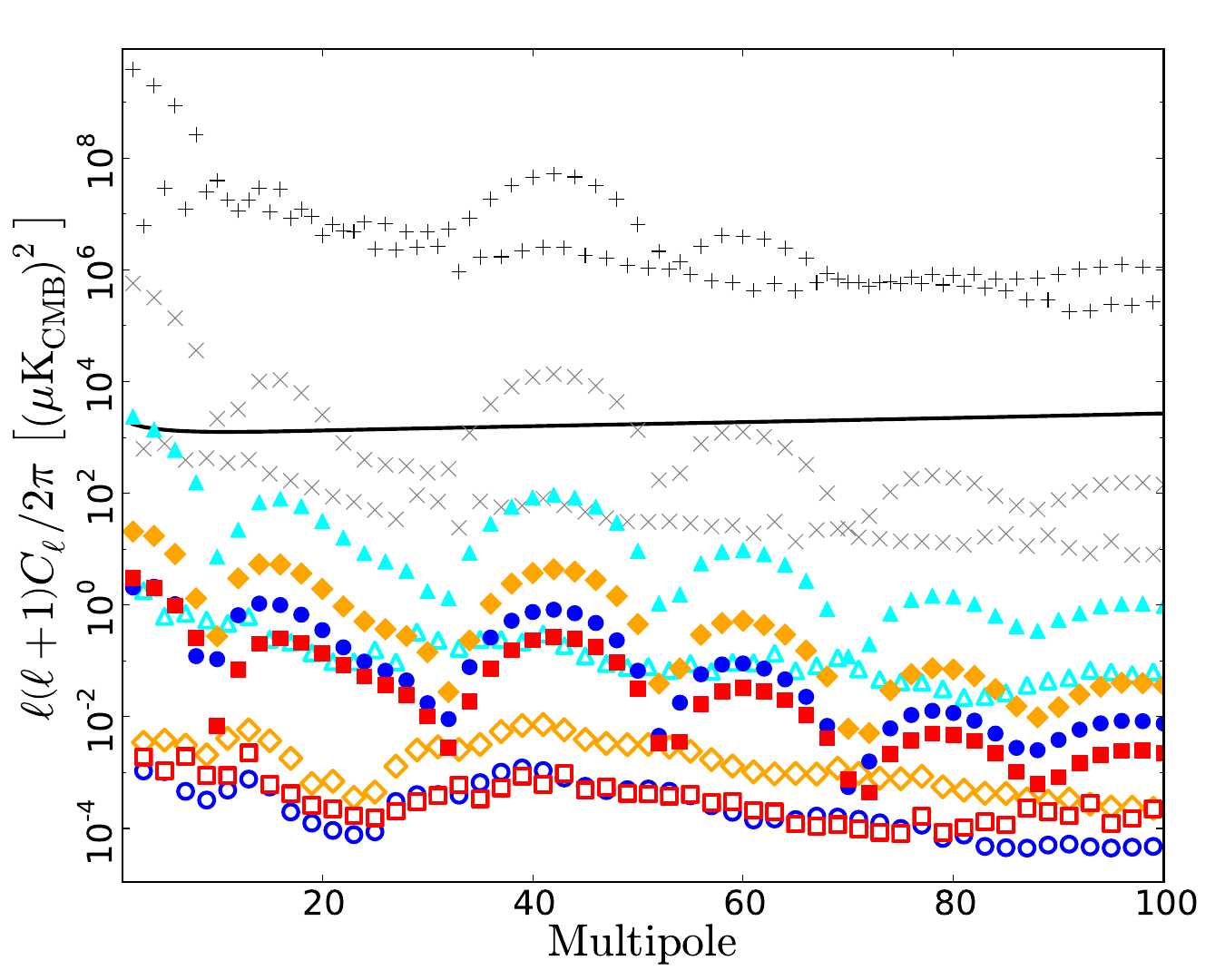}
   \caption{Power spectra of the zodiacal correction maps shown in Fig.~\ref{fig:MapCorrections}.
    Black plus signs = 857\,GHz; grey crosses = 545\,GHz; cyan triangles = 353\,GHz; orange diamonds = 217\,GHz; blue circles = 143\,GHz; and red squares = 100\,GHz.  For the ``CMB channels'' 100--353\,GHz, even multipoles are shown with filled symbols, odd multipoles with empty symbols. The ``even-odd'' pattern is a consequence of the symmetry around the ecliptic plane -- odd multipoles are almost absent, as they would indicate structure in the maps that was anti-symmetric about the ecliptic plane. The best-fit $\Lambda$CDM CMB temperature anisotropy spectrum using the ``Planck+WP+highL+BAO'' data combination from~\cite{planck2013-p11}, is shown as the solid line roughly half-way down the plot, orders of magnitude above the zodiacal spectrum in the \Planck\ CMB channels.}
   \label{fig:Cls}
  \end{figure}

  Figure~\ref{fig:Cls} shows the power spectra of the zodiacal 
  correction maps, all in units of $\left(\mu\mathrm{K}_\mathrm{CMB}\right)^2$. 
  Here, the cloud is seen at multipoles
  of less than about 10, while the bands and other structures are seen in higher
  multipoles. 
  
  At 143\,GHz, the signal reaches a few $\mu\mathrm{K}_\mathrm{CMB}$ in the map, while the power spectrum has values of the order of one $\left(\mu\mathrm{K}_\mathrm{CMB}\right)^2$. The absence of power in the odd multipoles is a consequence of the north-south symmetry of the signal. While this pattern is reminiscent of the so-called ``hemispheric anomaly'' \citep{Eriksen2004,planck2013-p09}, we emphasize that the zodiacal emission is essentially symmetric about the ecliptic plane, while the anomalies exhibit anti-symmetries about the plane. Thus, standard zodiacal emission cannot be evoked to explain the anomalies. This is consistent with the conclusions of~\cite{Dikarev2008}.
  
  The spectra in Fig.~\ref{fig:Cls} can be compared with the CMB temperature anisotropy spectrum, which is shown as the black line about halfway between the top and bottom of the plot. The zodiacal emission correction spectra are orders of magnitude smaller than the CMB spectrum. The zodiacal emission therefore cannot compromise \Planck's cosmological results. 
  
 \section{Conclusion}\label{sec:conclusion}
 
  Zodiacal emission has long been an important foreground for searches for 
  the extragalactic background at infrared wavelengths. With the 
  ever-increasing sensitivity of CMB experiments, 
  it will soon become important to account for at longer wavelengths
  as well. 
  
  The K98 model does fairly well in modelling the diffuse zodiacal cloud emission at \Planck\ wavelengths, as long as appropriate emissivities are assumed. It does less well, however, in modelling the other features. Because they appear to be more emissive at these frequencies than the cloud, the bands contribute more to the zodiacal emission relative to the Diffuse Cloud at CMB frequencies (i.e., near 143\,GHz = 2.1\,mm). The 2013 \Planck\ release includes both maps that have had zodiacal emission removed, and maps that have not had zodiacal emission removed. 
  
  We note that \cite{planck2013-p06b} and \cite{planck2013-XVII} found better results when using the Planck maps that had been cleaned of zodiacal emission, validating to some extent the zodiacal emission removal done here. \cite{planck2013-p06}, on the other hand, used maps that had not had the zodiacal emission removed to make estimations of the CMB and other astrophysical components in the HFI maps. The component separation methods used there naturally correct for a large amount of zodiacal emission, as it is spectrally similar to Galactic dust emission in the \Planck\ CMB channels. As might be expected, the differences between the dust maps obtained with the two different methods are explained by a consistent accounting of the zodiacal emission.
  
  Improvements in modelling of the Circumsolar Ring, Earth-Trailing Feature, 
  and the dust bands, as well as inclusion
  of fainter and partial bands~\citep[e.g.,][]{Espy2009} 
  should be done to make truly ``clean'' CMB maps. As these
  bands are believed to be the products of asteroid collisions, 
  further study of the bands at these wavelengths may also inform 
  us about the nature of the intermediate-sized particles created during the
  destruction of the associated asteroids. We may hope to learn not only
  more about the size distribution, but also the differences between,
  for example, the results of collisions involving siliceous and
  carbonaceous asteroids. 
  
  Just as material from asteroid collisions contributes to zodiacal emission, 
  material shed from comets must also contribute. As part
  of the HFI data reduction process, we mask out Solar System objects
  that would cause ``noise'' in the final sky maps. We searched for
  comets as part of this process, but found only one (Christensen; see Appendix~\ref{sec:Ast}). 
  We have not yet detected extended tails of comets. 
    
  One of the primary goals of the next stage of analysis, once this ``nearby''
  IPD has been completely removed, will be to search for or set limits on dust associated
  with the Kuiper Belt. This will require total power maps rather than the differenced data used here, as the Kuiper Belt is much
  farther away than the dust considered here and the amount of signal removed in the differencing
  process would be prohibitive.  However, a beneficial side effect of such analysis may be better limits on the IPD discussed here. 
  Assuming that the noise after removing such effects is perfectly Gaussian and that
  Galactic contamination is mastered, Planck
  should be able to reduce the uncertainties in emissivities by a factor of roughly three for the 
  Diffuse Cloud and Band 2, and by a factor of roughly eight for Bands 1 and 3. There would be only modest
  gains for the Circumsolar Ring, and the Earth-Trailing Feature would not be improved at all by 
  moving to total power fits. 
  
  The full-mission \Planck\ data release will include polarization information.
  While polarized zodiacal emission is not expected, limits will be put on possible contamination 
  of the polarization of the CMB by such emission. 
  
  Work is now under way to address all these points for the next \Planck\ data release. 
  While the signal is quite small -- at CMB wavelengths the signal we are discussing
  is orders of magnitude smaller than the primary CMB anisotropies -- it is 
  detectable and should be subtracted from the data. 
  There will be improvements in dust modelling, improvements in satellite modelling, 
  and additions to address polarization. The ultimate goal will be simultaneous
  analyses with \textit{IRAS}, \textit{COBE}, \textit{Akari}, and other data sets 
  to understand the large-scale zodiacal emission from the near-infrared to the microwave. 
  
 \begin{acknowledgements}
  This paper benefited from exchanges with Dale Fixsen, Tom Kelsall and Janet
  Weiland. We acknowledge the IN2P3 Computer Center (\url{http://cc.in2p3.fr}) 
  for providing a significant amount of the computing resources and services needed for this work.
  We acknowledge the use of the Legacy Archive for Microwave
  Background Data Analysis (LAMBDA). Support for LAMBDA is provided by
  the NASA Office of Space Science.

  The development of \Planck\ has been supported by: ESA; CNES and CNRS/INSU-IN2P3-INP (France); 
  ASI, CNR, and INAF (Italy); NASA and DoE (USA); STFC and UKSA (UK); 
  CSIC, MICINN, JA and RES (Spain); Tekes, AoF and CSC (Finland); 
  DLR and MPG (Germany); CSA (Canada); DTU Space (Denmark); SER/SSO (Switzerland); 
  RCN (Norway); SFI (Ireland); FCT/MCTES (Portugal); and PRACE (EU). 
  A description of the \Planck\ Collaboration and a list of its members, 
  including the technical or scientific activities in which they have been involved, 
  can be found at \url{http://www.sciops.esa.int/index.php?project=planck&page=Planck_Collaboration}.
 \end{acknowledgements}

 \bibliographystyle{aa}
 \bibliography{Planck_bib,Zodi,SSO4ken/SSO_extra}

 \appendix
 
 \section{Solar System Objects}\label{sec:Ast}

\Planck\ has detected many moving Solar System objects, mostly asteroids, but also one comet.  Most are masked at an early stage of the 
analysis so as not to affect the maps of the sky, thus they cannot be easily extracted from the delivered products. We
therefore present them here. 

Solar System objects are located using the
JPL Horizons\footnote{\url{http://ssd.jpl.nasa.gov/?horizons}}
\citep{giorgini1996} system programmed with the \Planck\ orbit.
Table \ref{table:SSOSeasons} shows the epochs of observation, and Tables
\ref{table:SSOEphem} and \ref{table:SSOEphem2} show the distance from the Sun and \Planck\ 
at the time of observation, and the position on the sky.

During the standard HFI timeline processing, these objects are flagged and not 
included in the standard HFI maps~\citep{planck2013-p03}, which makes these maps
an excellent tool for removing the background of these moving objects.

5\,Astraea, Christensen, and 128\,Nemesis were not flagged in the maps, 
so their flux densities reported here have been adjusted for the fact that a fraction
of their emission would have still been in the maps used to remove the background 
before estimating their fluxes.

We select time-ordered data within 0\pdeg5 of the source 
and recalibrate into MJy\,sr$^{-1}$ (using \textit{IRAS}-conventions).   
We project the pointing of \Planck\ into coordinates relative to the predicted position of the moving object.  
A synthetic background timeline is estimated  by resampling the 
HEALPix-gridded~\citep{gorski2005} \Planck\ maps using cubic spline interpolation.  

\begin{figure*}[!ht]
\centerline{\includegraphics[width=17cm]{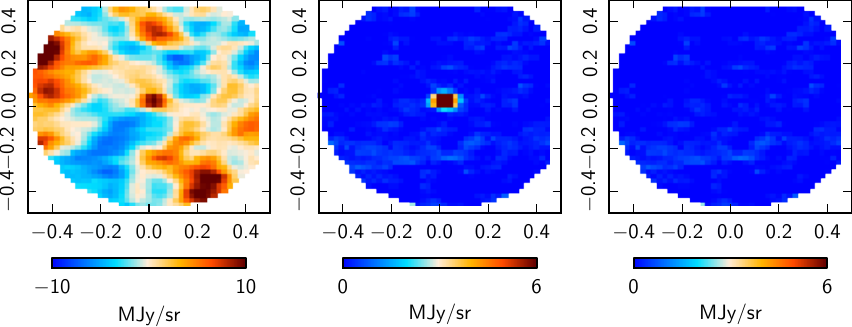}}
\caption{Example PSF fitting.  Background removal and fitting are done in the time domain, bolometer by bolometer.  These
  stacked maps were created only for visualization.
  \textit{Left}: Stacked data for the second observation season of 1\,Ceres at 857\,GHz, before background removal. 
  \textit{Center}: Stacked data after background removal. 
  \textit{Right}: Stacked data after background and source removal.}
 \label{fig:ProcessingExample}
\end{figure*}

We fit the main-beam template
\citep{planck2013-p03c} for each bolometer to the time-ordered data.
There are seven free parameters in the fit: $x_0$ and $y_0$, 
corresponding to the centroid of
the object; a rotation angle $\psi$; an amplitude $A$; and three
parameters describing a linear slope in $x$ and $y$ of any residual
background. The amplitude times the solid angle of the beam model gives the
flux density. Figure~\ref{fig:ProcessingExample} shows an example.

We find negligible difference between a fit assuming a Gaussian template for the 
beam instead of the PSF, as reported in \cite{planck2013-p03c}, 
so here we report the PSF-fit flux densities in Table~\ref{table:SSOFluxDensity2}.
We also tried aperture photometry, but the results were noisier and
inconsistent from season-to-season, which might be expected, as the Ecliptic
plane is somewhat under-sampled by \Planck/HFI when using only a single season of data.

\subsection{Notes}

The residual map variance can change with different backgrounds. 
During its first observation period, for example, 1\,Ceres was in a region of high foregrounds, and 
so was difficult to detect at 545 and 857\,GHz, the bands most susceptible to foregrounds. 

\begin{table*}[tmb]
\begingroup
\newdimen\tblskip \tblskip=5pt
\caption{Dates of observation of Solar System objects detected by \Planck.} 
\label{table:SSOSeasons} 
\nointerlineskip
\vskip -3mm
\footnotesize
\setbox\tablebox=\vbox{
   \newdimen\digitwidth 
   \setbox0=\hbox{\rm 0} 
   \digitwidth=\wd0 
   \catcode`*=\active 
   \def*{\kern\digitwidth}
   \newdimen\signwidth 
   \setbox0=\hbox{+} 
   \signwidth=\wd0 
   \catcode`!=\active 
   \def!{\kern\signwidth}
   \halign{\hbox to 3cm{#\leaderfil}\tabskip 2em&
    \hfil#\hfil&
    \hfil#\hfil&
    \hfil#\hfil&
    \hfil#\hfil&
    \hfil#\hfil\tabskip 0pt\cr   
    \noalign{\doubleline}
\omit&\multispan5\hfil\sc Dates of Observation\hfil\cr
\noalign{\vskip -3pt}
\omit&\multispan5\hrulefill\cr
\omit\hfil\sc Object\hfil&Season 1&Season 2&Season 3&Season 4&Season 5\cr
    \noalign{\vskip 3pt\hrule\vskip 5pt}
    1\,Ceres&24/03/10--25/03/10&10/09/10--11/09/10&03/07/11--04/07/11&03/12/11--04/12/11&\ldots\cr
2\,Pallas&23/02/10--25/02/10&30/07/10--31/07/10&02/05/11--03/05/11&13/10/11--15/10/11&\ldots\cr
3\,Juno&10/12/09--12/12/09&30/12/10--01/01/11&27/05/11--28/05/11&\ldots&\ldots\cr
4\,Vesta&24/11/09--26/11/09&29/04/10--01/05/10&28/04/11--29/04/11&26/10/11--28/10/11&\ldots\cr
5\,Astraea&05/10/09--06/10/09&30/07/10--31/07/10&02/01/11--03/01/11&27/12/11--28/12/11&\ldots\cr
6\,Hebe&27/06/10--30/06/10&13/12/10--15/12/10&21/12/11--22/12/11&\ldots&\ldots\cr
8\,Flora&12/06/10--15/06/10&30/11/10--02/12/10&06/01/12--07/01/12&\ldots&\ldots\cr
9\,Metis&30/01/10--31/01/10&11/07/10--13/07/10&23/04/11--25/04/11&10/10/11--11/10/11&\ldots\cr
10\,Hygiea&12/11/09--14/11/09&18/04/10--19/04/10&23/02/11--23/02/11&04/08/11--05/08/11&\ldots\cr
11\,Parthenope&01/10/09--02/10/09&10/03/10--11/03/10&23/01/11--25/01/11&03/07/11--05/07/11&\ldots\cr
12\,Victoria&20/02/10--21/02/10&15/08/10--16/08/10&09/09/11--11/09/11&21/09/11--21/09/11&\ldots\cr
13\,Egeria&21/03/10--22/03/10&09/09/10--10/09/10&17/07/11--18/07/11&24/12/11--25/12/11&\ldots\cr
15\,Eunomia&29/03/10--30/03/10&15/09/10--16/09/10&27/08/11--28/08/11&\ldots&\ldots\cr
16\,Psyche&18/10/09--19/10/09&16/09/10--17/09/10&01/03/11--01/03/11&23/12/11--24/12/11&\ldots\cr
17\,Thetis&03/09/09--04/09/09&12/02/10--13/02/10&30/11/10--01/12/10&02/05/11--03/05/11&\ldots\cr
18\,Melpomene&21/01/10--23/01/10&06/01/11--07/01/11&02/06/11--04/06/11&\ldots&\ldots\cr
19\,Fortuna&14/09/09--16/09/09&02/03/10--03/03/10&26/01/11--27/01/11&02/07/11--04/07/11&\ldots\cr
20\,Massalia&13/12/09--15/12/09&28/12/10--30/12/10&03/06/11--05/06/11&\ldots&\ldots\cr
29\,Amphitrite&02/04/10--03/04/10&21/09/10--22/09/10&10/08/11--11/08/11&\ldots&\ldots\cr
41\,Daphne&27/11/09--28/11/09&28/08/10--28/08/10&03/02/11--04/02/11&05/11/11--07/11/11&\ldots\cr
45\,Eugenia&19/04/10--20/04/10&06/10/10--07/10/10&06/08/11--07/08/11&11/01/12--12/01/12&\ldots\cr
52\,Europa&23/09/09--24/09/09&05/03/10--06/03/10&14/01/11--15/01/11&21/06/11--23/06/11&\ldots\cr
88\,Thisbe&08/11/09--09/11/09&25/09/10--26/09/10&03/03/11--03/03/11&06/12/11--06/12/11&16/12/11--16/12/11\cr
128\,Nemesis&24/09/09--25/09/09&05/03/10--06/03/10&05/01/11--07/01/11&06/06/11--08/06/11&\ldots\cr
324\,Bamberga&03/10/09--04/10/09&12/03/10--14/03/10&02/01/11--04/01/11&25/05/11--26/05/11&\ldots\cr
511\,Davida&19/02/10--20/02/10&31/07/10--01/08/10&01/04/11--02/04/11&07/09/11--08/09/11&\ldots\cr
704\,Interamnia&18/02/10--19/02/10&03/08/10--04/08/10&16/04/11--17/04/11&01/10/11--02/10/11&\ldots\cr
Christensen&29/09/09--30/09/09&18/04/10--19/04/10&11/09/10--11/09/10&23/03/11--24/03/11&19/08/11--20/08/11\cr

    \noalign{\vskip 3pt\hrule\vskip 5pt}
   }
}
\endPlancktablewide
\endgroup
\end{table*}

\begin{table*}[tmb] 
\begingroup
\newdimen\tblskip \tblskip=5pt
\caption{Location of Solar System objects detected by \Planck.}
\label{table:SSOEphem} 
\nointerlineskip
\vskip -3mm
\footnotesize
\setbox\tablebox=\vbox{
   \newdimen\digitwidth 
   \setbox0=\hbox{\rm 0} 
   \digitwidth=\wd0 
   \catcode`*=\active 
   \def*{\kern\digitwidth}
   \newdimen\pointwidth 
   \setbox0=\hbox{.} 
   \signwidth=\wd0 
   \catcode`;=\active 
   \def;{\kern\pointwidth}
   \newdimen\signwidth 
   \setbox0=\hbox{+} 
   \signwidth=\wd0 
   \catcode`!=\active 
   \def!{\kern\signwidth}
\halign{\hbox to 3cm{#\leaderfil}\tabskip 2em&
  \hfil#\hfil&
  \hfil#\hfil& 
  \hfil#\hfil& 
  \hfil#\hfil\tabskip 0.5em& 
  \hfil$#$\hfil\tabskip 0pt\cr 
\noalign{\doubleline}
\omit&&&&\multispan2\hfil\sc Ecliptic Coordinates\hfil\cr
\noalign{\vskip -3pt}
\omit&&&&\multispan2\hrulefill\cr
\omit&&\sc Solar Range&\Planck\ \sc Range&Longitude& \omit\hfil Latitude \hfil\cr
\omit\hfil\sc Object\hfil&\sc Season&[AU]&[AU]&[deg]& \omit\hfil[deg]\hfil\cr
\noalign{\vskip 3pt\hrule\vskip 5pt}
1\,Ceres&      1&2.774&2.536&271.1&*!2.17\cr
1\,Ceres&      2&2.899&2.624&263.4&*-4.66\cr
1\,Ceres&      3&2.982&2.595&359.3&-11.99\cr
1\,Ceres&      4&2.938&2.65*&348.2&-11.24\cr
2\,Pallas&     1&2.632&2.233&231.9&!26.18\cr
2\,Pallas&     2&2.994&2.807&218;*&!35.55\cr
2\,Pallas&     3&3.377&3.244&313.5&!32.56\cr
2\,Pallas&     4&3.403&3.137&297.9&!24.6*\cr
3\,Juno&       1&2.051&1.667&356.8&*-9.55\cr
3\,Juno&       2&2.588&2.21*&178.5&*-3.66\cr
3\,Juno&       3&2.959&2.615&165.4&*!3.95\cr
4\,Vesta&      1&2.472&2.236&151.6&*!3.15\cr
4\,Vesta&      2&2.325&1.872&143.3&*!8.00\cr
4\,Vesta&      3&2.174&1.952&310.1&*!0.22\cr
4\,Vesta&      4&2.318&1.966&311.1&*-6.73\cr
5\,Astraea&    1&3.065&2.7**&294.2&*!0.92\cr
5\,Astraea&    2&2.877&2.461&*24.4&*-4.22\cr
5\,Astraea&    3&2.601&2.376&*13.7&*-5.69\cr
5\,Astraea&    4&2.085&1.634&173.1&*!0.40\cr
6\,Hebe&       1&2.062&1.611&356.3&*-2.80\cr
6\,Hebe&       2&1.945&1.537&**0.6&-18.1*\cr
6\,Hebe&       3&2.648&2.239&166.3&*!1.85\cr
8\,Flora&      1&2.089&1.695&345.6&*-3.20\cr
8\,Flora&      2&1.865&1.416&349;*&*-7.85\cr
8\,Flora&      3&2.363&1.986&186.3&*!5.42\cr
9\,Metis&      1&2.448&2.017&207.6&*!6.15\cr
9\,Metis&      2&2.619&2.42*&199.1&*!2.71\cr
9\,Metis&      3&2.635&2.463&305.8&*-3.37\cr
9\,Metis&      4&2.468&2.067&298.4&*-6.37\cr
10\,Hygiea&    1&3.245&3.09*&141.5&*-1.42\cr
10\,Hygiea&    2&3.07*&2.675&132;*&*-3.25\cr
10\,Hygiea&    3&2.795&2.455&235.1&*-4.04\cr
10\,Hygiea&    4&2.783&2.455&229.5&*-2.32\cr
11\,Parthenope&1&2.531&2.303&*98.1&*-3.94\cr
11\,Parthenope&2&2.658&2.316&*89.0&*-1.31\cr
11\,Parthenope&3&2.633&2.222&200.7&*!4.42\cr
11\,Parthenope&4&2.489&2.255&192.2&*!5.08\cr
12\,Victoria&  1&2.299&1.906&232.1&*-4.83\cr
12\,Victoria&  2&1.895&1.585&233.6&*!5.21\cr
12\,Victoria&  3&2.386&2.181&*78.8&*!0.36\cr
12\,Victoria&  4&2.412&2.066&*80.9&*-0.10\cr
13\,Egeria&    1&2.686&2.452&269.3&*-7.48\cr
13\,Egeria&    2&2.781&2.525&260.9&-15.85\cr
13\,Egeria&    3&2.725&2.3**&10.81&-16.19\cr
13\,Egeria&    4&2.598&2.419&**1.3&*-6.08\cr
15\,Eunomia&   1&2.928&2.737&278.5&*-7.26\cr
15\,Eunomia&   2&2.625&2.3**&269.6&*-0.21\cr
15\,Eunomia&   3&2.146&1.877&*63.7&!13.24\cr
16\,Psyche&    1&2.616&2.197&309.1&*-0.06\cr
16\,Psyche&    2&2.59*&2.309&*79.7&*-3.43\cr
16\,Psyche&    3&2.785&2.534&*74.1&*-2.76\cr
16\,Psyche&    4&3.167&2.794&168.9&*!0.02\cr

\noalign{\vskip 3pt\hrule\vskip 5pt}
}
}
\endPlancktablewide 
\endgroup
\end{table*}

\begin{table*}[tmb]
\begingroup
\newdimen\tblskip \tblskip=5pt
\caption{Location of Solar System objects detected by \Planck.}
\label{table:SSOEphem2}
\nointerlineskip
\vskip -3mm
\footnotesize
\setbox\tablebox=\vbox{
   \newdimen\digitwidth 
   \setbox0=\hbox{\rm 0} 
   \digitwidth=\wd0 
   \catcode`*=\active 
   \def*{\kern\digitwidth}
   \newdimen\pointwidth 
   \setbox0=\hbox{.} 
   \signwidth=\wd0 
   \catcode`;=\active 
   \def;{\kern\pointwidth}
   \newdimen\signwidth 
   \setbox0=\hbox{+} 
   \signwidth=\wd0 
   \catcode`!=\active 
   \def!{\kern\signwidth}
\halign{\hbox to 3cm{#\leaderfil}\tabskip 2em&
  \hfil#\hfil&
  \hfil#\hfil& 
  \hfil#\hfil& 
  \hfil#\hfil\tabskip 0.5em& 
  \hfil$#$\hfil\tabskip 0pt\cr 
\noalign{\doubleline}
\omit&&&&\multispan2\hfil\sc Ecliptic Coordinates\hfil\cr
\noalign{\vskip -3pt}
\omit&&&&\multispan2\hrulefill\cr
\omit&&\sc Solar Range&\Planck\ \sc Range&Longitude& \omit\hfil Latitude \hfil\cr
\omit\hfil\sc Object\hfil&\sc Season&[AU]&[AU]&[deg]& \omit\hfil[deg]\hfil\cr
\noalign{\vskip 3pt\hrule\vskip 5pt}
17Thetis&     1&2.707&2.379&*64.0&*-6.28\cr
17Thetis&     2&2.799&2.62*&*53.7&*-4.61\cr
17Thetis&     3&2.643&2.394&155.5&*!0.79\cr
17Thetis&     4&2.433&1.987&145.6&*!4.74\cr
18Melpomene&  1&1.894&1.632&*31.2&-11.75\cr
18Melpomene&  2& 2.66&2.279&184.3&*!2.54\cr
18Melpomene&  3&2.792&2.47*&169.8&*!7.48\cr
19Fortuna&    1&2.067&1.709&*77.4&*-0.50\cr
19Fortuna&    2&2.263&1.938&*77.8&*-1.70\cr
19Fortuna&    3&2.757&2.354&203.5&*-0.86\cr
19Fortuna&    4&2.828&2.612&192;*&*!0.04\cr
20Massalia&   1&2.301&1.983&357.8&*-0.01\cr
20Massalia&   2&2.154&1.743&177.3&*-0.77\cr
20Massalia&   3&2.384&2.015&171.7&*-0.19\cr
29Amphitrite& 1&2.725&2.532&283.5&*-6.53\cr
29Amphitrite& 2&2.638&2.3**&276.7&*-5.86\cr
29Amphitrite& 3&2.406&2.121&*44.2&*!2.69\cr
41Daphne&     1&3.226&2.903&345.7&*-1.72\cr
41Daphne&     2&3.516&3.223&*56.7&-11.57\cr
41Daphne&     3&3.461&3.335&*43.1&-14.57\cr
41Daphne&     4&2.986&2.639&122.8&-17.39\cr
45Eugenia&    1&2.551&2.362&300.8&*!5.51\cr
45Eugenia&    2&2.679&2.279&296.1&*!1.39\cr
45Eugenia&    3&2.901&2.661&*41.1&*-5.7*\cr
45Eugenia&    4&2.945&2.607&*30.9&*-7.41\cr
52Europa&     1&2.847&2.62*&*88.9&*-7.08\cr
52Europa&     2&2.771&2.462&*82.8&*-3.53\cr
52Europa&     3&2.906&2.525&191.7&*!5.93\cr
52Europa&     4&3.065&2.841&183.3&*!7.74\cr
88Thisbe&     1&2.391&1.964&329.2&*!6.14\cr
88Thisbe&     2&2.901&2.667&*89.4&*!2.67\cr
88Thisbe&     3&3.108&2.838&*78.6&*-0.01\cr
88Thisbe&     4&3.216&2.981&159.3&*-3.99\cr
88Thisbe&     5&3.213&2.828&160.4&*-4.32\cr
128Nemesis&   1&2.493&2.23*&*89.5&*-1.29\cr
128Nemesis&   2&2.683&2.356&*83.0&*!3.43\cr
128Nemesis&   3&3.022&2.671&184.5&*!7.05\cr
128Nemesis&   4&3.091&2.801&172.6&*!6.19\cr
324Bamberga&  1&1.993&1.675&*98.0&!13.12\cr
324Bamberga&  2&2.572&2.204&*92.5&*!7.13\cr
324Bamberga&  3&3.398&3.059&181;*&*-3.52\cr
324Bamberga&  4&3.566&3.227&165.4&*-6.94\cr
511Davida&    1&3.422&3.071&228.5&!17.37\cr
511Davida&    2&3.626&3.496&217.5&!13.4*\cr
511Davida&    3&3.756&3.595&280.3&*!6.68\cr
511Davida&    4&3.717&3.349&268;*&*!1.34\cr
704Interamnia&1&3.498&3.201&230.6&-17.46\cr
704Interamnia&2&3.381&3.248&219.7&-12.38\cr
704Interamnia&3&3.053&2.907&297.9&*-0.61\cr
704Interamnia&4&2.806&2.43*&289.8&!10.38\cr
Christensen&  1&3.235&2.864&290.7&!22.74\cr
Christensen&  2&4.137&4.038&300.2&-10.42\cr
Christensen&  3&5.066&4.911&261.8&-24.26\cr
Christensen&  4&6.385&6.318&273.7&-34.56\cr
Christensen&  5&7.408&7.222&245.3&-40.48\cr

\noalign{\vskip 3pt\hrule\vskip 5pt}
}
}
\endPlancktablewide 
\endgroup
\end{table*}

\begin{table*}[tmb]
\begingroup
\newdimen\tblskip \tblskip=5pt
\caption{Flux density of Solar System objects detected at $\ge3\sigma$ significance, using PSF fitting. Uncertainties include statistical errors of the PSF fit only, and should be
added in quadrature to a calibration error of 1\,\%, 5\,\%, and 5\,\% at 353, 545,
and 857\,GHz, respectively.}
\label{table:SSOFluxDensity2} 
\nointerlineskip
\vskip -3mm
\footnotesize
\setbox\tablebox=\vbox{
   \newdimen\digitwidth 
   \setbox0=\hbox{\rm 0} 
   \digitwidth=\wd0 
   \catcode`*=\active 
   \def*{\kern\digitwidth}
   \newdimen\signwidth 
   \setbox0=\hbox{+} 
   \signwidth=\wd0 
   \catcode`!=\active 
   \def!{\kern\signwidth}
\halign{\hbox to 3cm{#\leaderfil}\tabskip 2em&
	\hfil#\hfil&
	\hfil$#$\hfil&
	\hfil$#$\hfil&
	\hfil$#$\hfil&
	\hfil$#$\hfil\tabskip 0pt\cr  
\noalign{\doubleline}
\omit&&\multispan4\hfil{\sc Flux Density} [Jy; {\it IRAS} convention]\hfil\cr
\noalign{\vskip -3pt}
\omit&&\multispan4\hrulefill\cr
\omit\hfil\sc Object\hfil&\omit\hfil\sc Frequency\hfil&\omit\hfil Season 1 \hfil&\omit\hfil Season 2 \hfil&\omit\hfil Season 3\hfil&\omit\hfil Season 4\hfil\cr
\noalign{\vskip 3pt\hrule\vskip 5pt}
1\,Ceres&       857&      \dots&17.1*\pm0.1*&17.7*\pm0.1*&16.1*\pm0.1*\cr
1\,Ceres&       545&      \dots&*6.72\pm0.11&*7.51\pm0.08&*6.68\pm0.11\cr
1\,Ceres&       353&2.6*\pm0.5*&*2.52\pm0.15&*2.99\pm0.09&*2.5*\pm0.1*\cr
2\,Pallas&      857&8.9*\pm0.1*&*5.08\pm0.07&*3.05\pm0.09&*4.2*\pm0.1*\cr
2\,Pallas&      545&4.31\pm0.11&*2.27\pm0.15&       \dots&       \dots\cr
2\,Pallas&      353&1.7*\pm0.2*&       \dots&       \dots&       \dots\cr
3\,Juno&        857&2.96\pm0.07&*1.79\pm0.23&*0.99\pm0.13&       \dots\cr
3\,Juno&        545&1.13\pm0.36&       \dots&       \dots&       \dots\cr
4\,Vesta&       857&7.39\pm0.07&*9.32\pm0.07&10.1*\pm0.1*&*8.29\pm0.06\cr
4\,Vesta&       545&2.9*\pm0.1*&*3.55\pm0.09&*3.96\pm0.09&*3.18\pm0.08\cr
4\,Vesta&       353&1.3*\pm0.2*&*1.58\pm0.32&*1.8*\pm0.2*&*1.2*\pm0.2*\cr
6\,Hebe&        857&1.9*\pm0.1*&*1.52\pm0.09&       \dots&       \dots\cr
8\,Flora&       857&1.19\pm0.11&*1.78\pm0.09&       \dots&       \dots\cr
9\,Metis&       857&1.19\pm0.19&       \dots&       \dots&*1.06\pm0.24\cr
10\,Hygiea&     857&2.5*\pm0.1*&*4.03\pm0.11&*4.5*\pm0.1*&*4.56\pm0.09\cr
10\,Hygiea&     545&1.18\pm0.17&*1.61\pm0.24&*1.81\pm0.25&*2.11\pm0.15\cr
12\,Victoria&   857&      \dots&*1.11\pm0.23&       \dots&       \dots\cr
13\,Egeria&     857&      \dots&       \dots&       \dots&*1.17\pm0.16\cr
15\,Eunomia&    545&      \dots&       \dots&*1.7*\pm0.3*&       \dots\cr
16\,Psyche&     857&1.18\pm0.38&       \dots&       \dots&       \dots\cr
18\,Melpomene&  857&1.05\pm0.27&       \dots&       \dots&       \dots\cr
19\,Fortuna&    857&2.63\pm0.11&*1.47\pm0.21&*1.1*\pm0.2*&       \dots\cr
20\,Massalia&   857&      \dots&*0.93\pm0.22&       \dots&       \dots\cr
29\,Amphitrite& 857&      \dots&       \dots&*0.91\pm0.17&       \dots\cr
45\,Eugenia&    857&1.54\pm0.28&*1.21\pm0.19&       \dots&       \dots\cr
52\,Europa&     857&2.81\pm0.12&*3.38\pm0.26&*2.05\pm0.12&*1.57\pm0.11\cr
52\,Europa&     545&1.08\pm0.26&       \dots&       \dots&       \dots\cr
88\,Thisbe&     857&1.59\pm0.33&       \dots&       \dots&       \dots\cr
128\,Nemesis&   857&1.4*\pm0.3*&       \dots&       \dots&       \dots\cr
324\,Bamberga&  857&2.8*\pm0.1*&*1.79\pm0.13&       \dots&       \dots\cr
511\,Davida&    857&      \dots&       \dots&*1.07\pm0.32&       \dots\cr
704\,Interamnia&857&1.61\pm0.29&*1.42\pm0.17&*2.19\pm0.16&*1.88\pm0.15\cr
Christensen&    857&2.6*\pm0.2*&       \dots&       \dots&       \dots\cr

\noalign{\vskip 3pt\hrule\vskip 5pt}
}
}
\endPlancktablewide 
\endgroup
\end{table*}
 
 \subsection{Basic Behaviour}
 
\begin{figure*}[htbp]
   \centering
   \includegraphics{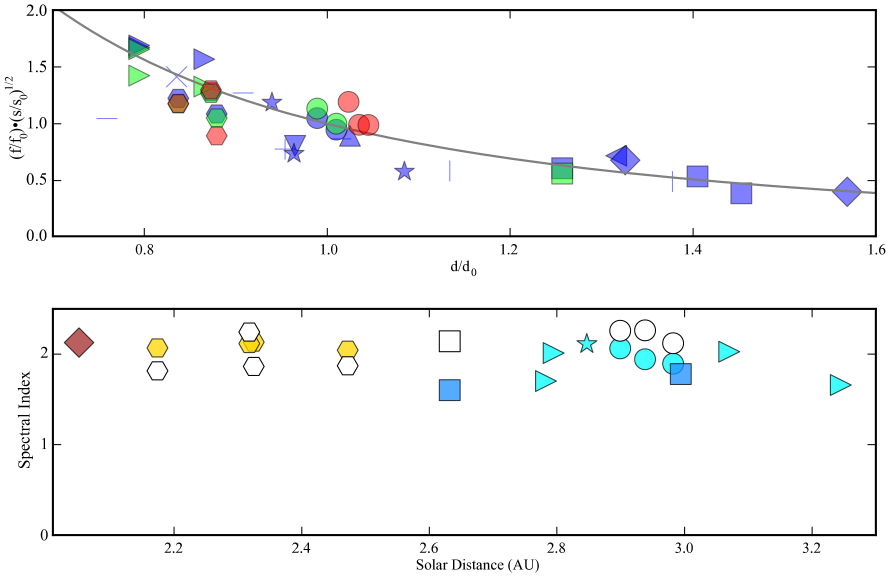}
   \caption{
    \textit{Top}: ratio of $f\,s^{1/2}$ of the first and subsequent measurements
    for those Solar System objects that were detected in more than one Survey. Measurements
    at 857\,GHz are shown in blue, measurements at 545\,GHz are shown in green, and measurements
    at 353\,GHz are shown in red. The grey line shows $\left(d/d_0\right)^{-2}$, which one would 
    expect for Rayleigh-Jeans objects. 
    \textit{Bottom}: spectral index for those asteroids detected in multiple frequency bands
    for a given Survey, defined as
    $\log\left(I_{545}/I_{857}\right)/\log\left(545/857\right)$. Juno and Vesta are shown in
    brown and yellow, to indicate that they have different spectral classifications than Ceres, Pallas,
    Hygiea and Europa, shown in shades of blue. The white symbols show the corresponding
    values for 353--545\,GHz, where they exist. 
    For both panels, the symbols used for each object are:
    1\,Ceres --- circles; 2\,Pallas --- squares; 3\,Juno --- diamonds; 4\,Vesta --- hexagons; 6\,Hebe --- +; 
    8\,Flora --- $\times$; 9\,Metis -- upward-pointing triangle; 10\,Hygiea -- right-pointing triangles; 
    19\,Fortuna -- vertical lines; 45\,Eugenia -- downward-pointing triangles; 52\,Europa -- stars; 
    324\,Bamberga -- left-pointing triangle; and 704\,Interamnia -- horizontal lines.}
   \label{fig:Asts}
  \end{figure*}

  Figure~\ref{fig:Asts} demonstrates the basics of the asteroid flux density measurements: $f$ is flux density, $d$ is the distance between \Planck\ and the object, and $s$ is the distance between the Sun and the object.  Zero subscripts identify the first measurement for a given asteroid.  The top panel shows asteroids detected in multiple Surveys.  Assuming the temperature of the object at any time goes as $s^{-1/2}$,
we expect 
  \begin{equation}
   f\,s^{1/2} \propto \frac{1}{d^2},
  \end{equation}
  which is roughly seen in the data.
   
  In the bottom panel, we show the 545-to-857\,GHz spectral indices, as well as the 
  545-to-353\,GHz spectral indices for Ceres and Vesta. 
  
 \raggedright 
\end{document}